\documentclass[aip,pof,amsmath, amssymb, groupedaddress, reprint]{revtex4-1}
\usepackage{comment}
\usepackage{amssymb,wasysym}

\usepackage{natbib}
\usepackage{graphicx}
\usepackage{dcolumn}
\usepackage{bm}
\usepackage{color}
\usepackage{subfigure}
\usepackage{url}
\definecolor{light-gray}{gray}{0.75}
\definecolor{dark-gray}{gray}{0.45}
\definecolor{mycopper}{RGB}{184,115,0} 
\definecolor{mygray}{RGB}{159,182,205} 
\font\smallfont=cmsy10 at 10truept
\mathchardef\bigCircle="280D

\font\bigfont=cmsy10 at 14.4truept
\mathchardef\tiMes="2902        %

\font\Bigfont=cmsy10 at 17.28truept
\mathchardef\DiaMond="2A05        %
\mathchardef\cirCle="2A0E
\mathchardef\BigCircle="2A0D

\font\Bbigfont=cmsy10 at 24.88truept
\mathchardef\buLLet="2B0F

\def\bigCirc{\raise 0.3ex\hbox{$\bigCircle$}\nobreak$\,$}
\def\Bullet{\raise-0.35ex\hbox{$\buLLet$}\nobreak$\,$}

\def\triangledown{\raise 0.2em\hbox{$\bigtriangledown$}\nobreak$\,$}

\def\minisquare{\hbox{${\vcenter{
               \hrule height 0.3pt \kern-0.4pt
               \hbox{\vrule width  0.3pt height 3.0pt \kern 2.6pt
               \vrule width  0.3pt height 3.0pt} \kern-0.4pt
               \hrule height 0.3pt}}$}}
\def\ssquare{\raise 0.175ex\hbox{${\vcenter{
               \hrule height 0.5truept       \kern-0.25truept
               \hbox{\vrule width 0.5truept height 3.0truept \kern 2.75truept
                     \vrule width 0.5truept height 3.0truept} \kern-0.25truept
               \hrule height 0.5truept}}$}\nobreak$\,$}
\def\squarex{\raise 0.175ex\hbox{${\vcenter{
               \hrule height 0.8truept       \kern-1.80truept
          \hbox{\vrule width 0.8truept height 8.0truept \kern-1.95truept
                \raise 0.8truept\hbox{$\tiMes$}     \kern-6.70truept
                \vrule width 0.8truept height 8.0truept} \kern-0.80truept
               \hrule height 0.8truept}}$}\nobreak$\,$}

\def\sqbull{\raise0.175ex\hbox{\vrule height 1.4ex width 1.6ex depth 0.2ex}\nobreak$\,$}
\def\smsqbull{\raise0.175ex\hbox{\vrule height 0.8ex width 0.9ex depth 0.2ex}\nobreak$\,$}

\def\Diamondplus{${\vcenter{\vcenter{\DiaMond} \kern-10truept
                            \hbox{\vrule width .4truept}\kern -3truept
                            \hrule height .4truept}}$\nobreak$\,$}
\newcount\ndots

\def\drawline#1#2{\raise 2.5truept\vbox{\hrule width #1truept height #2truept}}
\def\moonspace#1{\hskip #1truept}

\def\shortchain{\drawline{6.0}{0.75}}

\def\shortchainspace{\shortchain\moonspace{2}}

\def\Dashy{\drawline{4.00}{1.00}}     
\def\dashy{\drawline{4.00}{0.75}}     
\def\thindashy{\drawline{4.00}{0.25}}     
\def\dashyspace{\dashy\moonspace{2}}
\def\Dashyspace{\Dashy\moonspace{2}}
\def\thindashyspace{\thindashy\moonspace{2}}
\def\longdashy{\drawline{8.00}{0.75}} 
\def\thinlongdashy{\drawline{8.00}{0.25}} 
\def\longdashyspace{\longdashy\moonspace{2}}
\def\thinlongdashyspace{\thinlongdashy\moonspace{2}}
     
\def\dotty{\drawline{1.00}{0.75}}

\def\dottyspace{\dotty\moonspace{2}}

\def\Gcirc{\raise 0.15ex\hbox{$\circ$}\nobreak$\,$}
\def\Gdiamond{\raise 0.25ex\hbox{$\diamond$}\nobreak$\,$}
\def\Gsquare{\raise  0.05ex\hbox{$\square$}\nobreak$\,$}
\def\Gblacksquare{\raise  0.05ex\hbox{$\blacksquare$}\nobreak$\,$}
\def\Gtimes{\raise  0.25ex\hbox{$\times$}\nobreak$\,$}
\def\Gast{\raise  0.25ex\hbox{$\ast$}\nobreak$\,$}
\def\Gplus{\raise  0.00ex\hbox{$+$}\nobreak$\,$}
\def\Gbullet{\raise  0.20ex\hbox{$\bullet$}\nobreak$\,$}
\def\Gtriangleright{\raise  0.10ex\hbox{$\triangleright$}\nobreak$\,$}
\def\Gblacktriangleright{\raise  0.10ex\hbox{$\blacktriangleright$}\nobreak$\,$}
\def\Gsolid{\drawline{7}{0.75}\nobreak$\,$}  
\def\solid{\drawline{24}{0.75}\nobreak$\,$}

\def\dotbox{\hbox{\dottyspace}}

\def\dotted{\hbox{\leaders\dotbox\hskip 24truept}\nobreak$\,$}  
\def\dashbox{\hbox{\dashyspace}}  
\def\Dashbox{\hbox{\Dashyspace}}  
\def\dashed{\hbox {\ndots=0 \loop\ifnum\ndots<3\advance\ndots by 1
        \dashbox\repeat\dashy}\nobreak$\,$}       
\def\Dashed{\hbox {\ndots=0 \loop\ifnum\ndots<3\advance\ndots by 1
        \Dashbox\repeat\Dashy}\nobreak$\,$}       
\def\thindashbox{\hbox{\thindashyspace}}  
\def\thindashed{\hbox {\ndots=0 \loop\ifnum\ndots<3\advance\ndots by 1
        \thindashbox\repeat\thindashy}\nobreak$\,$}       
\def\thindash{\hbox {\ndots=0 \loop\ifnum\ndots<3\advance\ndots by 1
        \thindashbox\repeat\thindashy}\nobreak$\,$}       
\def\longdashbox{\hbox{\longdashyspace}}  
\def\thinlongdashbox{\hbox{\thinlongdashyspace}}  
\def\longdash{\hbox {\ndots=0 \loop\ifnum\ndots<3\advance\ndots by 1
        \longdashbox\repeat\longdashy}\nobreak$\,$}       
\def\thinlongdash{\hbox {\ndots=0 \loop\ifnum\ndots<3\advance\ndots by 1
        \thinlongdashbox\repeat\thinlongdashy}\nobreak$\,$}       
\def\dotdashed{\hbox{\shortchainspace\dottyspace\shortchain}\nobreak$\,$}      
\usepackage[normalem]{ulem}

\font\smallfont=cmsy10 at 10truept
\mathchardef\bigCircle="280D

\font\bigfont=cmsy10 at 14.4truept
\mathchardef\tiMes="2902        %

\font\Bigfont=cmsy10 at 17.28truept
\mathchardef\DiaMond="2A05        %
\mathchardef\cirCle="2A0E
\mathchardef\BigCircle="2A0D

\font\Bbigfont=cmsy10 at 24.88truept
\mathchardef\buLLet="2B0F

\newcount\ndots

\def\drawline#1#2{\raise 2.5truept\vbox{\hrule width #1truept height #2truept}}
\def\moonspace#1{\hskip #1truept}

\def\shortchain{\drawline{6.0}{0.75}}

\def\shortchainspace{\shortchain\moonspace{2}}

\def\Dashy{\drawline{4.00}{1.00}}     
\def\dashy{\drawline{4.00}{0.75}}     
\def\thindashy{\drawline{4.00}{0.25}}     
\def\dashyspace{\dashy\moonspace{2}}
\def\Dashyspace{\Dashy\moonspace{2}}
\def\thindashyspace{\thindashy\moonspace{2}}
\def\longdashy{\drawline{8.00}{0.75}} 
\def\thinlongdashy{\drawline{8.00}{0.25}} 
\def\longdashyspace{\longdashy\moonspace{2}}
\def\thinlongdashyspace{\thinlongdashy\moonspace{2}}
     
\def\dotty{\drawline{1.00}{0.75}}

\def\dottyspace{\dotty\moonspace{2}}

\def\solid{\drawline{24}{0.75}\nobreak$\,$}

\def\dotbox{\hbox{\dottyspace}}

\def\dotted{\hbox{\leaders\dotbox\hskip 24truept}\nobreak$\,$}  
\def\dashbox{\hbox{\dashyspace}}  
\def\Dashbox{\hbox{\Dashyspace}}  
\def\dashed{\hbox {\ndots=0 \loop\ifnum\ndots<3\advance\ndots by 1
        \dashbox\repeat\dashy}\nobreak$\,$}       
\def\Dashed{\hbox {\ndots=0 \loop\ifnum\ndots<3\advance\ndots by 1
        \Dashbox\repeat\Dashy}\nobreak$\,$}       
\def\thindashbox{\hbox{\thindashyspace}}  
\def\thindashed{\hbox {\ndots=0 \loop\ifnum\ndots<3\advance\ndots by 1
        \thindashbox\repeat\thindashy}\nobreak$\,$}       
\def\thindash{\hbox {\ndots=0 \loop\ifnum\ndots<3\advance\ndots by 1
        \thindashbox\repeat\thindashy}\nobreak$\,$}       
\def\longdashbox{\hbox{\longdashyspace}}  
\def\thinlongdashbox{\hbox{\thinlongdashyspace}}  
\def\longdash{\hbox {\ndots=0 \loop\ifnum\ndots<3\advance\ndots by 1
        \longdashbox\repeat\longdashy}\nobreak$\,$}       
\def\thinlongdash{\hbox {\ndots=0 \loop\ifnum\ndots<3\advance\ndots by 1
        \thinlongdashbox\repeat\thinlongdashy}\nobreak$\,$}       
\def\dotdashed{\hbox{\shortchainspace\dottyspace\shortchain}\nobreak$\,$}      
\newcommand{\beq}{\begin{eqnarray}}
\newcommand{\eeq}{\end{eqnarray}}

\begin{document}

%\preprint{APS/123-QED}

\title{Aspect ratio effect on particle transport in turbulent duct flows}
\author{A. Noorani}
\author{R. Vinuesa}
\author{L. Brandt}
\author{P. Schlatter}
\email{pschlatt@mech.kth.se}
\affiliation{Swedish e-Science Research Centre (SeRC) and Linn{\'e} FLOW Centre, KTH Mechanics, SE-100 44 Stockholm, Sweden}

\date{\today}

\begin{abstract}
The dynamics of dilute micron-sized spherical inertial particles in turbulent duct flows is studied by means of direct numerical simulations of the carrier phase turbulence with one-way coupled Lagrangian particles. The geometries are a square and a rectangular duct with width-to-height aspect ratio $AR$ of 3 operating at $Re_{\tau,c}=360$ (based on the centerplane friction velocity and duct half-height). \textcolor{black}{ The present study is designed to determine the effect of turbulence-driven secondary motion on the particle dynamics.} Our results show that a weak \textcolor{black}{ cross-flow secondary motion} significantly changes the cross-sectional map of the particle concentration, mean velocity and fluctuations. \textcolor{black}{ As the geometry of the duct is widened from $AR=1$ to 3, the secondary vortex on the horizontal wall significantly expands in the spanwise direction, and although the kinetic energy of the secondary flow increases close to the corner, it decays towards the duct centreplane in the $AR=3$ case so as the turbulent carrier phase approaches the behavior in spanwise-periodic channel flows, a fact that significantly affects the particle statistics. In the square duct the particle concentration in the viscous sublayer is maximum at the duct centreplane, whereas the maximum is found closer to the corner, at a distance of $|z/h|\approx1.25$ from the centreplane, in the $AR=3$ case. Interestingly the centreplane concentration in the rectangular duct is around 3 times lower than that in the square duct. Moreover, a second peak in the accumulation distribution is found right at the corners for both ducts. At this location the concentration increases with particle inertia. The secondary motion changes also the cross-stream map of the particle velocities significantly in comparison to the fluid flow statistics. These directly affect the particle velocity fluctuations such that multiple peaks appear near the duct walls for the particle streamwise and wall-normal velocity fluctuations.}
\end{abstract}

%\pacs{47.27.nb}
%47.27.nb, 47.27.ek, 47.27.er, 47.50.Ef, 47.80.Cb}

\maketitle

\section{Introduction}\label{sec:introduction}
The transport of micron-sized inertial particles in turbulent wall-bounded flows has a wide range of industrial, environmental and biological  applications. Droplets suspended in turbulent flow in an engine combustion chamber or pollen particles transported by the atmospheric boundary layer are some typical examples. These particle-laden flows are mainly characterised by two specific phenomena: \emph{preferential accumulation} or small-scale clustering and \emph{turbophoresis}. In the latter, heavy particles accumulate near the wall due to the gradients in the turbulent kinetic energy profile,\citep{reeks_1983,young_leeming_1997} while in the former, heavier-than-fluid particles aggregate in regions characterised by higher values of the turbulent kinetic energy dissipation rate and zero acceleration. Turbophoresis occurs in wall-dominated turbulence; however, preferential accumulation appears in both homogeneous and inhomogeneous particle-laden turbulent flows (see for instance \citet{balachandar_eaton_2010} or \citet{toschi_bodenschatz_2009} for a review). 

In the context of wall-bounded flows, particle dispersion has been mainly studied in canonical flows (channels, pipes and boundary layers) despite the fact that internal flows relevant to industrial applications often contain parts with geometrical complexities which promotes complex flow phenomena such as secondary flows, separation, \emph{etc}. It is therefore fundamental to investigate the particulate phase behaviour in more complex non-canonical geometries where a slight modification of the characteristics of the flow can induce a significant change in the vortical structures and consequently in particle clustering and near-wall turbophoresis. %The specific objective of this study is, therefore, to examine the effect of secondary motions on particulate dispersion in wall-bounded flows in comparison to the more studied canonical channel and pipe flows. 

Among the principal mechanisms driving secondary motions the \emph{lateral curvature} of the main flow and the \emph{gradients of the Reynolds stresses} are the most common. The former is generally found in arched geometries where the imbalance between the cross-stream pressure gradient and centrifugal forces causes a secondary motion. This skew-induced cross-flow motion (due to an inviscid process) is usually referred to as Prandtl's secondary motion of first kind. In the latter mechanism mean streamwise vortices are formed as a result of the local variation of Reynolds stresses. These are known as Prandtl's secondary motion of second kind (or stress-induced vortices) and are observed in, for instance, turbulent flows through straight ducts with noncircular cross-section. \citep[see][]{bradshaw_1987} The effect of the geometry-induced centrifugal forces on particle dynamics was recently addressed by Noorani et al.,\citep{noorani_etal_2014, noorani_etal_2015} who performed direct numerical simulations (DNS) of particle-laden turbulent flow in straight, mildly curved and strongly bent pipes. 

The focus of the present paper is therefore on the effect of turbulence-driven secondary motion on particle dynamics in turbulent duct flows with different width-to-height aspect ratios. The in-plane flow in turbulent ducts consists of four pairs of counter-rotating vortices (one on each quadrant of the duct). The magnitude of this cross-flow motion only amounts to around $2-3\%$ of the bulk velocity $u_{b}$, and its effect on particle transport is usually ignored in practical applications. The objective of the current study is to carefully quantify the impact of such small in-plane flow on the transport and dynamics of micron-sized heavy inertial solid spherical particles, which turns out to be significant.

The turbulent carrier phase through a square duct has been analysed in a number of numerical and experimental studies, which are reviewed in \citet{duct_jot_1,duct_jot_2} Limited number of studies also analysed the dispersion and deposition of heavy inertial particles in a turbulent square duct. \citet{winkler_rani_vanka_2004} studied the preferential concentration of inertial point particles in a downward square duct using large-eddy simulation (LES) of the fluid flow at $Re_\tau=180$ ($Re_{\tau}$ is defined in terms of the duct half-height $h$ and the mean friction velocity $u_{\tau}$). They assessed the usefulness of different vortex identification methods for analysing preferential accumulation of particles in four locations within the square duct cross-section. These authors also used the same geometry to study the wall-deposition of particles in various locations of the duct.\citep{winkler_rani_vanka_2006} They reported that deposition is least likely in the duct corner and most likely in the duct centre. The role of the secondary flows in the mixing and dispersion of one-way coupled inertial point particles in a turbulent square duct flow at $Re_\tau=180$ was investigated by \citet{sharma_phares_2006} and Phares and Sharma.\citep{phares_sharma_2006} Analysing statistics of single-particle and particle-pair dispersion, it was found that lateral mixing is increased for low-inertia particles due to the outward advective transport absent in the canonical straight pipe and channels flows. Heavy inertial particles accumulate close to the wall, and tend to mix more efficiently in the streamwise direction due to the fact that a large number of the particles spend more time in a region where the mean fluid velocity is smaller than the bulk flow. \citet{yao_fairweather_2010} studied particle re-suspension in a turbulent square duct flow by means of LES and Lagrangian particle tracking (LPT) of one-way coupled inertial particles. They reported that turbulence-driven secondary flows within the duct play an important role in the re-suspension process of particles. These authors also analysed the particle deposition under the same condition and their result confirmed the previous findings of \citet{winkler_rani_vanka_2006} and Sharma and Phares.\citep{sharma_phares_2006} Recently \citet{zhang_etal_2015} investigated the effect of collisions on the particle deposition in turbulent square duct flows. This group suggested that the inter-particle collisions increase the particle diffusion in the wall-normal direction, which makes the particles distribution more uniform close to the wall.

The flow through a rectangular duct of aspect ratio ${AR}=W/H$ (where $W$ is the total duct width and $H$ is the duct full height) adds an additional degree of complexity to the physics of wall-bounded turbulence, since the corner bisector symmetry is lost, which significantly influences the secondary flow pattern. It is known from studies on the carrier phase that the flow characteristics near the duct centerplane approach those of canonical spanwise-periodic turbulent channels when the aspect ratio increases. The response of the particulate phase dynamics to such changes in the aspect ratio of the duct has not been explored yet. The aim of this study is then to evaluate the Eulerian statistics of dispersed inertial micro-particles in square and non-square ducts and analyse the changes in solid phase dynamics when the duct becomes wider and approaches a canonical channel flow in the centerplane. To the authors' knowledge there are no studies in the literature regarding inertial point particles in non-square duct geometries, which makes the present study a first work to characterise the particle behaviour. The paper is organised as follows: in \S \ref{sec:method} we introduce computational methodology for the DNS and LPT which will be followed by a discussion of the basic flow features in \S \ref{sec:flow}.  The results and analysis of particulate phase data are presented in \S \ref{sec:part}. Finally, concluding remarks are given in \S \ref{sec:conclusions}.  

\section{Computational Methodology}\label{sec:method}
\subsection{Flow configuration and governing equations}
In the present study we consider two duct-flow cases: a square duct (${AR}=1$), and a rectangular duct with aspect ratio $3$. The $x$ coordinate denotes streamwise direction while $y$ and $z$ the vertical and spanwise wall-normal directions. The corresponding flow velocity components are $u_x$, $u_y$ and $u_z$. The streamwise length of the computational domain was chosen to be $L_{x}=25h$ in both cases. This length is long enough to capture the longest streamwise turbulent structures according to the pipe flow experiments by \citet{guala_et_al} and the DNSs of turbulent channel flows by \citet{jimenez_hoyas}. The same length was considered in our previous DNSs of straight pipe \citep{pipe}, curved pipe \citep{curved_pipe} and duct\citep{duct_jot_1,duct_jot_2} flows. Note that the flow is assumed to be fully-developed in the streamwise direction, and therefore periodicity is imposed in $x$. No-slip and no-penetration conditions are imposed at the four walls of the duct. The duct full height ($H$) is chosen to be $2h$ in the two cases, the length of the side walls (vertical walls). The duct width ($W$) is adjusted according to the aspect ratio under consideration. The dimension of the horizontal walls (bottom/top plates) is, therefore, $2h\times AR$. In this paper the vertical plane of symmetry is the horizontal wall bisector or duct centerplane ($z/h=0$), and the horizontal plane of symmetry indicates the vertical (side) wall bisector ($y/h=0$) in the rectangular geometry. In addition, a well-resolved channel flow DNS (with streamwise and spanwise periodicity) is performed in a standard domain of dimensions $25h \times 2h \times 10h$, with the same Cartesian coordinates as the ducts.  

For the carrier phase, the \textcolor{black}{ nondimensional} incompressible Navier--Stokes equations,
\begin{equation}\label{continuity_eq}
\nabla \cdot \mathbf{u}=0,
\end{equation}
\begin{equation}\label{momentum_eq}
\frac{\partial \mathbf{u}}{\partial t} + \left ( \mathbf{u} \cdot \nabla \right )  \mathbf{u}= - \nabla p + \frac{1}{Re_{b}} \nabla^{2} \mathbf{u},
\end{equation}
are integrated in time, where $p$ is the pressure, $\mathbf{u}$ is the carrier phase velocity vector and $Re_{b}=u_{b} h / \nu$ is the bulk Reynolds number, with $\nu$ and $u_b$ as the fluid kinematic viscosity and total bulk velocity in the duct cross-section.
 
\subsection{Numerical approach for the carrier phase}\label{sec:flowNumerical_approach} 
The spectral element code \texttt{nek5000}, developed by \citet{fischer_et_al} at the Argonne National Laboratory (ANL), is used in the present work to perform DNS of the two duct flows. This massively parallelised code is based on the spectral-element method (SEM), originally proposed by Patera,\citep{patera} in which the incompressible Navier--Stokes equations are discretised in space by means of the Galerkin approximation. In this study we considered the $\mathbb{P}_{N}-\mathbb{P}_{N-2}$ formulation by Maday and Patera,\citep{maday_patera} where the velocity is represented by Lagrange polynomials of order $N$, and the pressure by polynomials two degrees lower. The domain is divided into a number of hexahedral local spectral elements within which the solution is represented by the previously mentioned polynomials at the Gauss--Lobatto--Legendre (GLL) quadrature points. As concerns time integration, the nonlinear terms are treated explicitly by third-order extrapolation (EXT3), whereas the viscous terms are treated implicitly by a third-order backward differentiation scheme (BFD3) leading to a linear symmetric Stokes system for the basis coefficient vectors $\mathbf{u}^{n}$ and $p^{n}$ to be solved at every time step. To solve the final problem, velocity and pressure are decoupled and solved iteratively using conjugate gradient and the generalised minimal residual (GMRES) method. 

In order to simulate the flow, spectral elements were distributed uniformly along $x$, and the spacing was chosen so that $\Delta x^{+}_{max} < 10$. Here `+' denotes scaling with either the viscous length $\ell^{*}=\nu / u_{\tau,c}$, or the friction velocity $u_{\tau,c}$. Note that for the duct cases the viscous scaling is based on the friction velocity in the duct centerplane $u_{\tau,c}$. The mesh in the inhomogeneous directions was designed so that the maximum spacing $\Delta y ^{+}_{max}$ was below $5$ viscous units; moreover, we have 4 grid points below $y^{+}=1$, and 15 grid points below $y^{+}=10$. The spectral elements followed a Gauss--Lobatto--Chebyshev distribution close to the wall and they were uniformly distributed in the core region. The two distributions were blended respecting the two previous restrictions. In the channel flow simulation the spectral elements were uniformly distributed in the spanwise direction, and the spacing was such that $\Delta z^{+}_{max} < 5$. A summary of the details of the simulations, including number of spectral elements and grid-points, as well as mesh resolutions, is provided in table \ref{tab:tab1}.  

The carrier phase simulation procedure in terms of flow initialisation, tripping method and statistics is similar to the one described in \citet{duct_jot_1} As in the work by  Vinuesa et al.,\citep{duct_jot_1} we fix the conditions at the duct centreplane for increasing aspect ratios, so that it is possible to assess the effect of secondary flow and side-wall boundary layers on the flow in comparison to the corresponding channel data. Table \ref{tab:tab2} summarises the characteristics of both duct cases and the channel flow simulation. 

%%%%%%%%%%%%%%%%%%%%%%%%%%%%%%%%%%%%%%%%%%%%%%%%%%%%%%%%%%%%%%%%%%%
\begin{table} 
   \begin{center}
   \def~{\hphantom{0}}
\begin{tabular}{lccccc}
\hline
%\hline\noalign{\smallskip}
${  AR}$  & \# elements & \# grid points         & $\Delta x^+$   &  $\Delta y^+$ & $\Delta z^+$  \\[3pt]  
   \hline
$1$  &  $70\,848$       & $ 122.4  \times 10^{6}$ & $(1.99,9.91)$ &  $(0.15,4.67)$      & $(0.15,4.67)$ \\
$3$ &  $188\,928$      & $ 326.5  \times 10^{6}$ & $(2.03,10.07)$ &  $(0.16,4.75)$      & $(0.16,4.96)$ \\
$\infty$ &  $292\,248$      & $ 505.0  \times 10^{6}$ & $(2.01,9.98)$ &  $(0.15,4.12)$      & $(0.98,4.87)$ \\
%\noalign{\smallskip}\hline\end{tabular}
\hline
   \end{tabular}
\caption{Details of the present turbulent duct flow simulations. The channel flow simulation is denoted by $AR \rightarrow \infty$.} 
\label{tab:tab1} 
   \end{center}
\end{table}
%%%%%%%%%%%%%%%%%%%%%%%%%%%%%%%%%%%%%%%%%%%%%%%%%%%%%%%%%%%%%%%%%%%

%%%%%%%%%%%%%%%%%%%%%%%%%%%%%%%%%%%%%%%%%%%%%%%%%%%%%%%%%%%%%%%%%%%%%%%%%%
\begin{table}  
   \begin{center}
   \def~{\hphantom{0}}
   \begin{tabular}{lllccc}
\hline
 ${  AR}$ & $Re_{b,c}$ & $Re_{b}$ & $Re_{\tau,c}$ & $Re_{\tau}$ & Averaging time frame \\[3pt]
   \hline
 $1$ &$6253$ & $5693$   &  $357$    & $342$       &   $3000$\\
 $3$ &$6273$ & $5817$   &  $363$    & $336$       &   $3000$\\
 $\infty$ &$6240$ & $6240$   &  $360$    & $360$       &   $2200$\\
   \hline
   \end{tabular}
   \caption{Simulation parameters of the present study. While $Re_{b}=u_{b} h / \nu$ ($u_b$ being the total bulk velocity per cross-section), $Re_{b,c}$ is defined as $u_{b,c} h / \nu$ at the centreplane of the ducts where $u_{b,c}$ is the mean velocity (at $z/h=0$). The non-dimensional friction Reynolds number $Re_{\tau}$ is defined as $hu_{\tau}/\nu$ where the mean friction velocity $u_\tau$ is averaged over all the walls. The same variable in the centreplane of the duct is defined as $Re_{\tau,c}=hu_{\tau,c}/\nu$ where $u_{\tau,c}$ is the friction velocity of the wall bisector (horizontal wall bisector in the case of rectangular duct). The channel case is denoted with $AR \rightarrow \infty$. The time $t$ is normalised by $h/u_{b}$. }   
   \label{tab:tab2}
   \end{center}
\end{table}
%%%%%%%%%%%%%%%%%%%%%%%%%%%%%%%%%%%%%%%%%%%%%%%%%%%%%%%%%%%%%%%%%%%%%%%%%%%%

\subsection{Numerical approach for the particulate phase}\label{sec:partNumerical_approach} 
In the current research we are concerned with low volume fractions of spherical micro-size dilute heavier-than-fluid particles which are assumed to be smaller than the smallest spatial scales of the flow. The particulate phase is then described as point particles without considering feedback on the flow or inter-particle collisions (\emph{i.e.\ one-way} coupled). Considering large enough density ratios ($\rho_p/\rho_f \sim 10^3$; $\rho_p$ and $\rho_f$ being the particle and fluid density) the particles are only subjected to the non-linear Stokes drag. For the current investigation, the effect of gravitational acceleration is neglected to allow isolation of the effect of the turbulence-driven secondary cells on the particle behaviour. Hence the particle equations of motion read
\begin{eqnarray}
\frac{\mathrm{d} \boldsymbol{v}_p}{\mathrm{d} t} & = & \frac{\boldsymbol{u}(\boldsymbol{x}_p,t)-\boldsymbol{v}_p}{St_b} f(Re_p), \label{eq:part1}\\
\frac{\mathrm{d} \boldsymbol{x}_p}{\mathrm{d} t} & = & \boldsymbol{v}_p, \label{eq:part2}
\end{eqnarray}
where $\boldsymbol{x}_p$ is the particle position, $\boldsymbol{v}_p$ and $\boldsymbol{u}(\boldsymbol{x}_p,t)$ the particle velocity and the fluid velocity at the particle position. Equation (\ref{eq:part1}) expresses the particle acceleration ($\boldsymbol{a}_p$) according to the steady-state aerodynamic drag acting on the spherical solid particle in a uniform velocity field. The \emph{non-linear} correction term $f(Re_p)$ is due to the finite particle Reynolds number \citep{schiller_naumann_1933} and is expressed as $(1+0.15Re_p^{0.687})$. The instantaneous particle Reynolds number $Re_p$ is defined as $|\boldsymbol{v}_p-\boldsymbol{u}(\boldsymbol{x}_p,t)|d_p/\nu$. The bulk Stokes number $St_b$ is given as the ratio of particle relaxation time $\tau_p$ to the bulk flow residence time \textcolor{black}{($t_{b}=h/u_b$)}. The particle relaxation time $\tau_p=\rho_p d_p^2/(18\rho_f \nu)$ where $d_p$ is the particle diameter and finally the Stokes number in terms of viscous time-scale may be given as $St^+=\tau_p u_\tau^2 /\nu$. 

The fluid velocity at the particle position is evaluated using a spectrally-accurate interpolation scheme with an order of accuracy equal to the one of the spectral element method, {\it i.e.}, $N=11$ in the current simulations. The time integration was performed using the classical 3rd order multistep Adams--Bashforth (\emph{AB3}) scheme. \textcolor{black}{ As for the carrier phase solver, the time step ($\Delta t$) is fixed such that the Courant--Friedrichs--Lewy ($CFL$) condition is always below $0.5$, and the particle time step $\Delta t$ is set to be equal to that of the flow solver. Assessment of the maximum possible time step for the \emph{AB3} method and the relevant right-hand sides in the particle equations indicates that the mentioned choice of time step is always within the region of absolute stability of the \emph{AB3} scheme. Extensive validation and verification tests regarding the time-integration scheme of the particle-phase solver are reported by Noorani.\citep{noorani_thesis}} The wall-particle interaction is treated as purely \emph{elastic} collisions, \emph{i.e.}, the total kinetic energy is conserved in the collision process. Similar to the carrier phase, periodic boundary conditions were used for the dispersed phase in the axial direction (and spanwise direction of the channel case). This Lagrangian particle tracking (LPT) module was verified and validated for a classical particle-laden turbulent channel flow against  \citet{sardina_schlatter_brandt_picano_casciola_2012} and was used in the previous studies of \citet{noorani_etal_2014,noorani_etal_2015} The implementation, validation and verification of the LPT module are reported by \citet{noorani_lic}   

In the present study, seven particle populations with different $St_b$ are simulated. While the density ratio is fixed to $\rho_p/\rho_f\approx1000$, each particle population $St_b$ is defined by the radius of the particles. The bulk Stokes number of each population is fixed for the various configurations, which implies fixing the physical characteristics of particles (\emph{i.e.}\ the particle density and radius) when varying the aspect ratio of the duct. The $St_b$ of these populations are chosen to be identical to the particle-laden turbulent flow simulations of bent pipes by Noorani et al.,\citep{noorani_etal_2014,noorani_etal_2015} which makes it possible to compare the results directly the effect of the two different kinds of secondary motions on particle transport. The code names of the particle populations and their corresponding parameters are presented in table \ref{tab:tab3}. Hereinafter, \textit{heavier} or \textit{larger} particles are referred to as \textbf{Stp25} and \textbf{Stp50}, and \textit{smaller}, \textit{lighter} or \textit{less inertial} populations as \textbf{Stp1} and \textbf{Stp5}. The \textbf{Stp100} population is simulated in order to investigate the asymptotic ballistic behaviour of very heavy inertial particles. Note that the \textbf{Stp0} population corresponds to fluid tracers.
%%%%%%%%%%%%%%%%%%%%%%%%%%%%%%%%%%%%%%%%%%%%%%%%%%%%%%%%%%%%%%%%%%%%%%%%%%%%%%%%%%%%%%
\begin{table}
\begin{center}
\begin{tabular}{llllllll}
\hline
\emph{Case}     & $d_p/h$              & $St_b$     & $St^+_{channel}$ & $St^+_{(AR=1)}$ & $St^+_{(AR=3)}$ & \textcolor{black}{$Re_{p,max,channel}$} \\ [3pt]
\hline
\textbf{Stp0}   & n/a                   & $0.0$          & $0.0$      & $0.0$    & $0.0$ & $0.0$ \\
\textbf{Stp1}   & $ 3.727\times 10^{-4}$ &  $0.0451$  & $1.0$      & $0.9375$ & $0.8494$ & $0.02$ \\
\textbf{Stp5}   & $ 8.333\times 10^{-4}$ &  $0.2257$  & $5.0$      & $4.6915$ & $4.2510$ & $0.2$ \\
\textbf{Stp10}  & $ 1.179\times 10^{-3}$ &  $0.4514$  & $10.0$     & $9.3829$ & $8.5019$ & $0.5$ \\
\textbf{Stp25}  & $ 1.863\times 10^{-3}$ &  $1.1284$  & $25.0$     & $23.455$ & $21.253$ & $1.3$ \\
\textbf{Stp50}  & $ 2.635\times 10^{-3}$ &  $2.2569$  & $50.0$     & $46.912$ & $42.508$ & $2.2$ \\
\textbf{Stp100} & $ 3.727\times 10^{-3}$ &  $4.5139$  & $100.0$    & $93.827$ & $85.018$ & $3.6$ \\
\hline
\end{tabular}
\caption{Parameters for the particle populations: number of particles per population $N_p=1.28\times10^5$. The density ratio $\rho_p/\rho_f$ is fixed to $1000$ for all populations. The bulk Stokes number $St_b$ and particle diameter $d_p/h$ are fixed while the duct aspect ratio varies. \textcolor{black}{ The maximum particle Reynolds number in the turbulent channel is also reported for each population.} \textcolor{black}{ Note that n/a refers to not applicable, due to the fact that the \textbf{Stp0} population corresponds to fluid tracers, and that the } channel values are normalised for $Re_{\tau}=360$ and $\rho_p/\rho_f=1000$.}
\label{tab:tab3}
\end{center}
\end{table}
%%%%%%%%%%%%%%%%%%%%%%%%%%%%%%%%%%%%%%%%%%%%%%%%%%%%%%%%%%%%%%%%%%%%%%%%%%%%%%%%%%%%%%

In order to obtain Eulerian statistics from the Lagrangian particle data the computational domain is divided into wall-parallel slabs to form 2D cells in the duct cross-section. The slabs are distributed non-uniformly according to the function: $r=h(1-(\sinh \gamma \eta)/(\sinh \gamma)) $, where $\gamma=2$ is chosen as the stretching factor, and $\eta$ indicates a uniformly spaced grid. As a result, the binning is more resolved near the wall than near the the duct centreline. The particles are binned into these cells and averaged in time, streamwise direction and over all four quadrants to obtain the Eulerian 2D maps presented hereinafter. A total of $40\,000$ cells are used for the 2D representation of the statistics. The statistical analysis in the wall bisectors (vertical and horizontal planes of symmetry of the duct) is carried out with $100$ wall-parallel slabs (from the wall to the duct middle) with stretching factor of  $\gamma=2$ in the distribution function above and a constant cell-width of $10$ viscous units. Similarly, a coarser grid is used with the first near-wall cell width of $\Delta y=0.0185$ ($\approx 5$ wall units) and $1600$ equally distributed cells to cover the duct cross-sections. The profiles evaluated along the walls in the buffer layer and the viscus sublayer (\emph{e.g.}\ Fig.\ \ref{fig:fig10} \emph{(right)}) are obtained from the coarser binning. The Eulerian statistics of particulate phase in the channel are computed as for the wall bisector of the duct, with the difference that the spanwise homogeneity is exploited.

\section{Results and analysis}\label{sec:results}
The results of the carrier phase are briefly discussed first to characterise the general flow features of turbulence in square and rectangular duct geometries. This is mainly to provide a clear overview of the carrier phase turbulence before analysing the complex dynamics of the particulate phase. The main focus is then on characterising the Eulerian statistics of the particle phase. 

\subsection{Carrier phase}\label{sec:flow}
\begin{figure}
\centerline{
\includegraphics[width=0.263\textwidth]{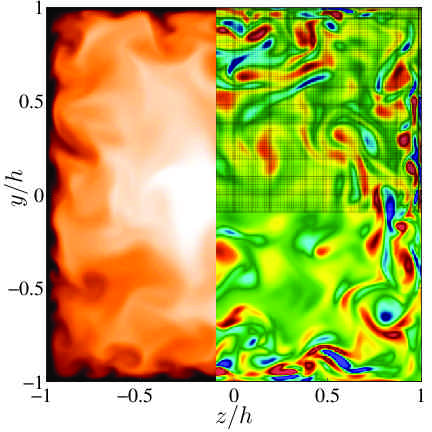}
\includegraphics[width=0.731\textwidth]{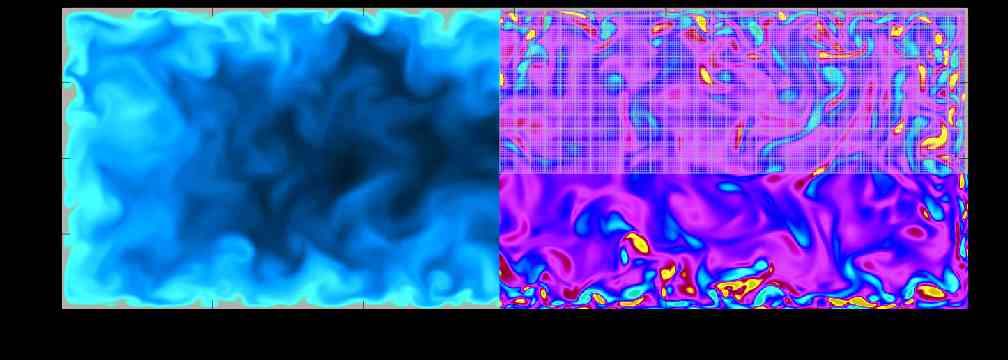}  
\put(-480,110){$(a)$}}
\centerline{\includegraphics*[width=0.99\textwidth]{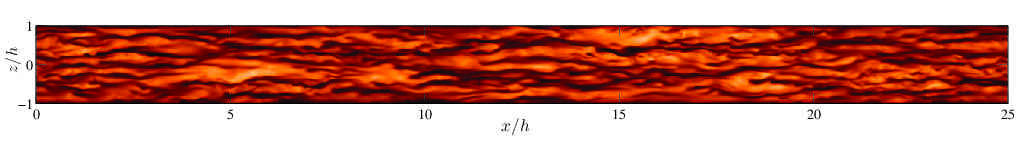}\put(-480,30){$(b)$}}
\centerline{\includegraphics*[width=0.99\textwidth]{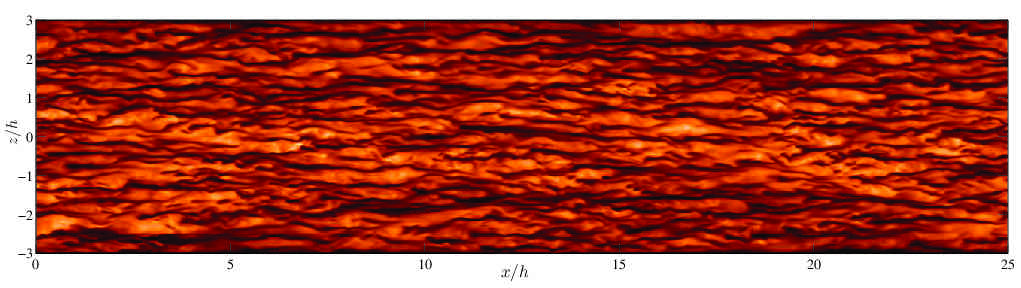}\put(-480,105){$(c)$}}
\caption{\emph{(a)} Pseudocolours of instantaneous streamwise velocity $u_x$ \emph{(left panel)} and instantaneous streamwise vorticity $\omega_x$ \emph{(right panel)} in a cross-section of the ducts with different aspect ratio, normalised with bulk values. Here, $\omega_x$ (scaled with $U_{b}$ and $h$) is plotted in the range of $(-7,7)$ with dark blue denoting minimum and dark red maximum values. For $u_x$ the colours vary from $0$ (black) to $1.3$ (white). A cross-sectional view of a quarter-section of the computational mesh is also shown. \emph{(b,c)} Top view of instantaneous streamwise velocity (at $y/h=-0.95$, $\approx 18^{+}$ from the wall) from the square and $AR=3$ ducts, respectively. In these two panels the flow direction is from left to right and \textcolor{black}{ the colours of $u_x$  vary from $0.25$ (black) to $1.3$ (white).}}
\label{fig:fig1}
\end{figure}

The level of detail obtained in the present simulations can be appreciated in Fig.\ \ref{fig:fig1}, where several instantaneous visualisations of the flow are shown. The cross-sectional view of the instantaneous streamwise velocity $u_x$ in both ducts (Fig.\ \ref{fig:fig1} \emph{a}) shows that the smallest turbulence structures close to the walls are properly resolved. As discussed by Huser and Biringen,\citep{huser_biringen} the corners are of special interest due to the interaction between the bursting events from the vertical and horizontal walls, which is the basic mechanism responsible for the appearance of the secondary flow. The instantaneous streamwise vorticity $\omega_{x}$, shown in the right panels, clearly displays the complexity of the turbulent vortices in the flow, as well as the emerging scale disparity even at this relatively low Reynolds number. Top views of the instantaneous streamwise velocity ($u_x$) are presented at $y/h=-0.95$ (a distance of $\approx 18^{+}$ from the wall) for the square and the $AR=3$ ducts in Fig.\ \ref{fig:fig1} \emph{(b,c)} to show the elongated near-wall streaky structures. Note that close to the corners the formation of streaks is reduced by the presence of the side wall. This was also observed at lower Reynolds numbers by \citet{pinelli_et_al} and \textcolor{black}{ \citet{duct_jot_1,duct_tsfp}}
%%%%%%%%%%%%%%%%%%%%%%%%%%%%%%%%%%%%%%%%%%%%%%%%%%%%%%%%%%%%%%%%%%%%%%%%%%%%%%%%%%%%%%
%%%%%%%%%%%%%%%%%%%%%%%%%%%%%%%%%%%%%%%%%%%%%%%%%%%%%%%%%%%%%%%%%%%%%%%%%%%%%%%%%%%%%%

\begin{figure}
  \centerline{\includegraphics*[width=0.32\textwidth]{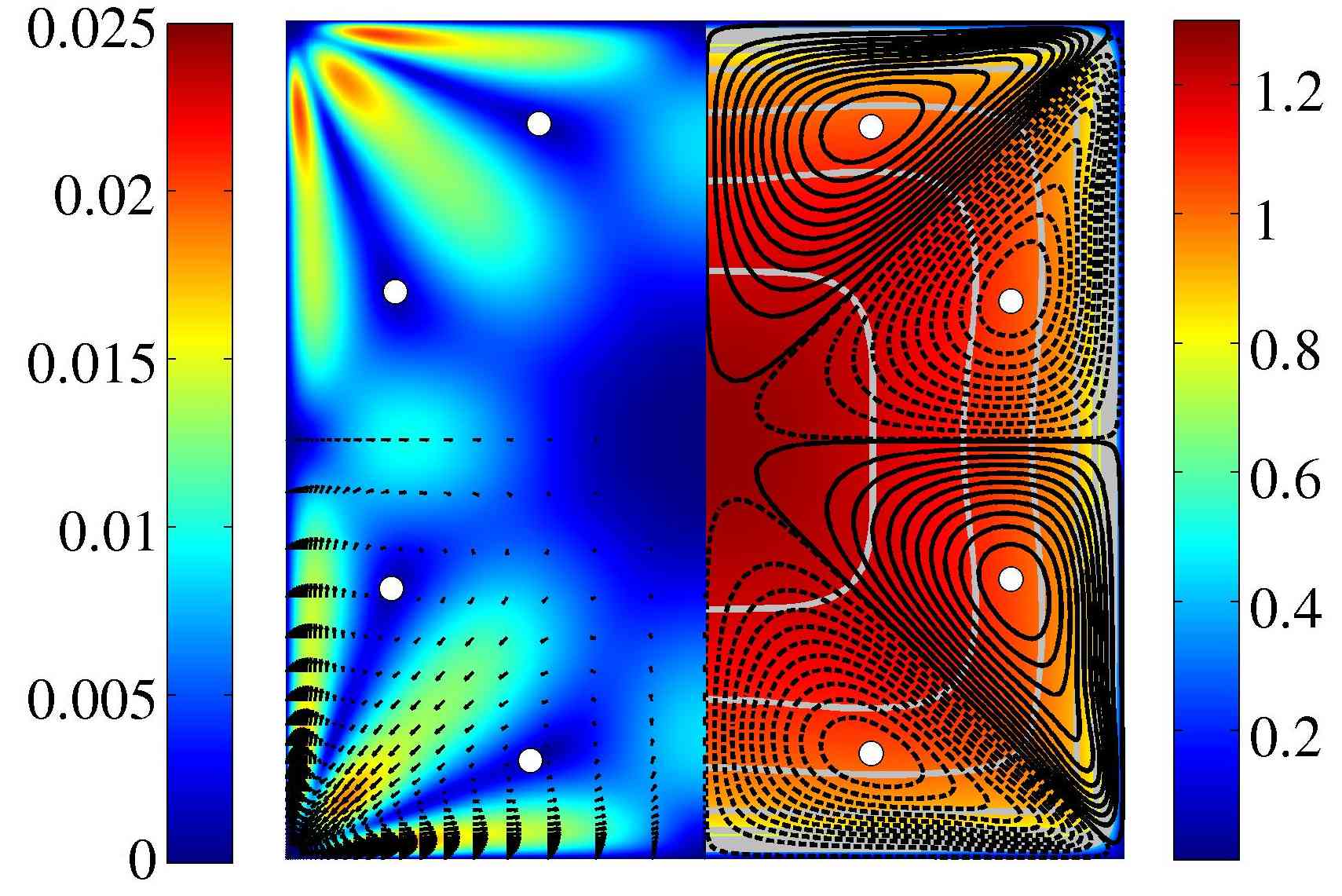}
              \includegraphics*[width=0.68\textwidth]{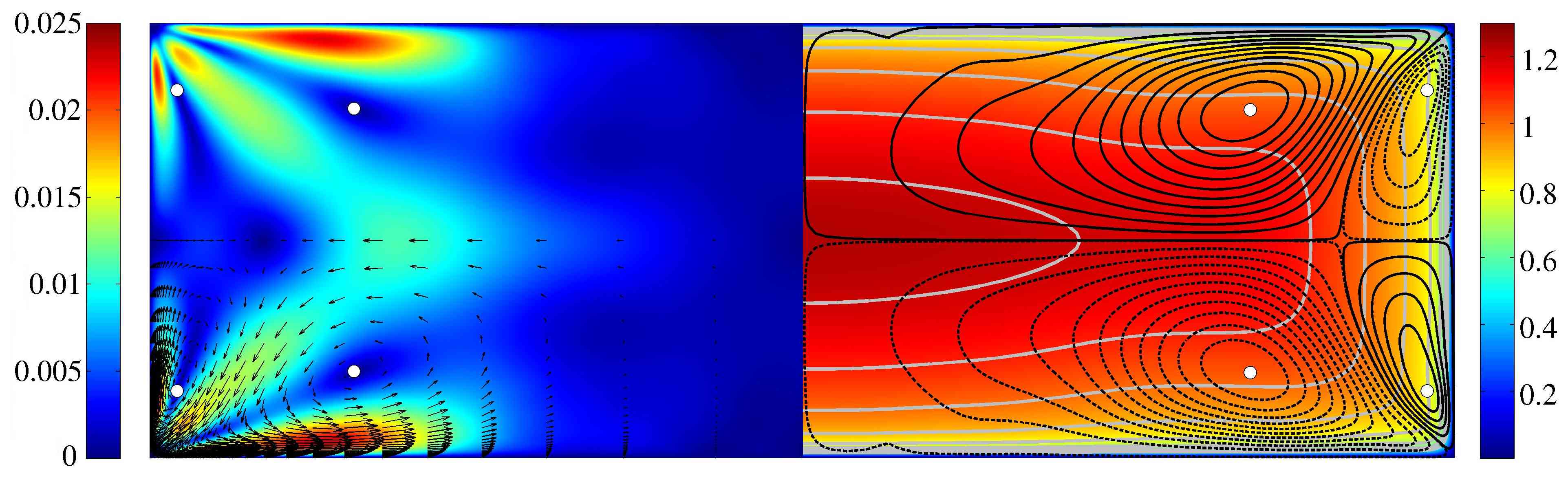}}
  \caption{(\emph{Left panel}) Pseudocolours of the magnitude of mean in-plane velocities of the carrier phase ($\sqrt(\langle u_{y} \rangle ^2+\langle u_{z}\rangle ^2)$) with vector plot in the two ducts under consideration. (\emph{Right panel}) Pseudocolours of the streamwise component of the mean fluid velocity and some of its isocontours \textcolor{black}{($\langle u_{x} \rangle \in [0.5,1.3]/0.1$)} in grey. Isocontours of the cross-flow stream function also projected on top where dashed lines represent negative values of $\psi$ (note that $(\psi_{min},\psi_{max})\times 10^{3}$ for the square duct is $(-2,2)$ and $(-1.5,4.5)$ for the duct with ${AR}=3$). White dots indicate the core of the secondary flow vortices. \textcolor{black}{ The velocity and length scales used in this figure are the bulk velocity $u_{b}$ and the duct half-height $h$, respectively.}}
\label{fig:fig2}
\end{figure}

Some selected statistics of the mean flow are shown in Fig.\ \ref{fig:fig2}. The left panel illustrates the magnitude of the in-plane velocity, $\sqrt(\langle u_{y} \rangle ^2+\langle u_{z}\rangle ^2)$, which represents the intensity of the secondary flow.  The mean over the streamwise direction and time is indicated by $\langle \rangle$. The vector plot shows the direction of the secondary flow. Such in-plane motion convects momentum from the core of the duct towards the duct bisectors, forming the characteristic eight vortex pattern which can also be observed in the stream function $\psi$; see plots in the right panel. While the magnitude of the secondary flow is largest adjacent to the walls and at the corner bisectors in the square duct, the portion of the secondary vortices near the longer side is modified when increasing the aspect ratio: these vortices stretch in the spanwise direction, and introduce an asymmetry in the flow. This asymmetry is reflected in the locations of the cores of the secondary vortices, as well as in the streamline pattern. Another manifestation of the effect of the secondary flow is shown by the isocontours of the streamwise velocity shown on the right panel of the figure. In the square duct case, the mean profile is lifted up at the centerplane (with the corresponding slight reduction in wall shear), and the isocontours are deformed towards the corner. The lack of symmetry in the rectangular duct leads to a slightly more complicated picture, although with a similar transfer of momentum from the core towards the corner of the duct.
%%%%%%%%%%%%%%%%%%%%%%%%%%%%%%%%%%%%%%%%%%%%%%%%%%%%%%%%%%%%%%%%%%%%%%%%%%%%%%%%%%%%%%
%%%%%%%%%%%%%%%%%%%%%%%%%%%%%%%%%%%%%%%%%%%%%%%%%%%%%%%%%%%%%%%%%%%%%%%%%%%%%%%%%%%%%%

\begin{figure}
  \centerline{\includegraphics*[width=0.3\textwidth]{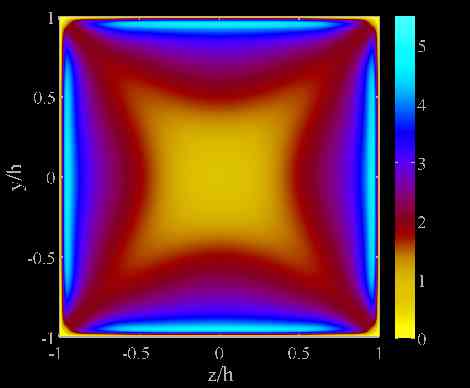}
              \includegraphics*[width=0.68\textwidth]{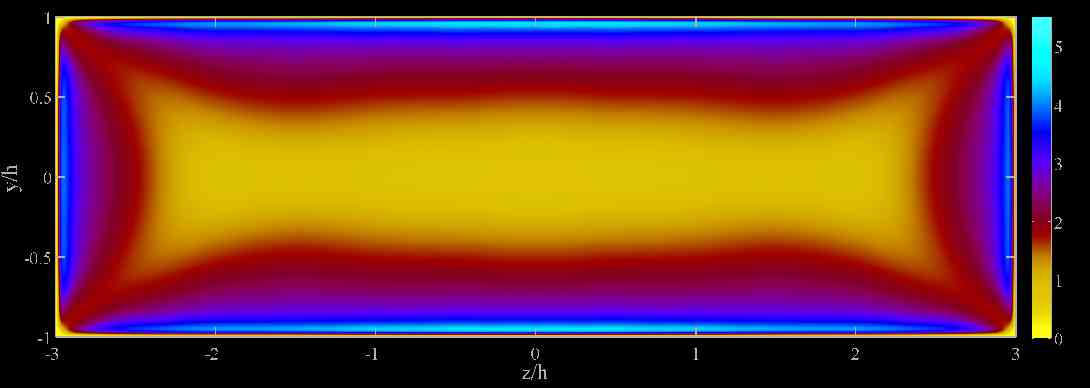}}
  \caption{Pseudocolours of turbulent kinetic energy $k$ of the carrier phase (normalised with the corresponding $u_{\tau,c}^2$) for each aspect ratio case.}
\label{fig:fig4}
\end{figure}

The turbulence kinetic energy (TKE) over the cross-sectional area is shown in Fig.\ \ref{fig:fig4}. As expected the maximum TKE is observed close to the four walls, which approximately corresponds to the location of the inner peak in the streamwise turbulence intensity, found at $y^{+} \simeq 15$. Note that the maximum TKE is slightly lower in the rectangular duct than in the square one. This can be explained by the fact that these plots are normalised with the corresponding centreplane friction velocity $u_{\tau,c}$, and as shown in table \ref{tab:tab2} the ${AR}=3$ case has slightly larger skin friction at $z/h=0$. It is also important to highlight that in the region around the duct corners the TKE is essentially zero, due to the inhibiting effect of the side wall on near-wall turbulence, also connected with the lack of streaks observed in this region in Fig.\ \ref{fig:fig1} \emph{(b,c)}. In addition, the TKE contours show low values along  the lines where the secondary vortices meet: in the case of the square duct, at the corner bisector, in the case of the ${  AR}=3$ this line has a different slope due to the deformation of the corner vortices. Note that the viscous diffusion process taking place along this line is produced by the interaction of the two corner vortices. \textcolor{black}{ The maximum kinetic energy of the secondary flow (defined as $K=1/2 \left ( \langle u_{y} \rangle ^2+\langle u_{z}\rangle ^2 \right )$ ) represents around $1.2\%$ of the maximum TKE in the square duct, and around $1.6\%$ in the $AR=3$ case. This shows that despite the relatively small magnitude of the secondary flow of second kind, its impact on the duct-flow physics is significant.}
%%%%%%%%%%%%%%%%%%%%%%%%%%%%%%%%%%%%%%%%%%%%%%%%%%%%%%%%%%%%%%%%%%%%%%%%%%%%%%%%%%%%%%
%%%%%%%%%%%%%%%%%%%%%%%%%%%%%%%%%%%%%%%%%%%%%%%%%%%%%%%%%%%%%%%%%%%%%%%%%%%%%%%%%%%%%%

\subsection{Particulate phase} \label{sec:part}
 \subsubsection{Temporal convergence}\label{sec:partConvergence}
\begin{figure}
  \centerline{\includegraphics*[width=0.5\textwidth]{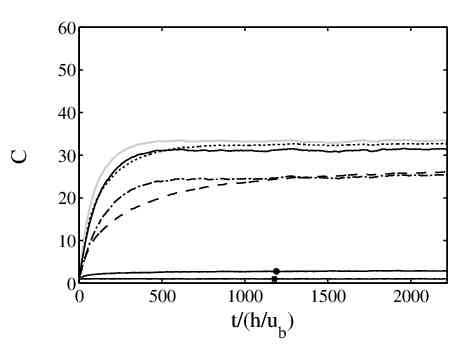}
              \includegraphics*[width=0.5\textwidth]{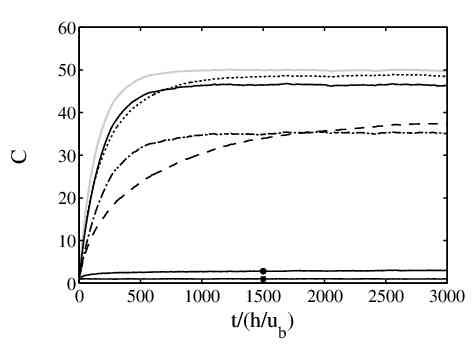}
              \put(-460,150){$(a)$}
              \put(-225,150){$(b)$}}
%  \centerline{\includegraphics*[width=0.5\textwidth]{convergence_R360AR3_Vwall.jpg}
%               \includegraphics*[width=0.5\textwidth]{convergence_R360AR3_Hwall.jpg}
%               \put(-378,120){$(c)$}
%               \put(-185,120){$(d)$}}
  \caption{Instantaneous particle concentration $C$ in the viscous sublayer ($y^+(z^+)\leq 5$) normalised with the mean particle concentration in the domain as a function of convection time: \emph{(a)} $AR=1$ \emph{(b)} $AR=3$ . ${\blacksquare}$ \textbf{Stp0}, $\bullet$ \textbf{Stp1}, \dashed \textbf{Stp5}, \dotted \textbf{Stp10}, {\color{light-gray} \solid} \textbf{Stp25}, \solid \textbf{Stp50}, \dotdashed  \textbf{Stp100}.}
\label{fig:fig5}
\end{figure}

The Lagrangian particle tracking is initialised by introducing the particle populations randomly into the fully developed turbulent flow with the initial velocity set equal to the local flow velocity at the particle positions. The instantaneous particle concentration $C$, defined as the number of particles per unit volume, normalised with the mean (bulk) concentration, is monitored in the viscous sublayer of all the walls (\emph{i.e.}\ $y (z) \leq 0.01415$ or $y^{+} (z^{+}) \leq 5$). Doing so, it is possible to identify when the particle dispersion reaches a statistically stationary stage. Usually, after injecting particles in canonical \textcolor{black}{ wall turbulence} the near-wall concentration increases due to turbophoresis until the statistically-steady state. Drifting particles with small momentum engage in a slow and diffusive process before segregating near the wall. According to \citet{sikovsky_2014} the time for this process is almost inversely proportional to the particle Stokes number, which makes the current observable (instantaneous $C$ in the viscous sublayer) a strict but suitable indicator for the steady state. The evolution of this indicator versus the convective time of the simulation is illustrated in Fig.\ \ref{fig:fig5}. For both duct flows, after an initial increase in $C$, its value levels off at $t>500h/u_b$ for large particles. This time is considerably longer for \textbf{Stp5} as the particle distribution reaches a steady state at about $1500h/u_b$ in the square duct and $2000h/u_b$ in the rectangular duct. %From figures \ref{fig:fig5} \emph{(a,b)} it appears that the integral concentration in the viscous sublayer all over the walls of the rectangular duct is $1.5$ times larger than in the square duct for almost all the populations. This will be further explained in \S \ref{sec:partConcentration}. 
%%%%%%%%%%%%%%%%%%%%%%%%%%%%%%%%%%%%%%%%%%%%%%%%%%%%%%%%%%%%%%%%%%%%%%%%%%%%%%%%%%%%%%
%%%%%%%%%%%%%%%%%%%%%%%%%%%%%%%%%%%%%%%%%%%%%%%%%%%%%%%%%%%%%%%%%%%%%%%%%%%%%%%%%%%%%%\emph{(c)} $AR=3$ only side plates \emph{(d)} $AR=3$ only bottom/top plates
% Also in comparison with the square duct the process takes relatively longer to reach a steady state in the rectangular geometry for almost all populations. The evolution of concentration in the viscous sublayer of horizontal and vertical plates of rectangular duct in time are separately plotted in figures \ref{fig:fig5} \emph{(c,d)}. This shows that the concentration near the horizontal walls are very similar to the overall integration values; however, the values associated with the viscous sublayer of side walls in rectangular geometry shows the same steady state behaviour as in square duct. 
% This rules out the wider geometry reasoning and shows that the process is further non-trivial. As One argument could be due to the geometrical difference. The rectangular duct is wider and provided that a the turbophoresis driving particles towards the side walls, this process may take a longer time in the $AR=3$ duct compared to the square geometry. Therefore, 

\subsubsection{Instantaneous particle distribution}\label{sec:partInstantDist}
\begin{figure}
  \centerline{\includegraphics*[width=0.21\textwidth]{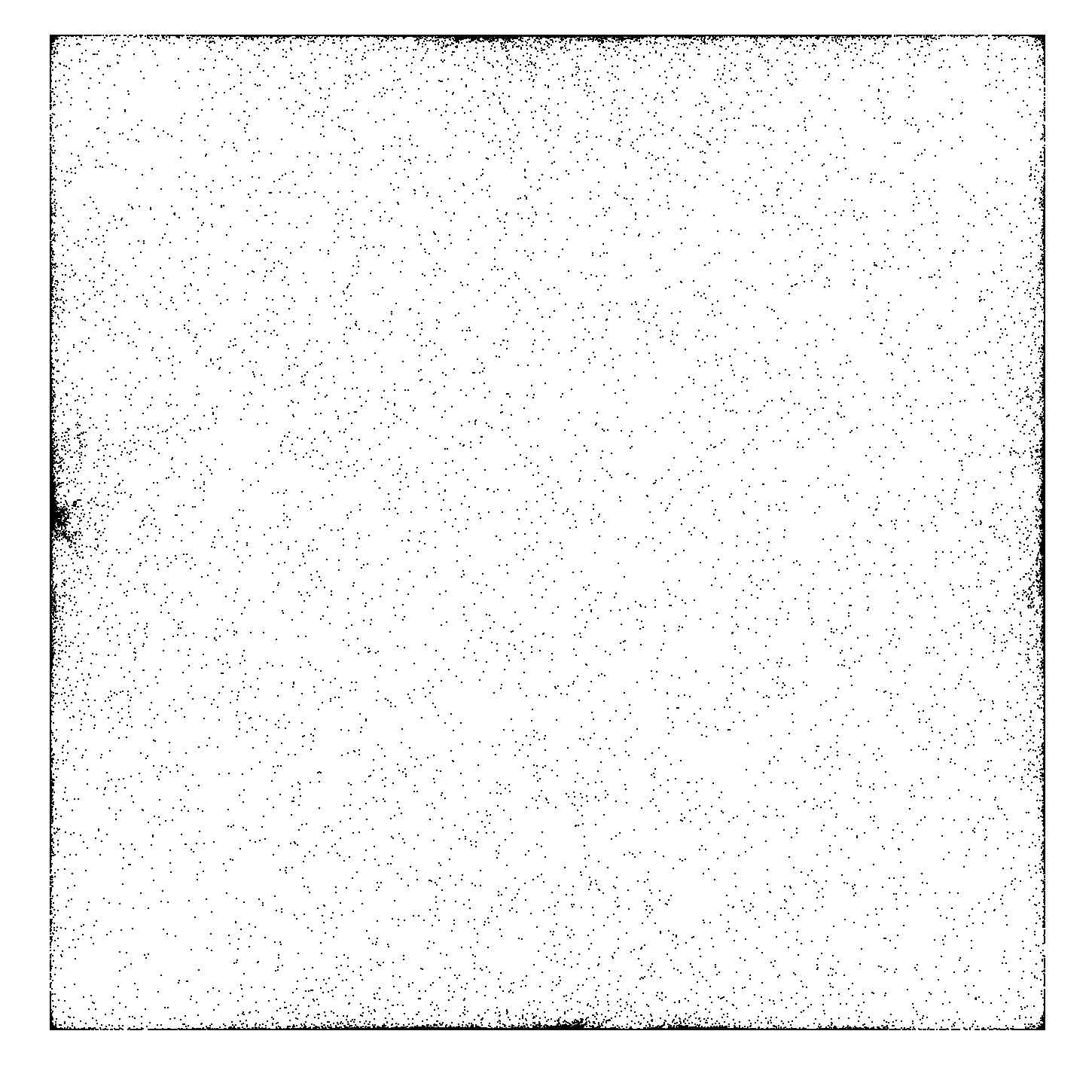}
              \includegraphics*[width=0.6\textwidth]{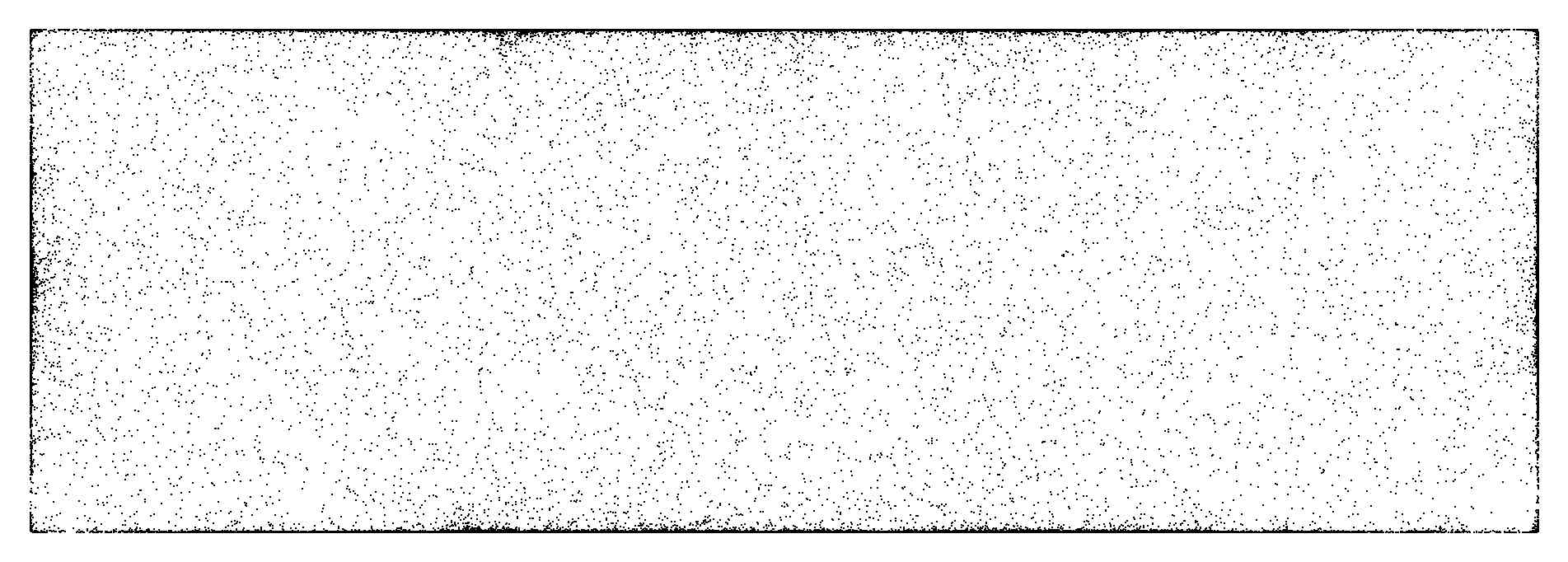}}
  \centerline{\includegraphics*[width=0.750\textwidth]{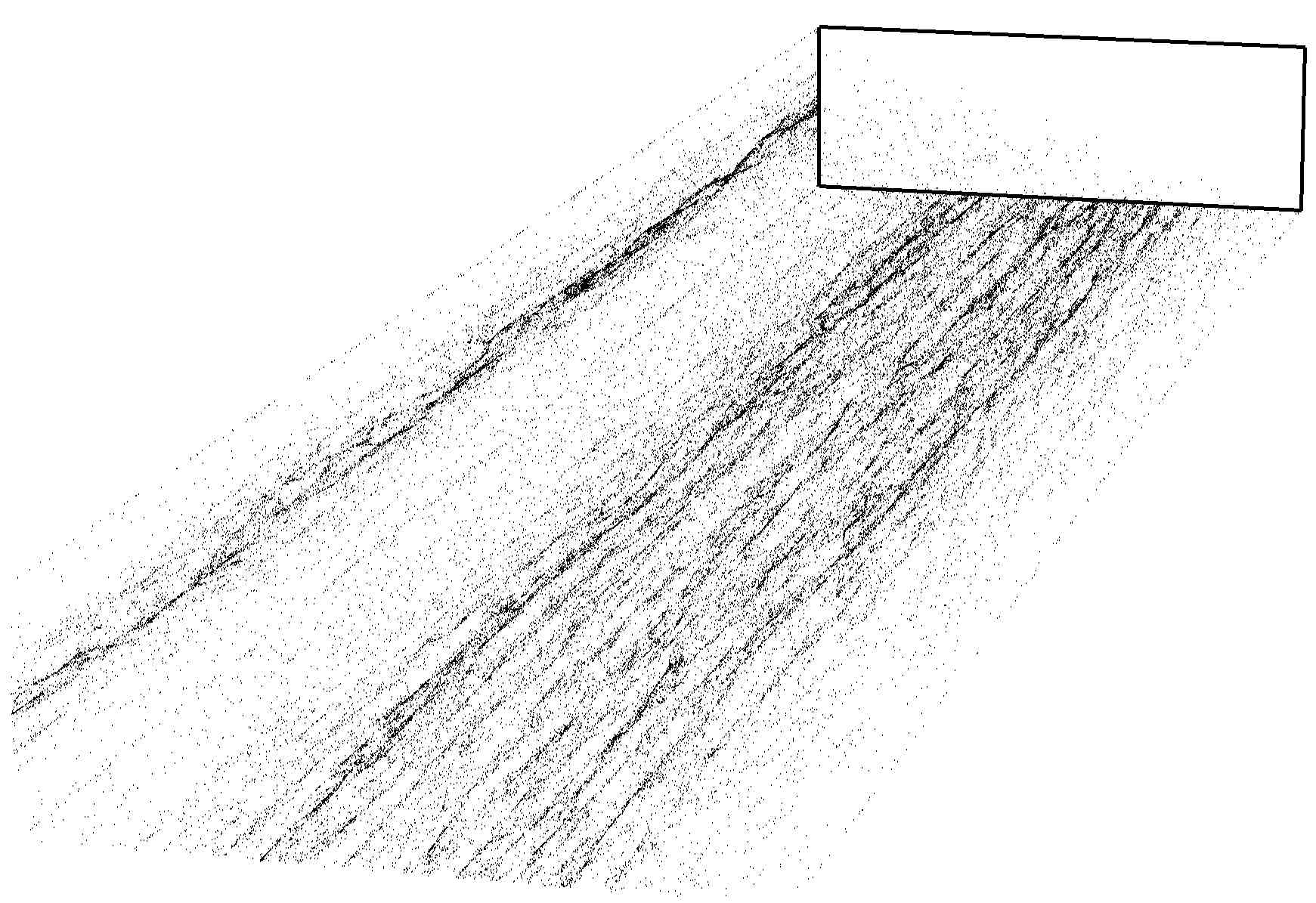}}
	\begin{picture}(0,0)
\put(130,60){\vector(2,3){20}}
\put(130,60){\vector(0,1){36}}
\put(130,60){\vector(1,0){36}}
\put(152,66){$z/h$}
\put(110,90){$y/h$}
\put(152,87){$x/h$}
\put(-5,258){\thicklines \circle{10}}
\put(-84,302){\thicklines \circle{10}}
\put(-193,49){\vector(4,3){20}}
\put(-123,-1){\vector(3,4){15}}
\end{picture}
  \caption{\emph{(Top)} Projected instantaneous front views of particles (\textbf{Stp25}) distributed in square duct and rectangular duct with $AR=3$. \emph{(Bottom)} A cut-away view of the particles (\textbf{Stp25}) distributed in one half of the rectangular duct (beneath the diagonal from top left to bottom right corner). \textcolor{black}{ Circles and arrows indicate sample particle clusters, and the duct frame of reference is also shown.}}
\label{fig:fig6}
\end{figure}
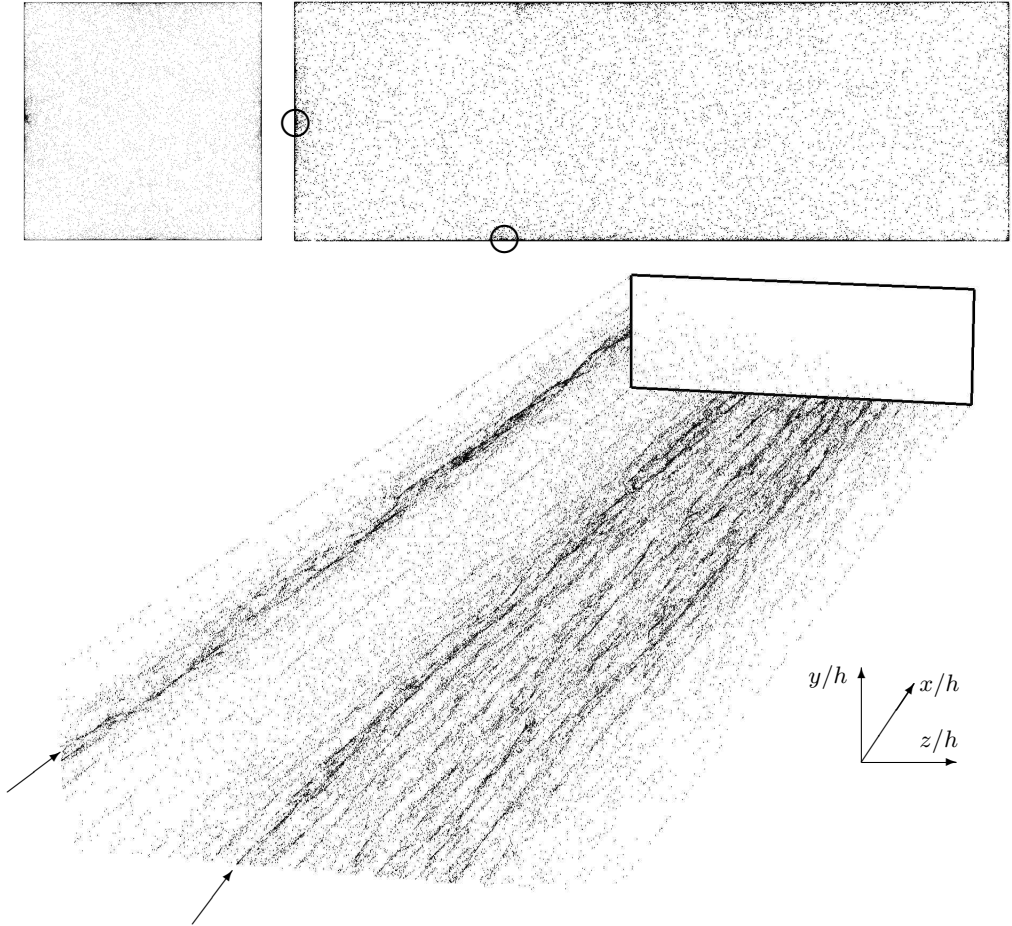

Fig.\ \ref{fig:fig6} displays the instantaneous particle distribution of the \textbf{Stp25} population for both configurations at the final time of the simulations. From the projected front view it can be clearly seen that the particles are largely segregated near the wall and have generated clusters. These streaky structures are due to the preferential localisation of particles with regards to near-wall carrier phase dynamics.\citep{marchioli_soldati_2002,sardina_schlatter_brandt_picano_casciola_2012} Clearly the particle clusters are not distributed homogeneously over the walls as less particles are found near the corner. This is more obvious from the cut-away view of the rectangular duct (Fig.\ \ref{fig:fig6} \emph{bottom}) where the particle streaks are mostly near the side-wall centre while hardly any clusters can be found near the corners. Visually, in terms of the particle aggregation, the centre region of the bottom plate ($AR=3$) is very similar to a channel; however, half-way towards the corners a larger amount of particle streaks is visible. A long thin and almost continuous streak of particles is also obvious right at the corners of the duct. These are the particles that are deposited at the corner by the secondary motion and are trapped due to the weakness of the secondary motions and flow perturbations in the corner region (\emph{c.f.}\ \citet{yao_fairweather_zhao_2014} and Fig.\ $11$ therein).  
%%%%%%%%%%%%%%%%%%%%%%%%%%%%%%%%%%%%%%%%%%%%%%%%%%%%%%%%%%%%%%%%%%%%%%%%%%%%%%%%%%%%%%
%%%%%%%%%%%%%%%%%%%%%%%%%%%%%%%%%%%%%%%%%%%%%%%%%%%%%%%%%%%%%%%%%%%%%%%%%%%%%%%%%%%%%%
 
Provided that the magnitude of the secondary motion in a duct is similar to a weakly curved pipe, the distribution of particle streaks on the duct wall is directly comparable to that in the mildly curved pipe (\emph{c.f.}\ Fig.\ $11$ \emph{(b)} in \citet{noorani_etal_2014}). The large accumulation of particle streaks near the saddle point of the secondary flow vortices on the wall bisector of the square duct and the side walls of the rectangular geometry is very similar to that near the stagnation point of the mean Dean cells in the inner side of the mildly curved pipe. Note that the TKE of the carrier phase is minimum in the stagnation point of mean Dean cells in the inner bend; however, in a duct, the TKE of the carrier phase is maximum in the stagnation point of secondary flow vortices on the wall bisector. Hence, the appearance of  such large particle segregation is mainly due to the effect of mean in-plane secondary motion cells.

The large-scale organisation of the particle streaks that were previously observed in channel flow by \citet{sardina_schlatter_brandt_picano_casciola_2012} is also clearly visible in the duct simulations. These large-scale motions modulate the smaller-scale accumulation and are of the order of the duct half-height $h$. These structures, however, are absent in the curved configurations where the particle streaks are rather elongated in the axial direction of the pipe inclined uniformly in the azimuthal direction to point towards the inner bend to generate helicoidal shapes. In fact, the streaks in the bent pipe form \emph{fish-bone} like structures and the inclination of these helices appears to be directly connected to the magnitude of in-plane motion of the carrier phase in bends. However, that inclination in the direction of the secondary flow was not found in the duct geometry. This could be related to the relatively weaker amplitude of the secondary flow in the duct, and the more restricted spatial extent of the region with significant amplitudes of the cross-flow components.

\subsubsection{Particle trajectories}\label{sec:partTrajectories}
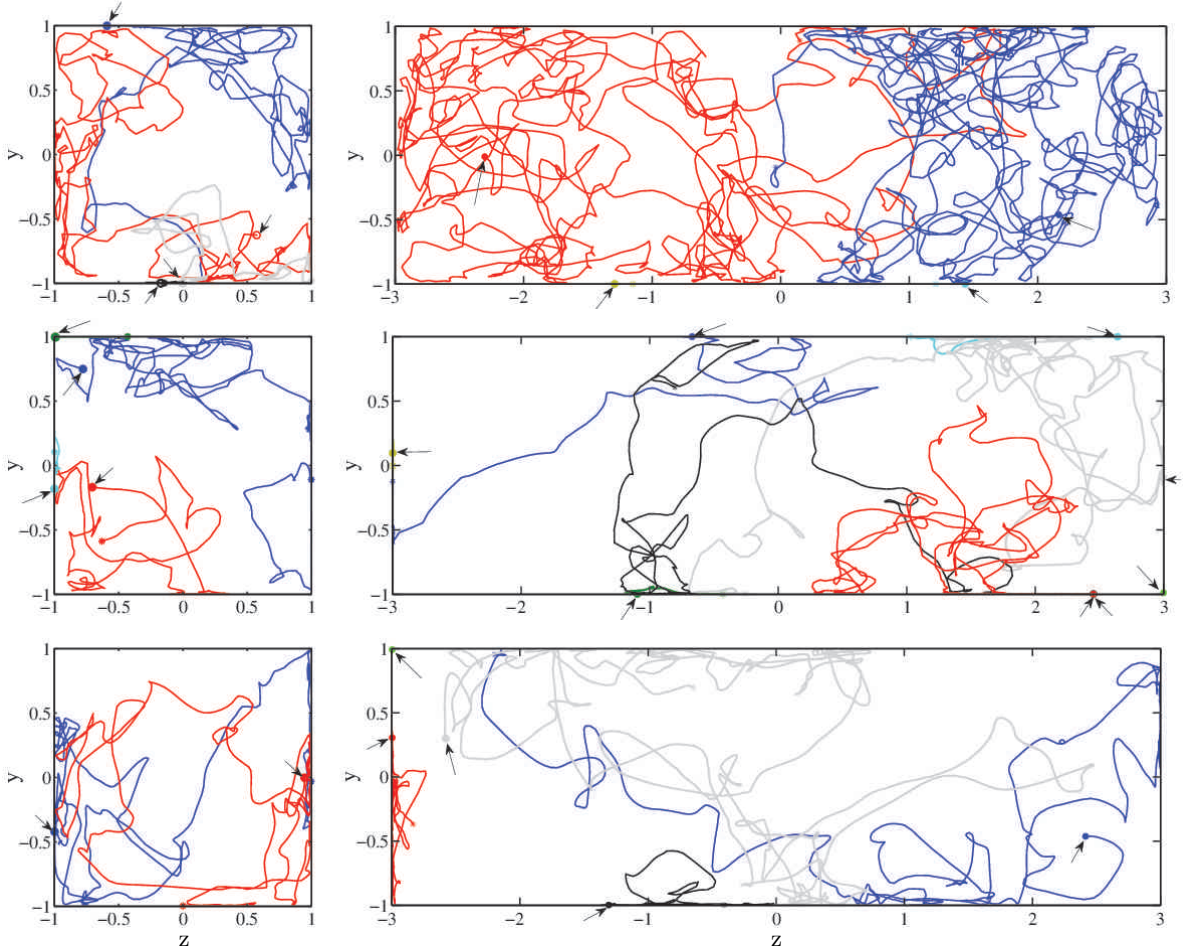
\begin{figure}
  \centerline{\includegraphics*[width=0.98\textwidth]{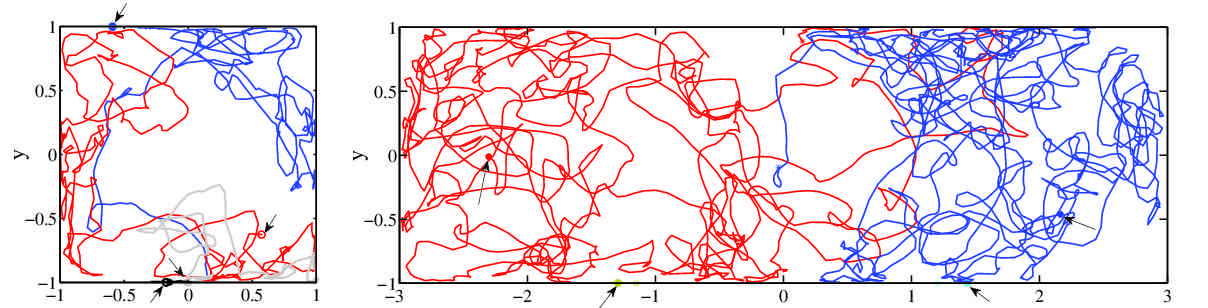}}
  \centerline{\includegraphics*[width=0.98\linewidth]{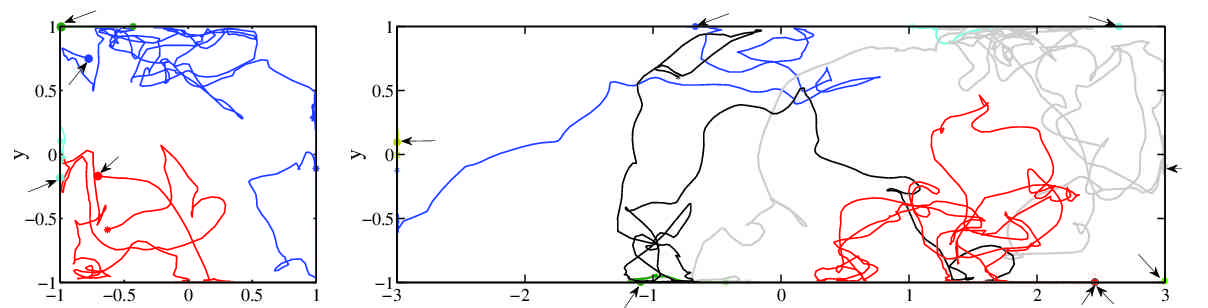}}
  \centerline{\includegraphics*[width=0.98\textwidth]{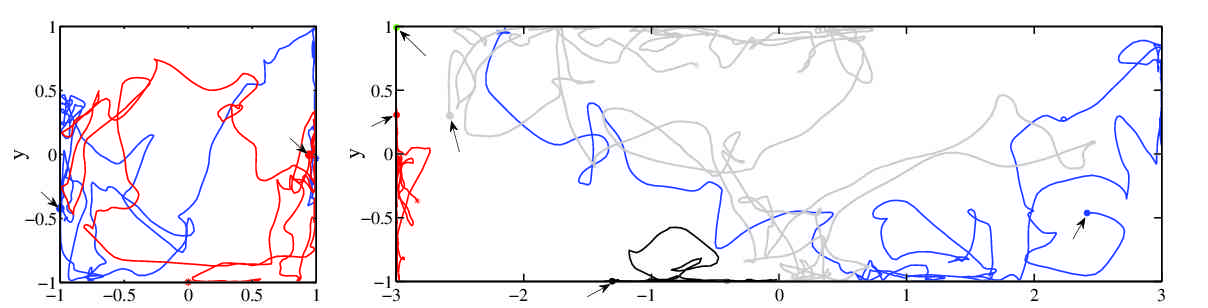}}
	  	\begin{picture}(0,0)
\put(66,5){z}
\put(-158,5){z}
	\end{picture}
  \vspace{.2 cm}
\caption{Some typical particle trajectories projected in the streamwise direction of square duct \emph{(left)} and  rectangular duct with $AR=3$ \emph{(right)}. \emph{From top:} \textbf{Stp5}, \textbf{Stp25} and \textbf{Stp100} population. The $\longrightarrow \bullet$ indicates the beginning of the track and $\ast$ shows the final time of the trajectory. Different colours indicate different particle tracks. All trajectories span a time interval of $1500$ convective units.}
\label{fig:fig7}
\end{figure}

Some typical trajectories of \textbf{Stp5}, \textbf{Stp25} and \textbf{Stp100} particles within the last $1500h/u_b$ convection time units of the simulation are shown in Fig.\ \ref{fig:fig7}. For both ducts, the lighter particles spend longer time away from the walls. They are erratically modulated by the flow turbulence unless they are trapped close to the wall. The trajectories of heavier particles indicate that the ones that enter the vicinity of a corner are almost immobilised. The particles trapped in the wall viscous sublayer also remain in this region for a long period. Even though the common assumption is that these inertial particles sample one single low-speed streak each, they are transported from the corner towards the wall bisector \emph{slowly} by the cross-flow motion. These large particles in the duct core are re-suspended by ejection towards the duct centre region. Heavier particles are also modulated by the turbulence; however, they roughly follow the mean secondary vortex performing a spiralling motion through the duct. The particles follow the mean vortical cells and as soon as they are ejected to the duct centre they are absorbed by another cell. Interestingly, most of the particles in the square duct and near the vertical plates of the rectangular duct eject near the plane of symmetry, whereas heavy particles near the horizontal plate of the rectangular duct typically are re-suspended before reaching the bisector where the secondary motion is most prominent.
%%%%%%%%%%%%%%%%%%%%%%%%%%%%%%%%%%%%%%%%%%%%%%%%%%%%%%%%%%%%%%%%%%%%%%%%%%%%%%%%%%%%%%

\subsubsection{Concentration statistics}\label{sec:partConcentration}
\begin{figure}
  \centerline{\includegraphics*[width=0.28\textwidth]{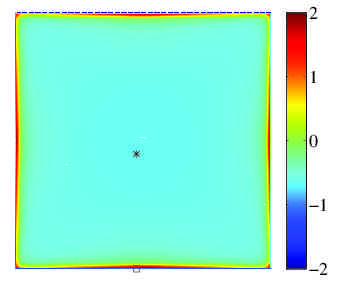}\put(-140,100){$(a)$}
              \includegraphics*[width=0.71\textwidth]{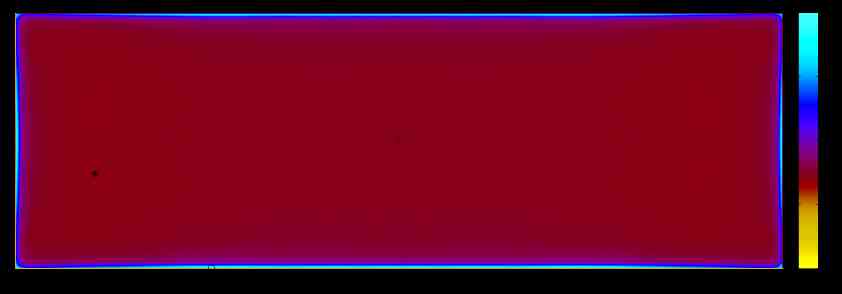}\put(-338,100){$(b)$}}
  \centerline{\includegraphics*[width=0.28\linewidth]{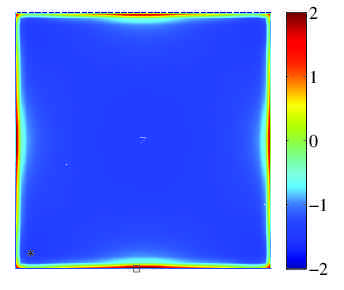}\put(-140,100){$(c)$}
              \includegraphics*[width=0.71\linewidth]{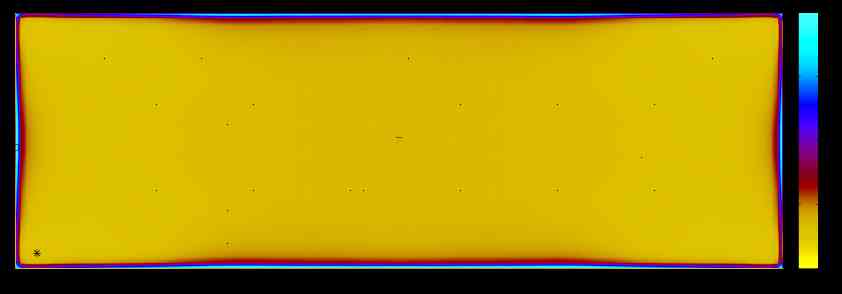}\put(-338,100){$(d)$}}
  \centerline{\includegraphics*[width=0.28\linewidth]{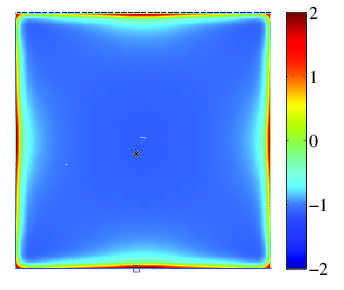}\put(-140,100){$(c)$}
              \includegraphics*[width=0.71\linewidth]{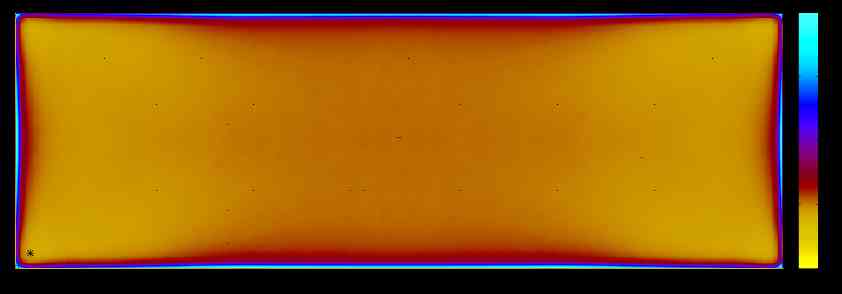}\put(-338,100){$(d)$}}
  \centerline{\includegraphics*[width=0.28\textwidth]{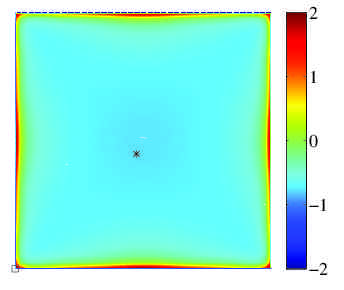}\put(-140,100){$(e)$}
              \includegraphics*[width=0.71\textwidth]{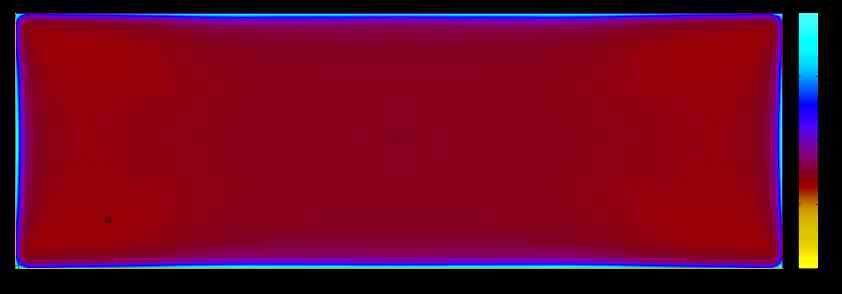}\put(-338,100){$(f)$}}
\caption{Logarithmic representation ($\mathrm{log}_{10}(C)$) of the mean normalised concentration map of particle dispersion in square and rectangular duct with $AR=3$. From \emph{top:} \textbf{Stp5}, \textbf{Stp25}, \textcolor{black}{ \textbf{Stp50}} and \textbf{Stp100}. \textcolor{black}{ The maximum and minimum particle concentrations are indicated with $\Box$ and $\ast$ respectively.}}
\label{fig:fig9}
\end{figure}

\begin{figure} 
   \begin{center}
    \centerline{\includegraphics*[width=0.43\textwidth]{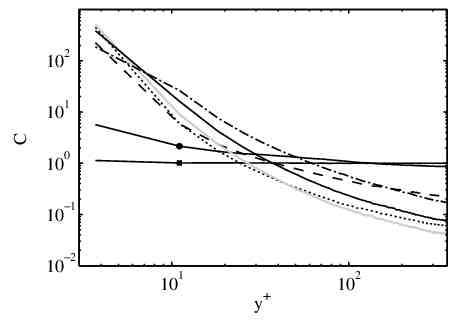}\put(-195,130){$(a)$}
                \includegraphics*[width=0.43\textwidth]{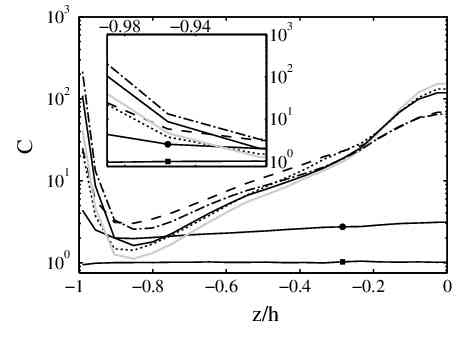}\put(-195,130){$(e)$}}
    \centerline{\includegraphics*[width=0.43\textwidth]{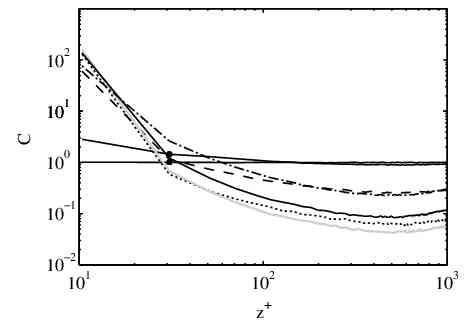}\put(-195,130){$(b)$}
                \includegraphics*[width=0.43\textwidth]{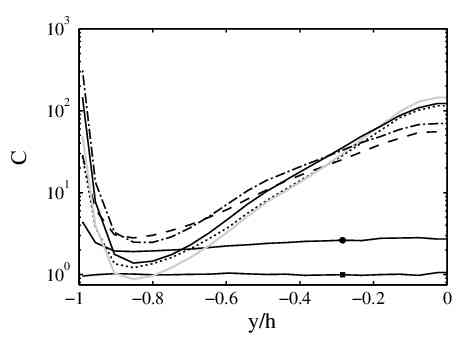}\put(-195,130){$(f)$}}
    \centerline{\includegraphics*[width=0.43\textwidth]{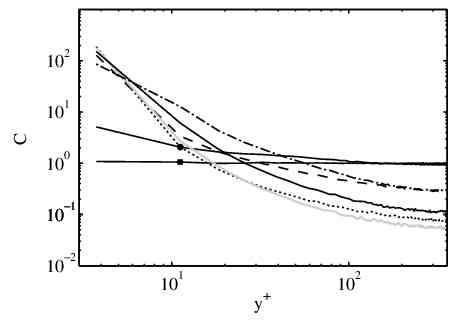}\put(-195,130){$(c)$}
                \includegraphics*[width=0.43\textwidth]{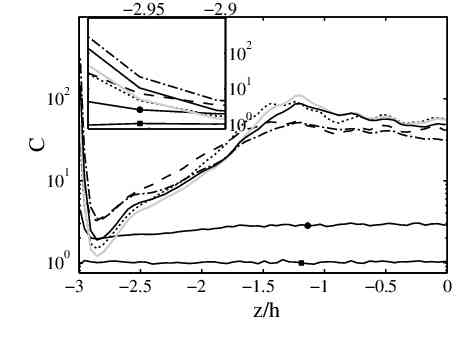}\put(-195,130){$(g)$}}
    \centerline{\includegraphics*[width=0.43\textwidth]{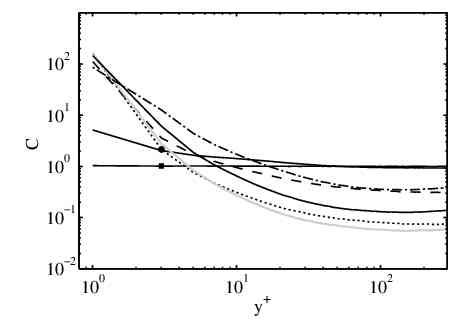}\put(-195,130){$(d)$}
                \includegraphics*[width=0.43\textwidth]{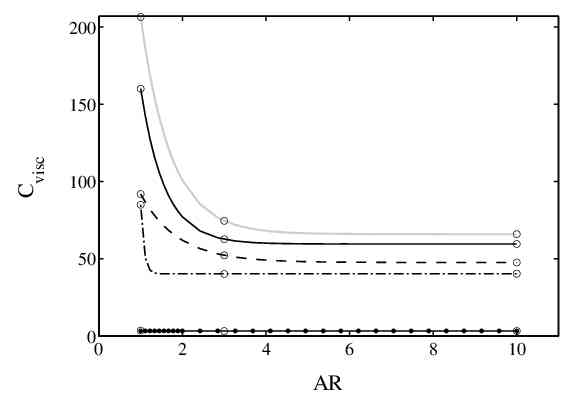}\put(-195,130){$(h)$}}
    \caption{\emph{Left}: Wall-normal distribution of the mean normalised particle concentration in \emph{(a)} vertical cut $AR=1$, \emph{(e)} \textcolor{black}{ \emph{(b)} horizontal cut $AR=3$,} \emph{(c)} vertical cut $AR=3$,  and \emph{(d)} spanwise periodic channel. \emph{Right}: mean normalised particle concentration close to the wall ($y^+ \lesssim 5$) of the square duct \emph{(e)}, \textcolor{black}{ the rectangular duct $(AR=3)$ along the side plate \emph{(f)} and bottom plate \emph{(g)}}. The corners are located at $z/h=-1$, $z/h=-3$ and $y/h=-1$. \emph{(h)}  Particle concentration at the viscous sublayer ($y < 0.01415$, $y^{+} < 5$) of the wall bisector for the square duct, the horizontal wall-bisector of the rectangular duct and the channel wall. The channel data are used for a hypothetical large aspect ratio duct ($AR=10$).   
             ${\blacksquare}$ \textbf{Stp0}, 
             $\bullet$ \textbf{Stp1}, 
             \dashed \textbf{Stp5}, 
             \dotted \textbf{Stp10}, 
             {\color{light-gray} \solid} \textbf{Stp25}, 
             \solid \textbf{Stp50}, 
             \dotdashed  \textbf{Stp100}.}
   \label{fig:fig10}     
\end{center}        
\end{figure}

To quantify particle accumulation in the duct cross-section the particle concentration $C$, normalized with the mean concentration, is computed. Fig.\ \ref{fig:fig9} shows this normalised mean particle concentration in logarithmic scale for three populations in the square and rectangular ducts. As in canonical wall-bounded geometries, the particles largely accumulate near the wall. The largest values for concentration are found in the corners and at the wall bisectors. These high concentration regions at the horizontal and vertical symmetry planes extend towards the duct centreline. This is due to the outward cross-flow motion of the secondary-flow vortices. \textcolor{black}{ Except for the heaviest populations (\textbf{Stp100} particles in square duct and \textbf{Stp50} and \textbf{Stp100} particles in the rectangular case), which have the largest accumulation in the corners, the peak on the wall bisector is generally larger than the one in the corner (\emph{c.f.}\ Fig.\ \ref{fig:fig10} \emph{e-g}). Interestingly, in the $AR=3$ duct the minimum concentration does not occur at the duct centreline but rather close to the corner bisector where the in-plane motion is most pronounced, an effect that is most noticeable in the \textbf{Stp25} and \textbf{Stp50} populations. On the other hand, in the square duct the minimum concentration is observed at the duct centreline (except for the \textbf{Stp25} particles), which is connected to the fact that the centreplane energy of the secondary flow is lower in the $AR=3$ duct than in the square case. In fact, a large portion in the middle of the rectangular duct cross-section ($|z/h|\lesssim1$) is similar to that of a spanwise-periodic channel  for all populations.} In both ducts, the largest difference between the concentration near the wall and that in the bulk can be found for \textbf{Stp25} which is four orders of magnitude.%%%%%%%%%%%%%%%%%%%%%%%%%%%%%%%%%%%%%%%%%%%%%%%%%%%%%%%%%%%%%%%%%%%%%%%%%%%%%%%%%%%%%%
%%%%%%%%%%%%%%%%%%%%%%%%%%%%%%%%%%%%%%%%%%%%%%%%%%%%%%%%%%%%%%%%%%%%%%%%%%%%%%%%%%%%%%

Fig.\ \ref{fig:fig10} \emph{(a-c)} displays the wall-normal distribution of the particle concentration at the horizontal and vertical symmetry planes of the ducts. As expected, the Lagrangian tracers (\textbf{Stp0}) are uniformly distributed ($C=1$), whereas the inertial heavy particles accumulate close to the wall and the maximum concentration appears at the wall. Qualitatively speaking, the trends in those three plots are similar to the wall-normal distribution of concentration in canonical channel flows (shown in Fig.\ \ref{fig:fig10} \emph{(d)}) where the maximum concentration at the wall appears for the \textbf{Stp25} population and the minimum concentration is in the centre for the same population (see also Young and Leeming,\citep{young_leeming_1997} Noorani et al.,\citep{noorani_etal_2014} Sardina et al., \citep{sardina_schlatter_brandt_picano_casciola_2012} Picano et al.\citep{picano_sardina_casciola_2009}).
%%%%%%%%%%%%%%%%%%%%%%%%%%%%%%%%%%%%%%%%%%%%%%%%%%%%%%%%%%%%%%%%%%%  
%%%%%%%%%%%%%%%%%%%%%%%%%%%%%%%%%%%%%%%%%%%%%%%%%%%%%%%%%%%%%%%%%%%

Further insight can be gained by measuring the distribution of the particle concentration in the viscous sublayer ($y^+ \lesssim 5$) along the walls (Fig.\ \ref{fig:fig10} \emph{e-g}). In the square duct and vertical wall of the $AR=3$ duct, the peak values of concentration occur at the corners and wall bisectors. Note that the trend of the maximum concentration for various populations at the corners differs from that in the centreplanes. The values at the corner increase with Stokes number as opposed to the above mentioned trend for the wall bisectors, which is similar to a canonical channel. In the square duct and the vertical wall of the rectangular geometry, traversing from the corner towards the duct centreplane, the concentration decreases to reach a minimum around $z/h (y/h) \approx-0.85$ before increasing again to reach another maximum at $z/h =0$. Along the horizontal wall of the rectangular duct, however, the second maximum is not at the wall bisector but rather half-way through at $z/h\approx-1.25$. This confirms the observations from the instantaneous snapshots in Fig.\ \ref{fig:fig6} \emph{(bottom)}. \textcolor{black}{ Except for the corner cluster, the current statistics clarify that heavy particles are densely clustered around the region of maximum outward motion of the secondary vortex on the horizontal wall. In Fig. \ref{fig:fig2} it can be observed that the maximum $V$ component of the horizontal vortex is located at a wall-normal distance of around $y/h \approx 0.3$ from the horizontal wall in the two ducts, and at spanwise distances from the centreplane of $z/h \approx 0$ and $z/h \approx 1.25$ in the square and $AR=3$ duct cases, respectively.} \textcolor{black}{ More importantly, in the horizontal wall of the rectangular case  (Fig.\ \ref{fig:fig10} \emph{g}),} moving \textcolor{black}{ farther} towards the wall bisector the concentration profiles decrease to reach a plateau. This is due to the fact that the secondary motion is significantly attenuated in this region and the flow progressively becomes \textcolor{black}{ more} similar to a channel flow. %Integrating $C$ in this plot along the walls will explain why the integral concentration in the viscous sublayer all over the walls of rectangular duct is larger than the square duct (\emph{c.f.}\ figure \ref{fig:fig10} \emph{d,e}).

In order to quantify the effect of aspect ratio on the wall accumulation of particles with different Stokes numbers we provide a comparison between channel and ducts with different aspect ratios in terms of maximum concentration in the viscous sublayer for various populations.  The results are shown in Fig.\ \ref{fig:fig10} \emph{(h)}. The data for the duct cases are obtained in the wall bisector (horizontal wall bisector for the rectangular duct) and the concentration is averaged over the viscous sublayer of each case ($y^+ (z^+) \leq 5$ or $y (z) < 0.01415$). Note that the width of the averaging cell is twice the height, and that spanwise periodicity is exploited in the channel case. According to \citet{duct_tsfp} when $AR\rightarrow10$ the flow characteristics approach the ones of a spanwise-periodic channel. Therefore, the channel data are shown for $AR=10$ on the abscissa of the Fig.\ \ref{fig:fig10} \emph{(h)}. A simple exponential curve fit of the data for each population is also projected on top of the figure. From Fig.\ \ref{fig:fig10} \emph{(h)} it is obvious that the lightest populations (\textbf{Stp1}) are most \emph{insensitive} to any changes in the duct aspect ratio and the values for maximum concentration are close to those for a channel. On the other hand, heavier particles exhibit much larger accumulations at the viscous sublayer of the square duct wall bisector compared to the rectangular case. The results smoothly converges to a channel with increasing the aspect ratio of the duct.
%%%%%%%%%%%%%%%%%%%%%%%%%%%%
%%%%%%%%%%%%%%%%%%%%%%%%%%%%
%%%%%%%%%%%%%%%%%%%%%%%%%%%%

\subsubsection{Particle velocities}\label{sec:partVelocities}
\begin{figure}
  \centerline{\includegraphics*[width=0.25\textwidth]{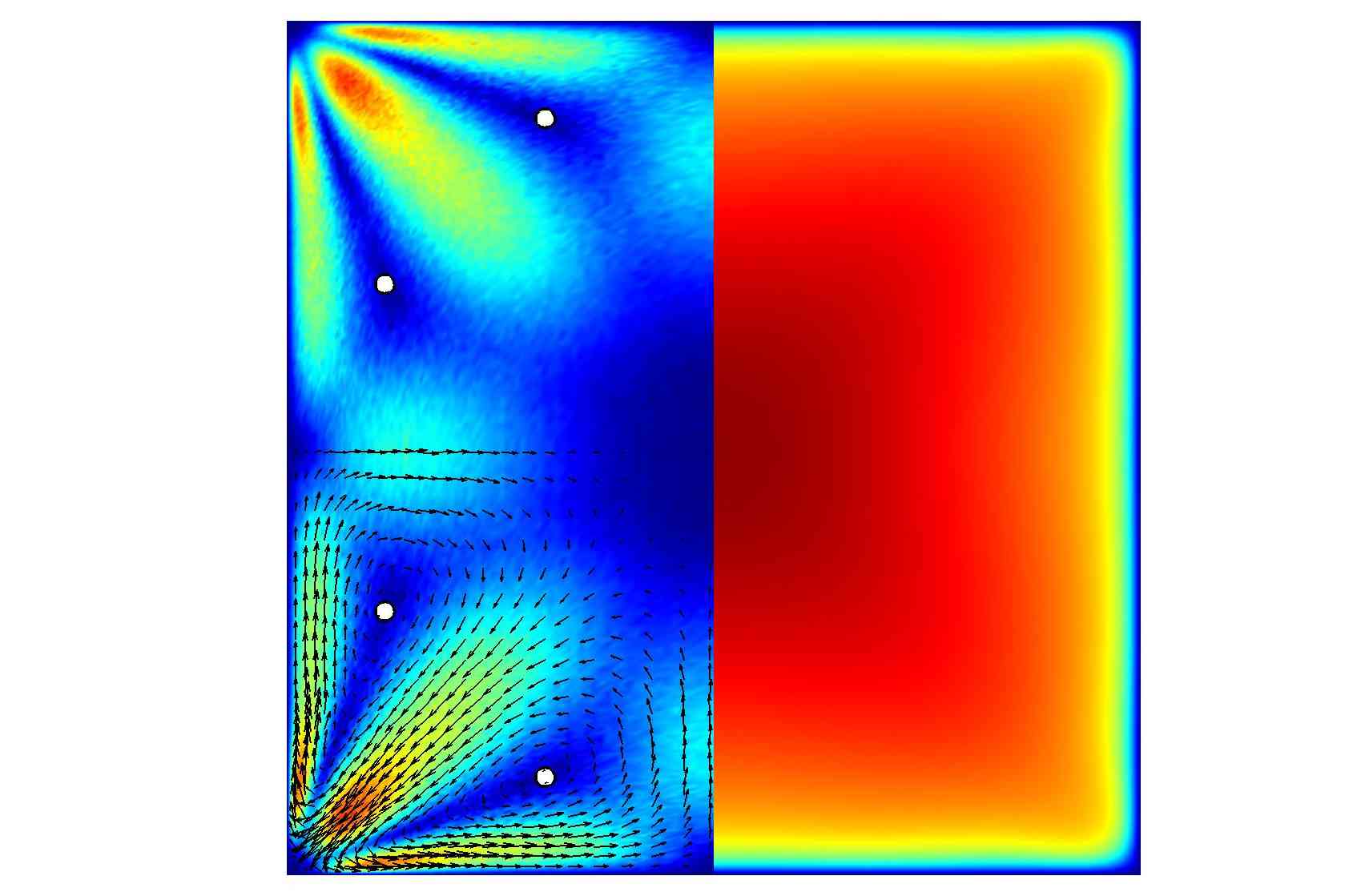}
              \includegraphics*[width=0.65\textwidth]{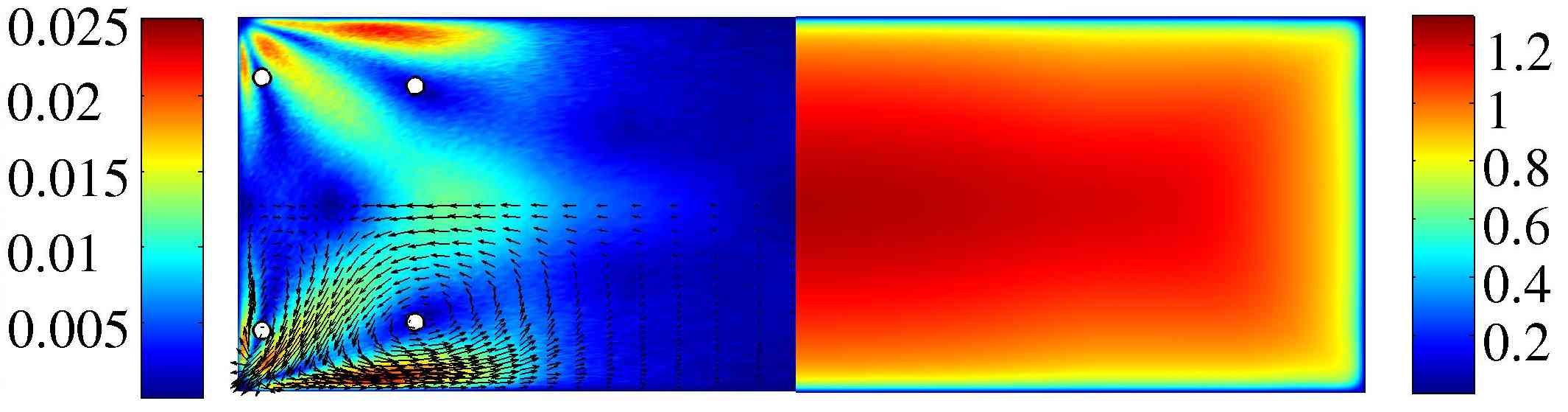}}
  \centerline{\includegraphics*[width=0.25\linewidth]{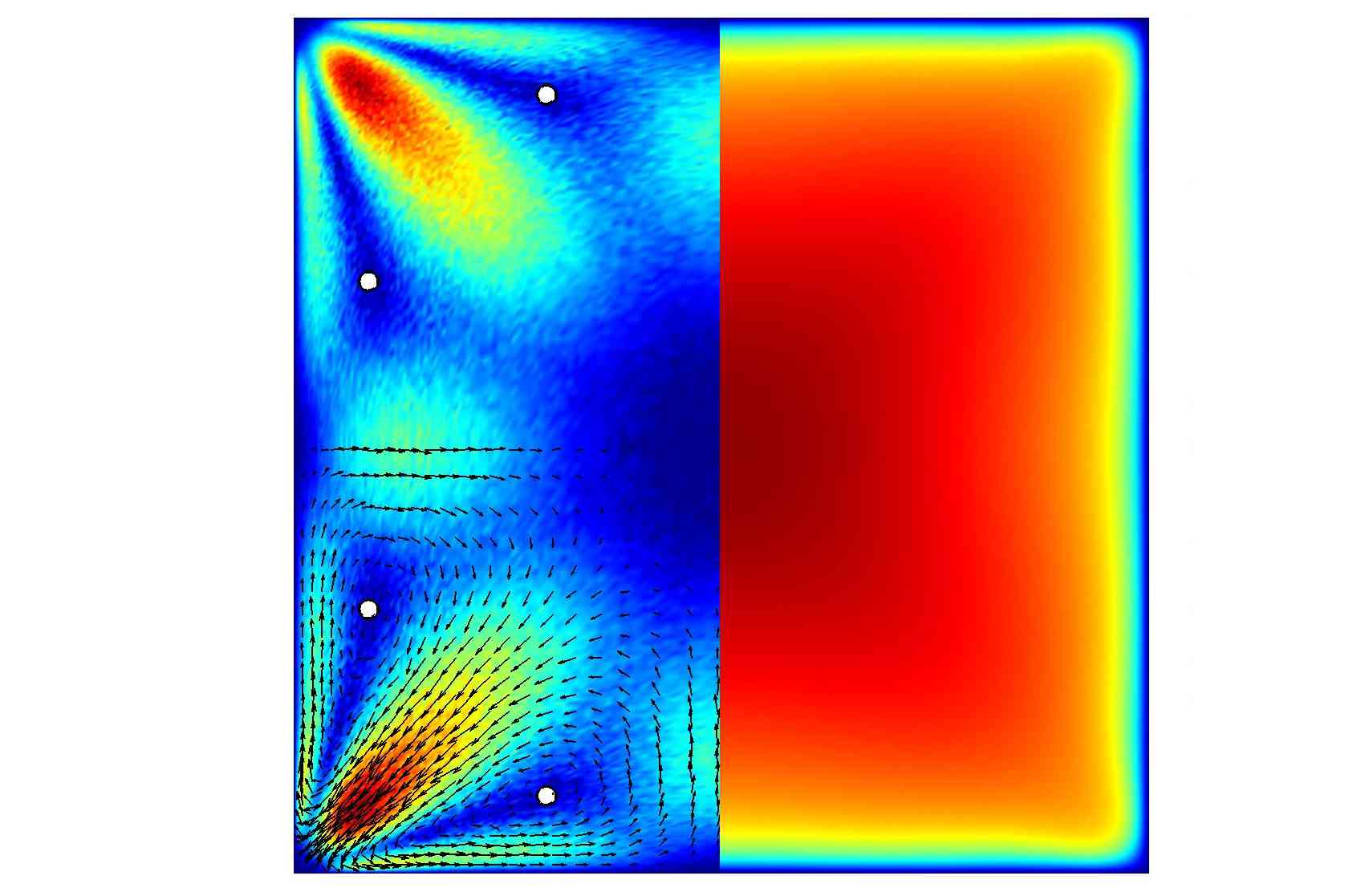}
              \includegraphics*[width=0.65\linewidth]{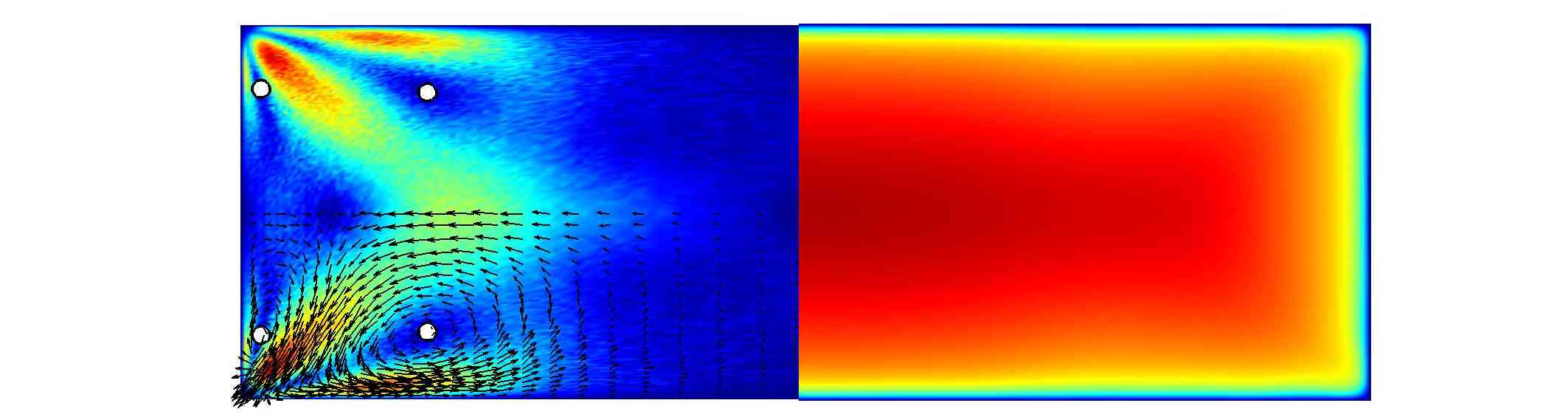}}
  \centerline{\includegraphics*[width=0.25\textwidth]{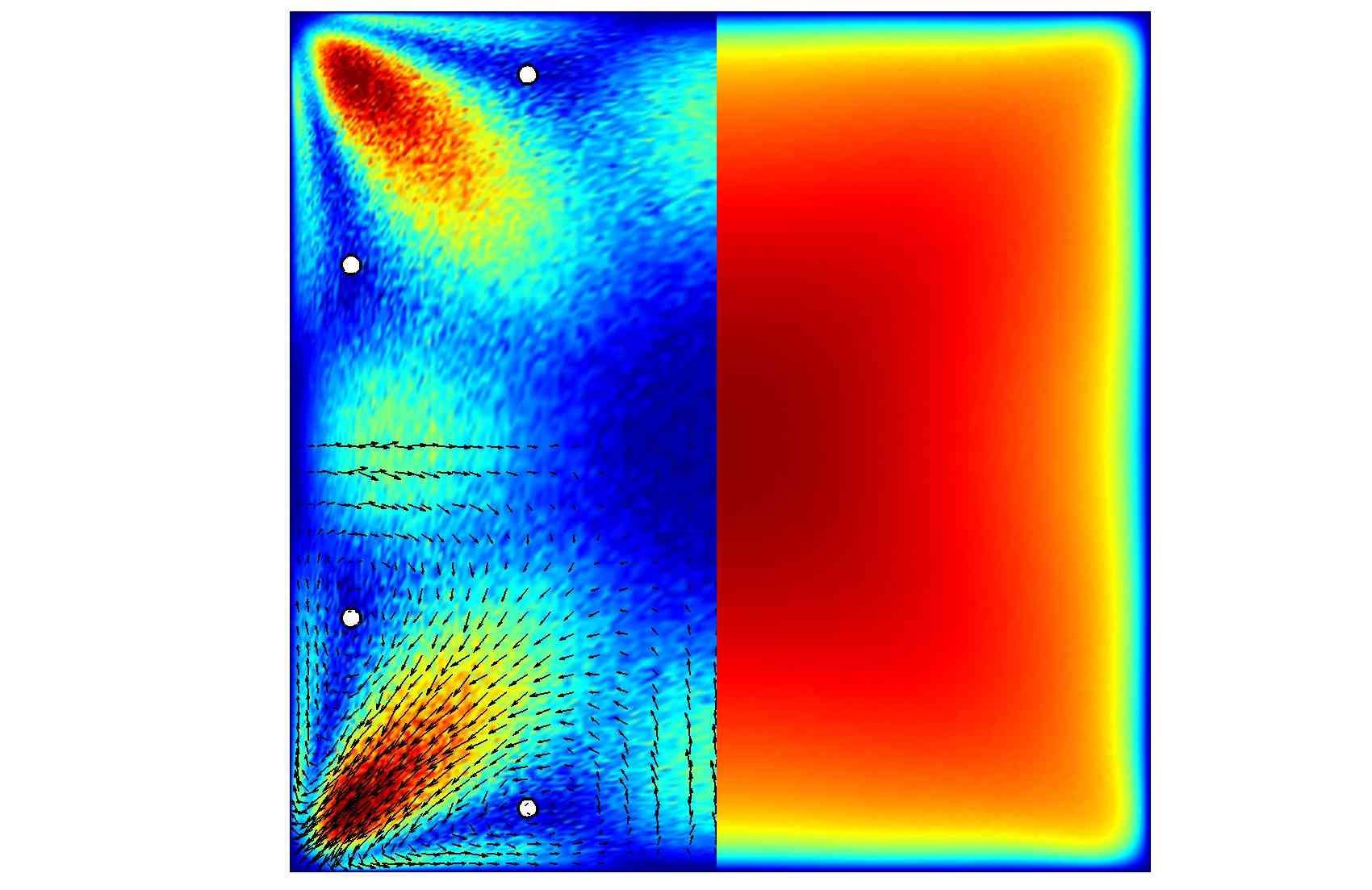}
              \includegraphics*[width=0.65\textwidth]{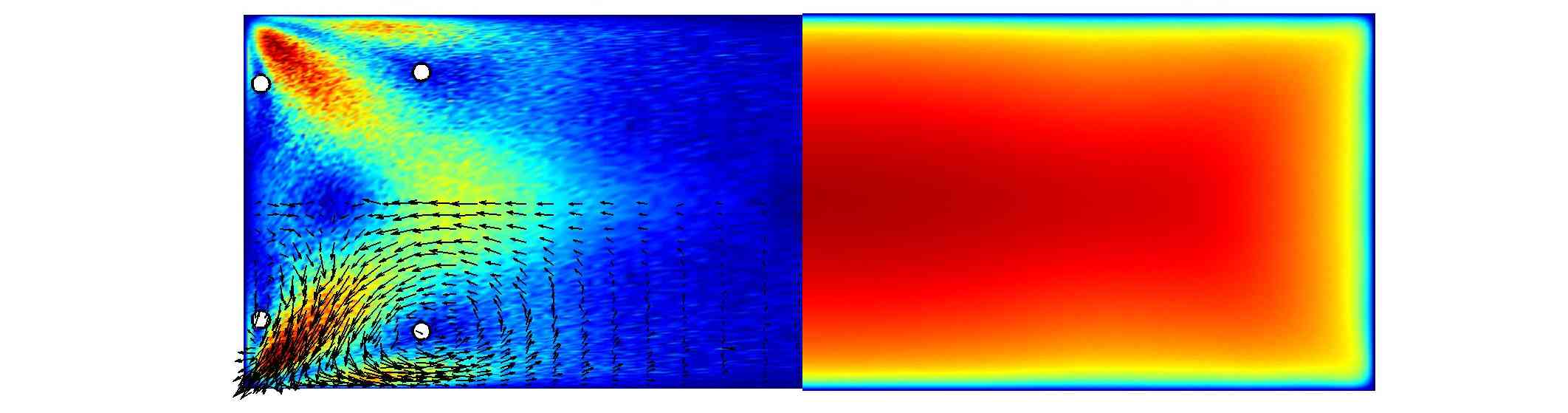}}
  \centerline{\includegraphics*[width=0.25\textwidth]{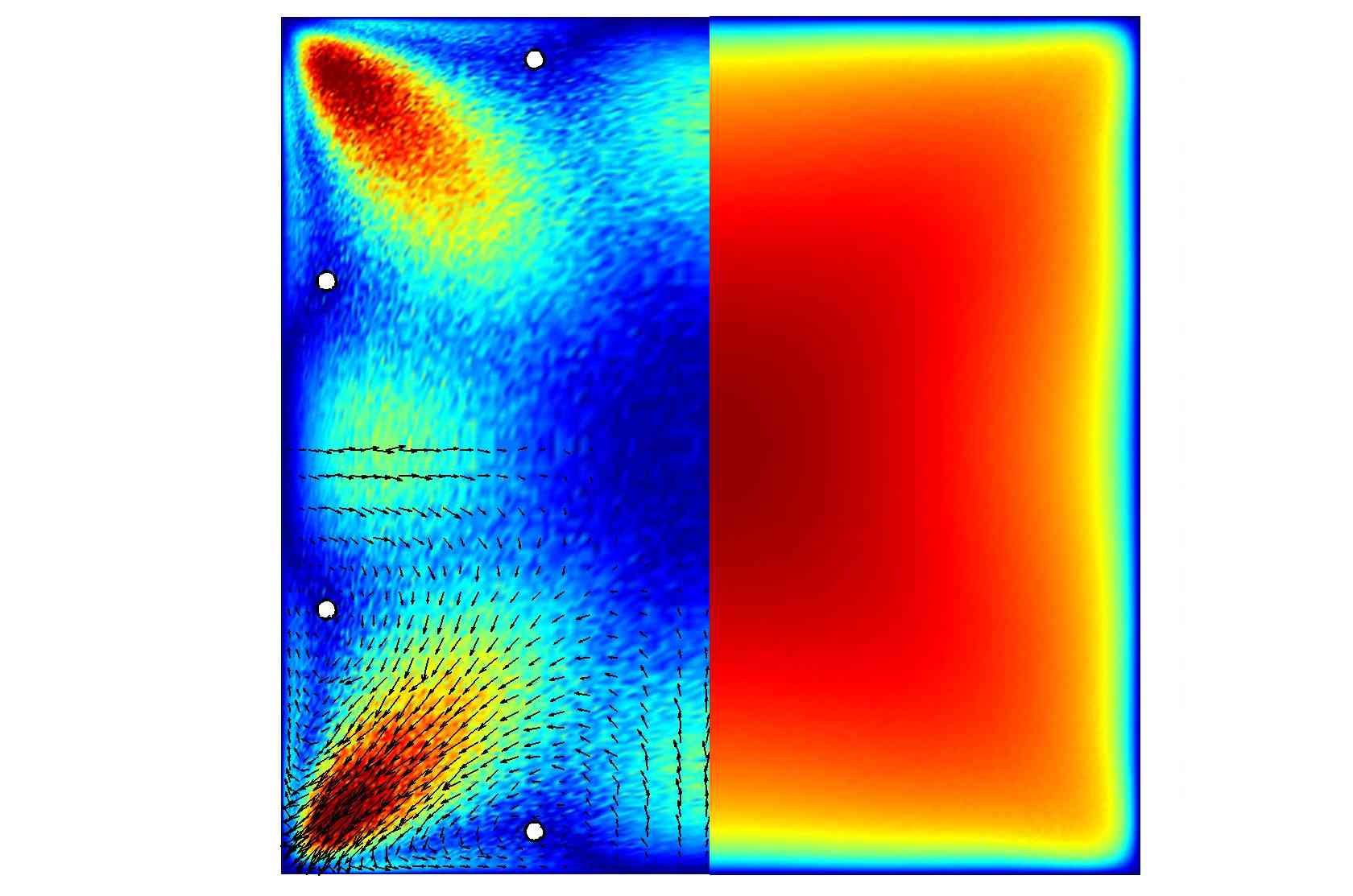}
              \includegraphics*[width=0.65\textwidth]{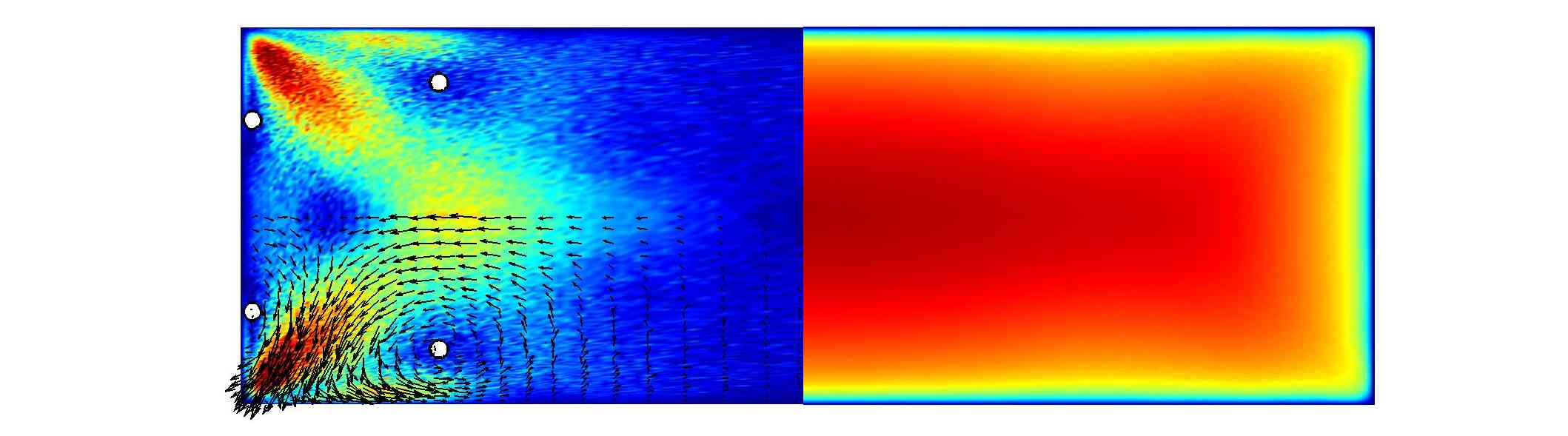}}
  \centerline{\includegraphics*[width=0.25\linewidth]{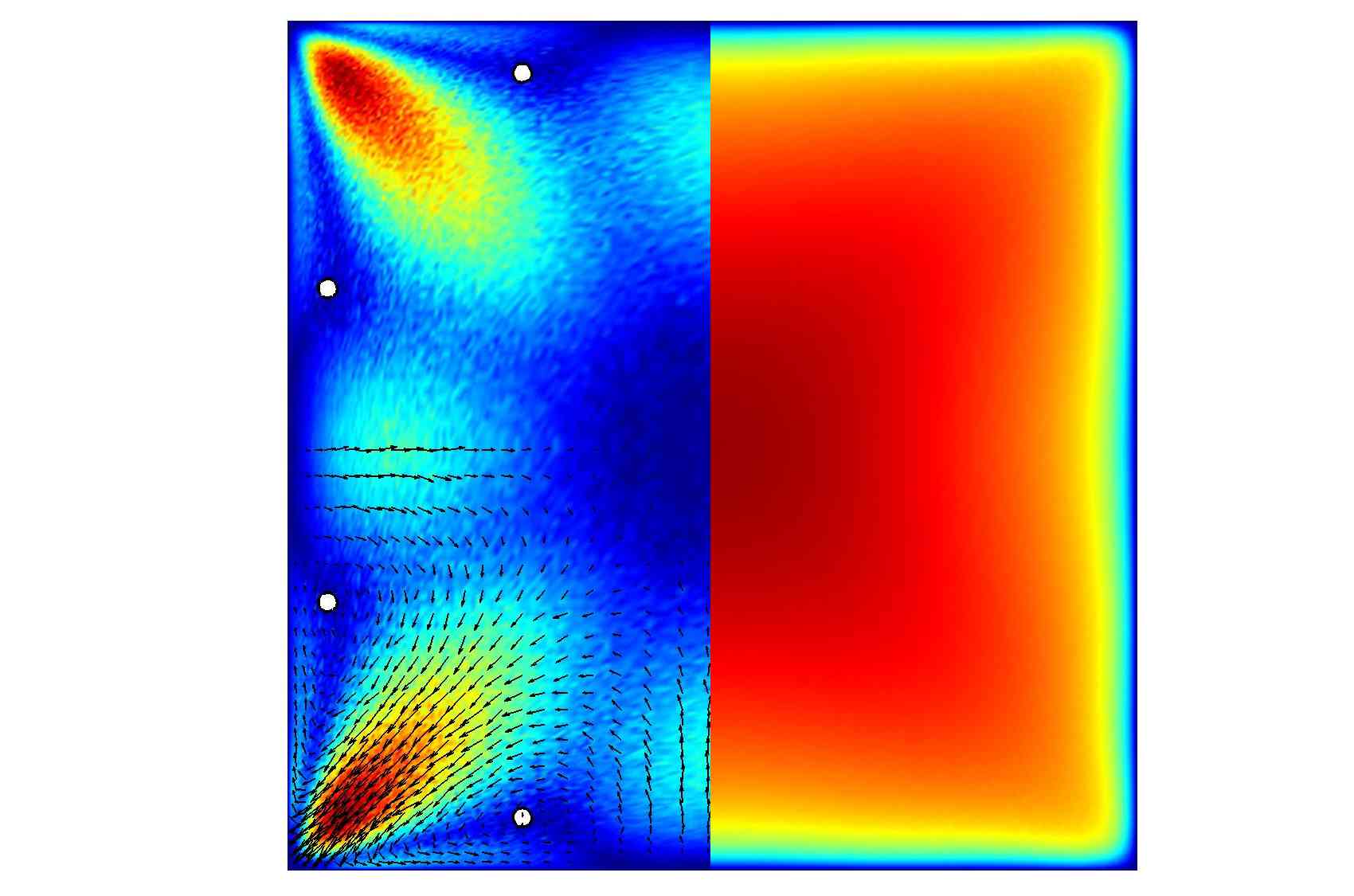}
              \includegraphics*[width=0.65\linewidth]{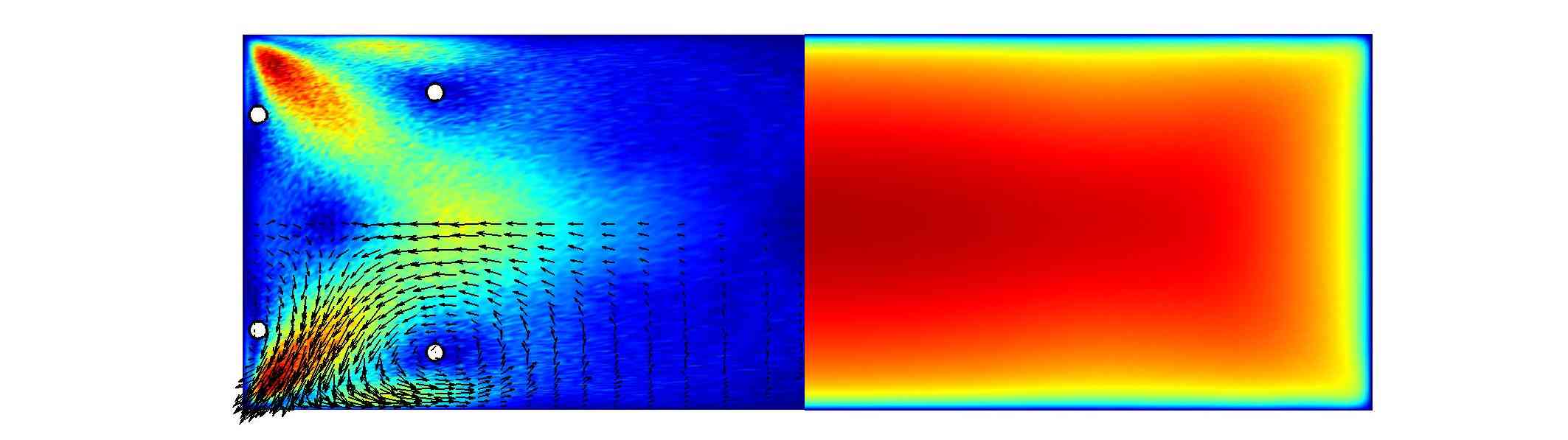}}
  \centerline{\includegraphics*[width=0.25\textwidth]{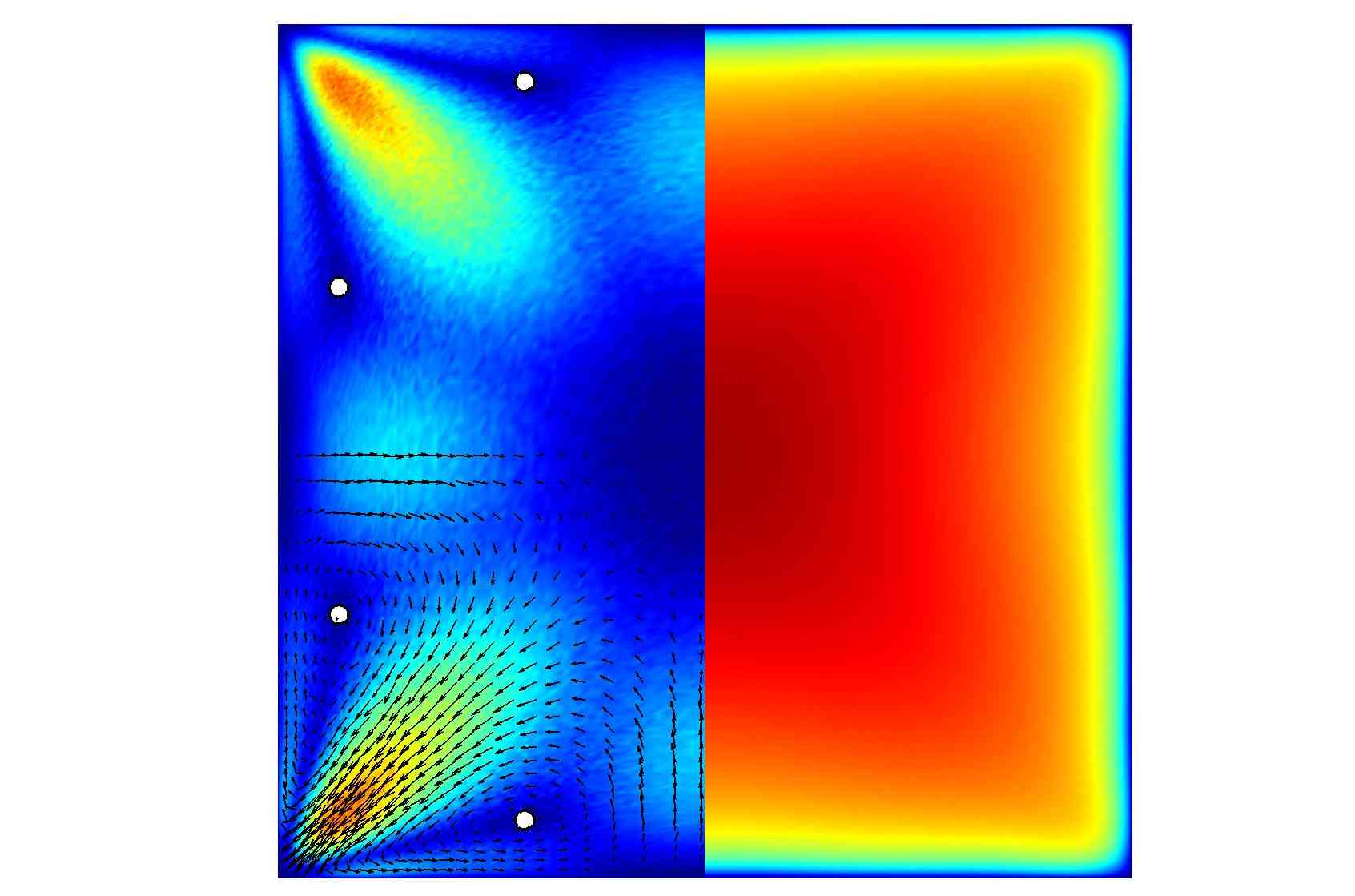}
              \includegraphics*[width=0.65\textwidth]{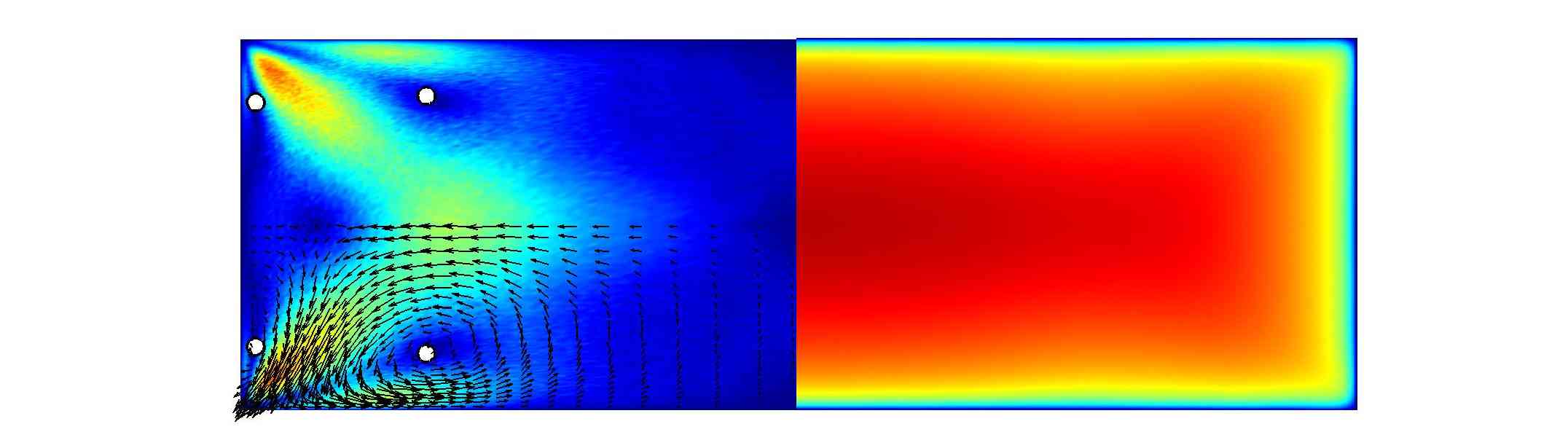}}
  \caption{ (\emph{Left panel}) Pseudocolours of the magnitude of mean in-plane velocities of the particulate phase in the same configuration with the vectors for the in-plane motion projected on top. (\emph{Right panel}) Pseudocolours of the streamwise component of the mean particle velocity. White dots indicate the core of the mean vortical structures associated with the particulate phase data. \emph{From top:}
    \textbf{Stp1},
    \textbf{Stp5},
    \textbf{Stp10},
    \textbf{Stp25},
    \textbf{Stp50},
    \textbf{Stp100}. \textcolor{black}{ The velocity and length scales used in this figure are the bulk velocity $u_{b}$ and the duct half-height $h$, respectively.}}
\label{fig:fig11}
\end{figure}

\begin{figure}
  \centerline{\includegraphics[width=0.167\textwidth,trim={0 -1.5in 0 0in}]{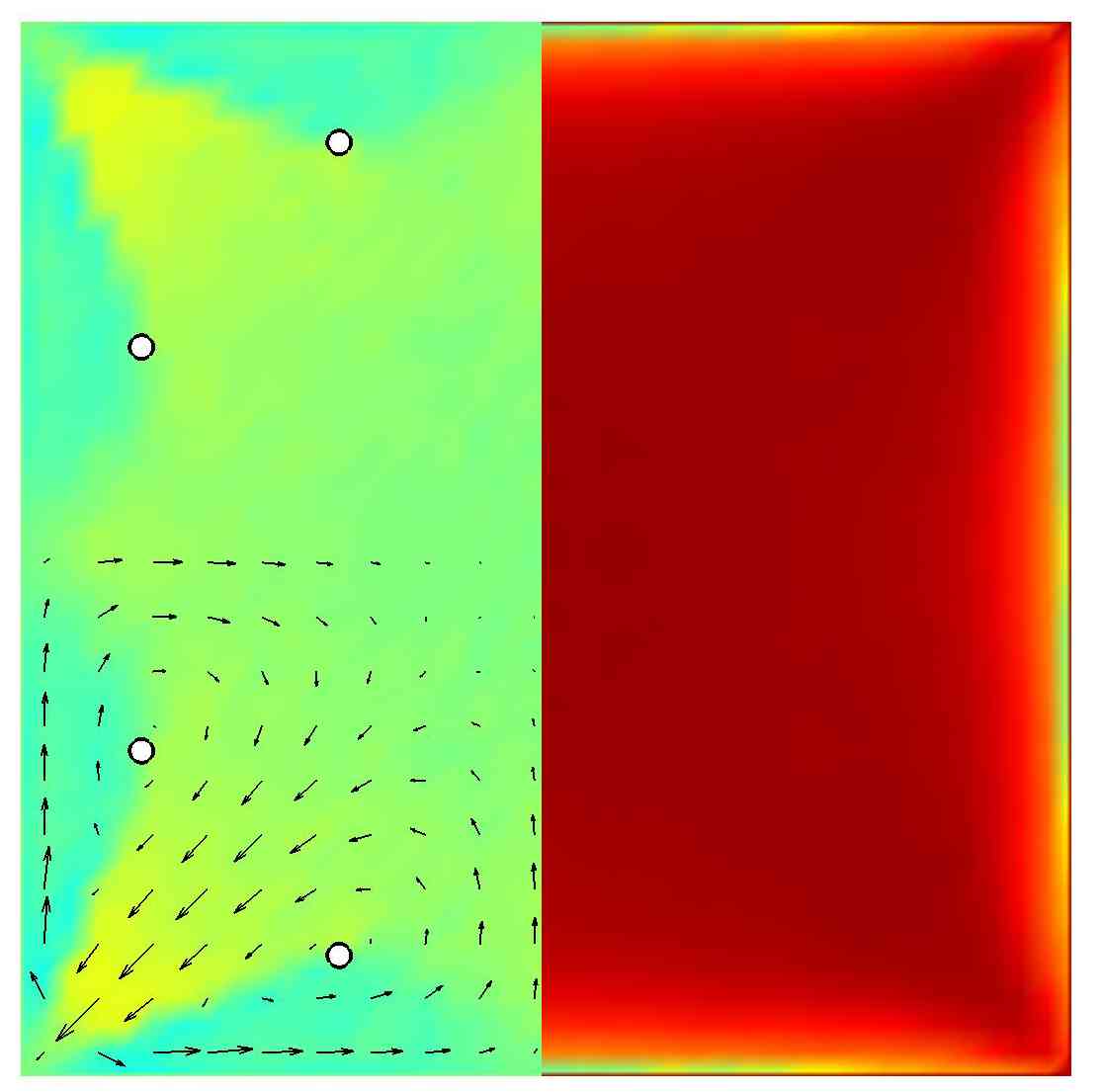}
              \includegraphics[width=0.6\textwidth]{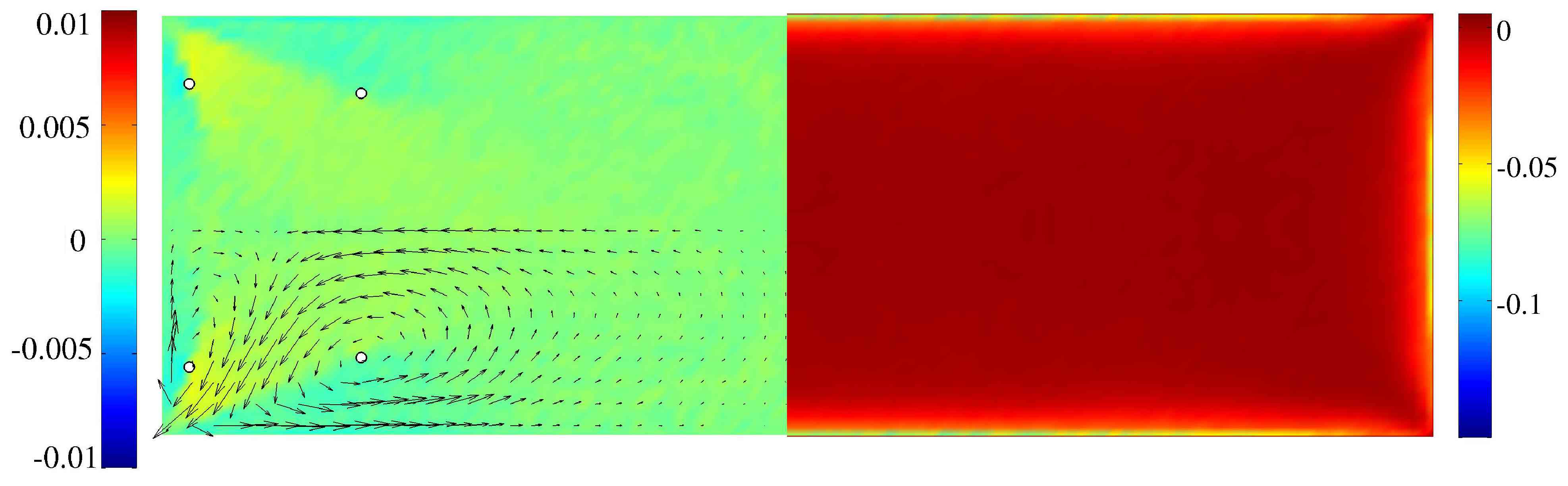}}
  \centerline{\includegraphics[width=0.167\linewidth,trim={0 -1.8in 0 0in}]{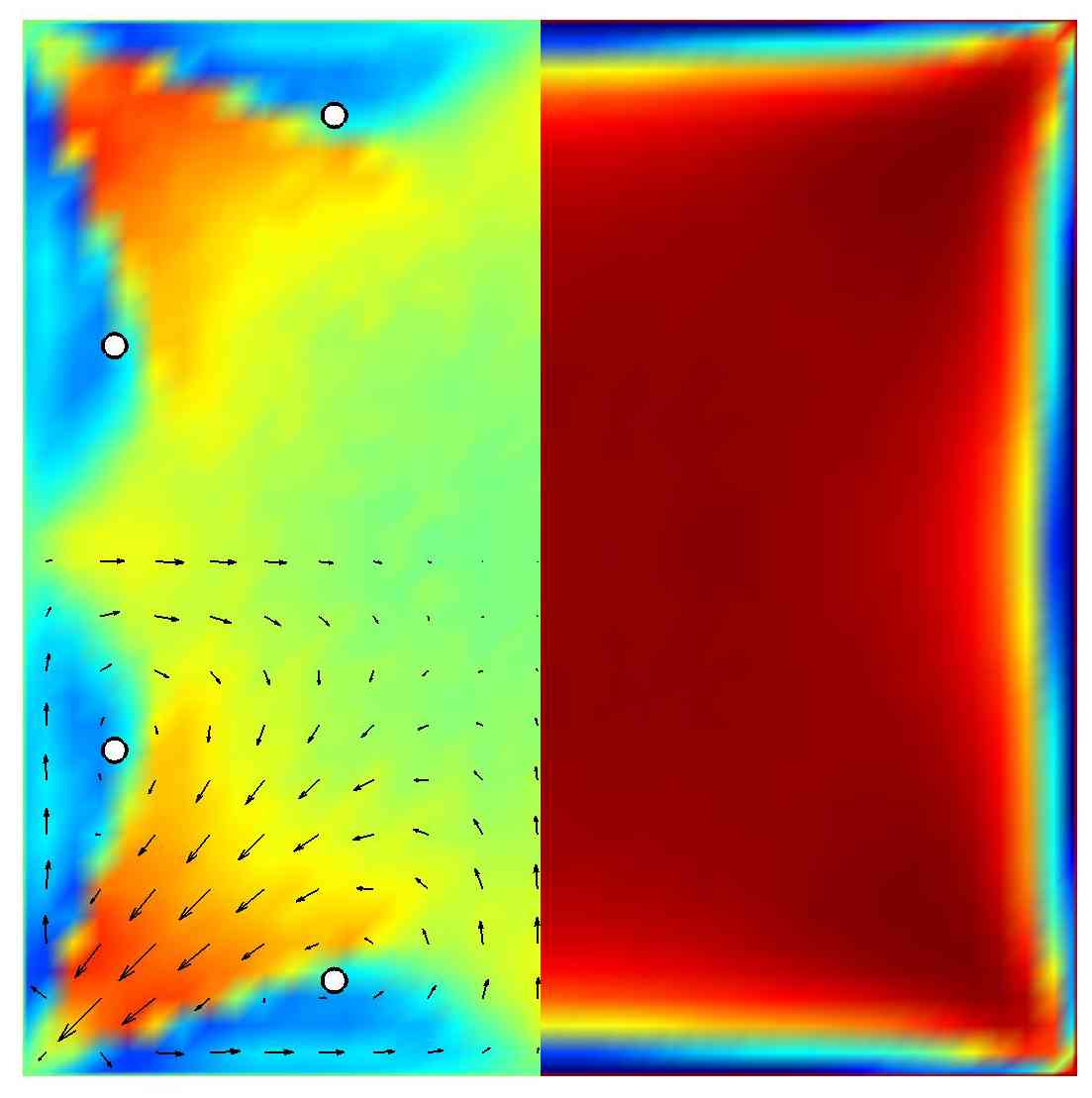}
              \includegraphics[width=0.6\linewidth]{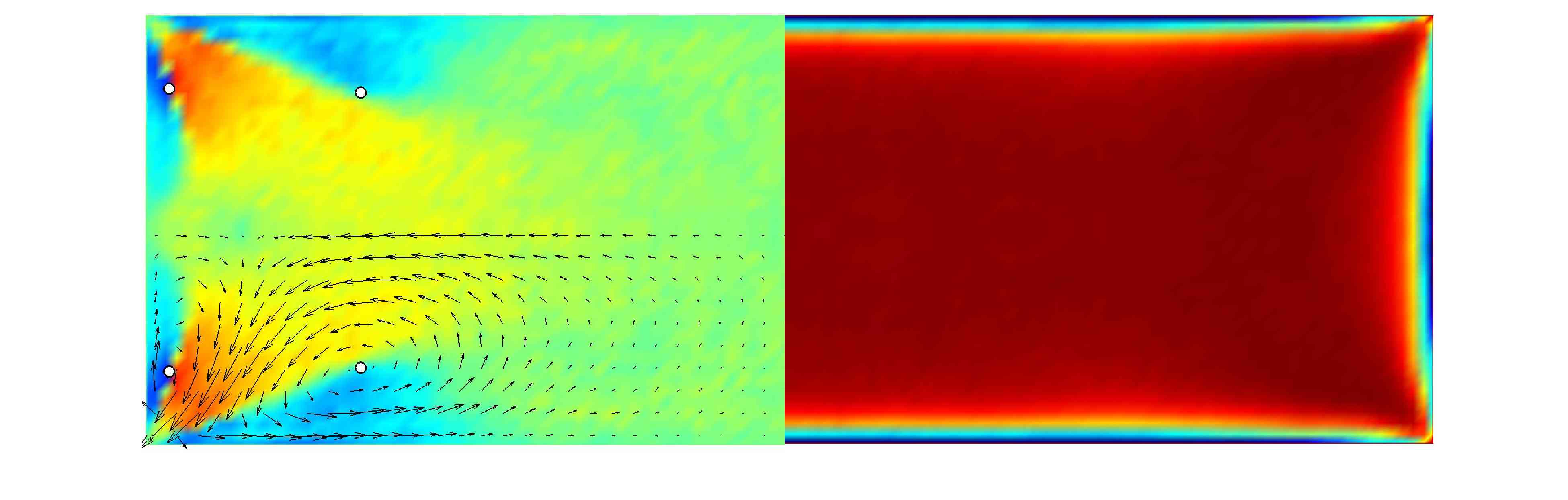}}
  \centerline{\includegraphics[width=0.167\textwidth,trim={0 -1.5in 0 0in}]{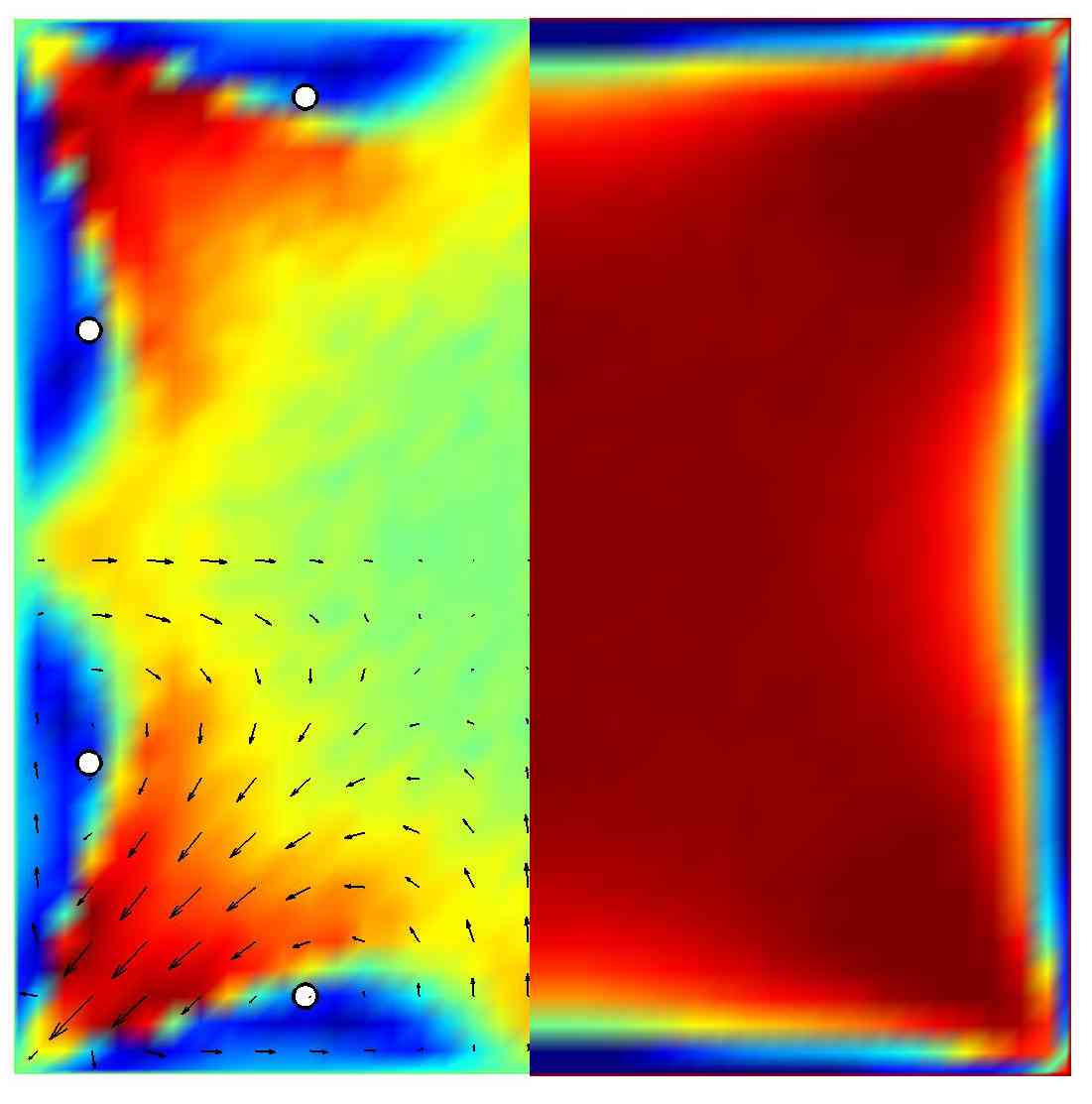}
              \includegraphics[width=0.6\textwidth]{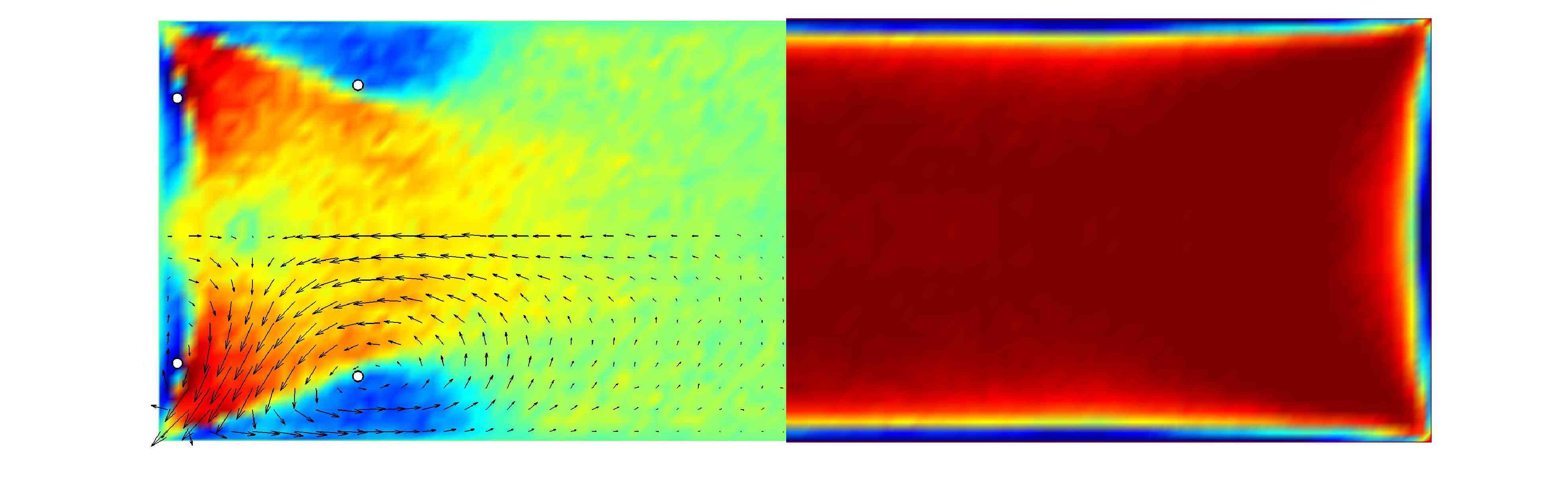}}
  \centerline{\includegraphics[width=0.167\textwidth,trim={0 -2.2in 0 0in}]{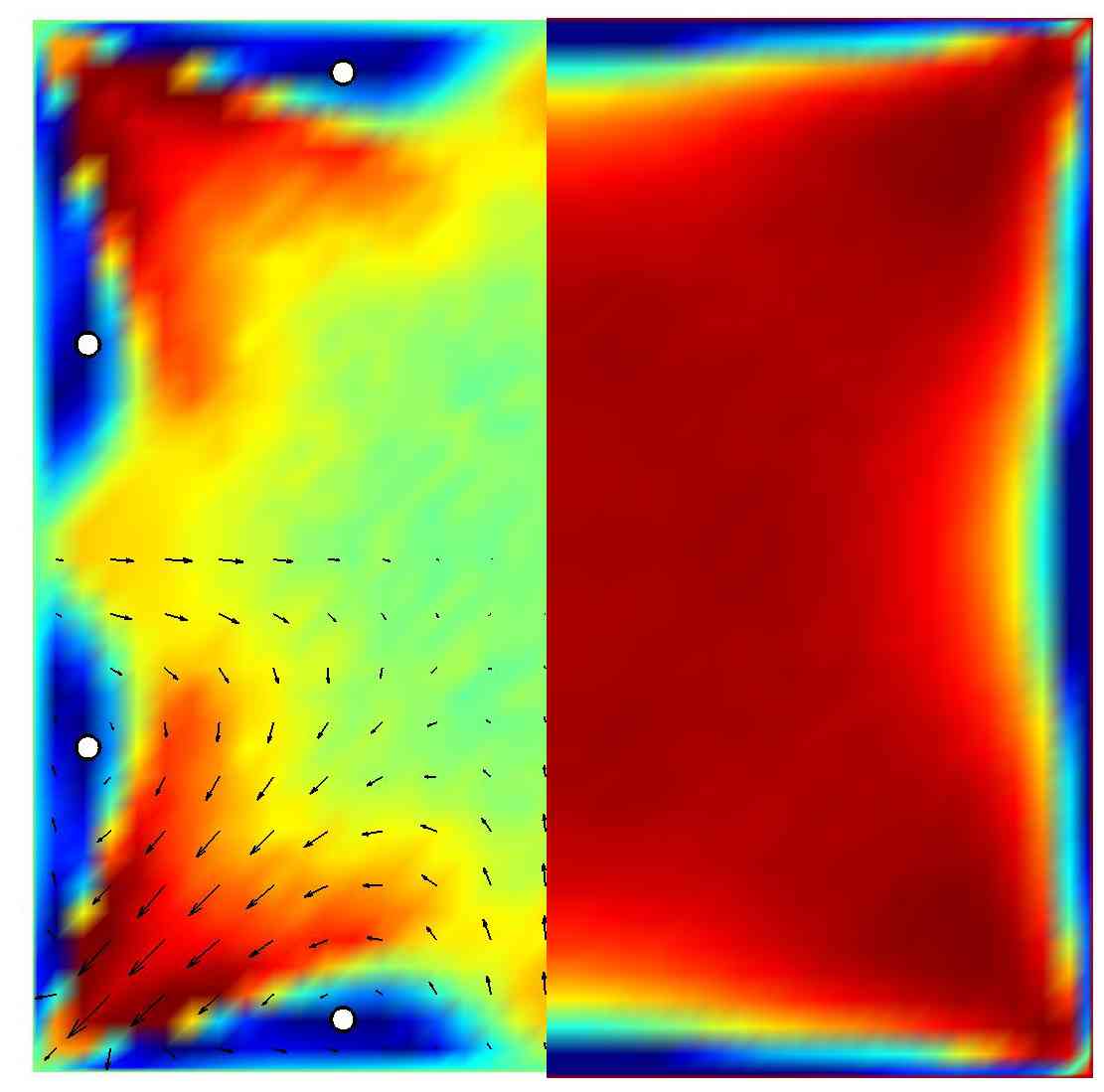}
              \includegraphics[width=0.6\textwidth]{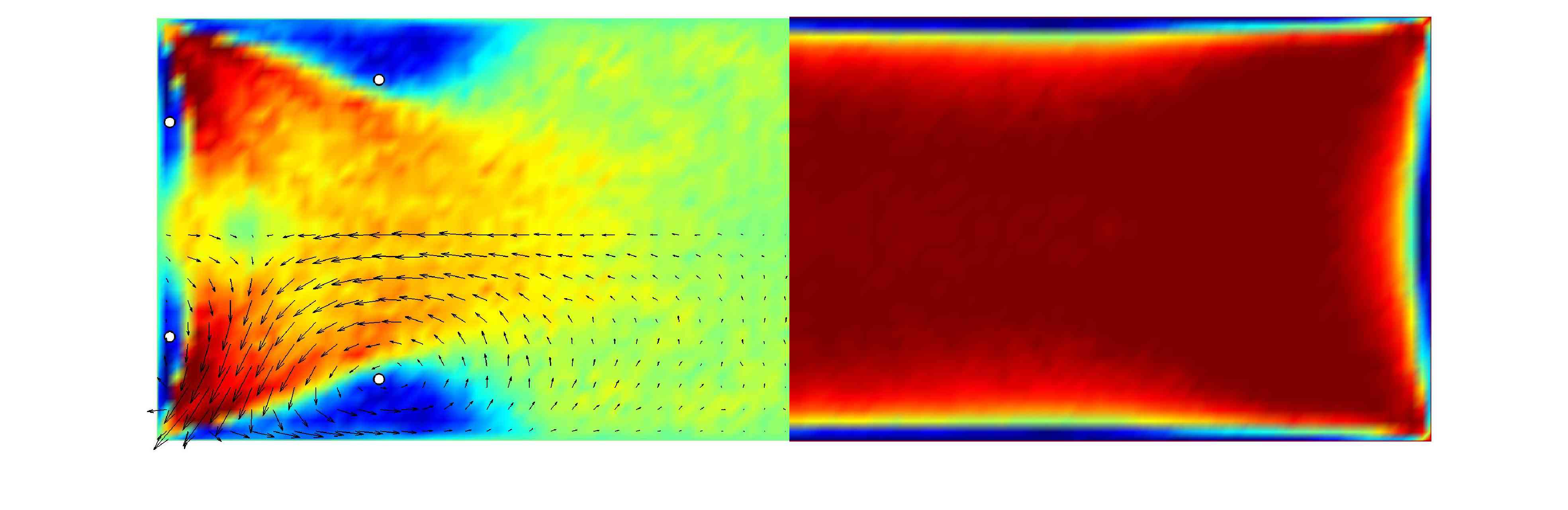}}
  \centerline{\includegraphics[width=0.167\linewidth,trim={0 -2.5in 0 0in}]{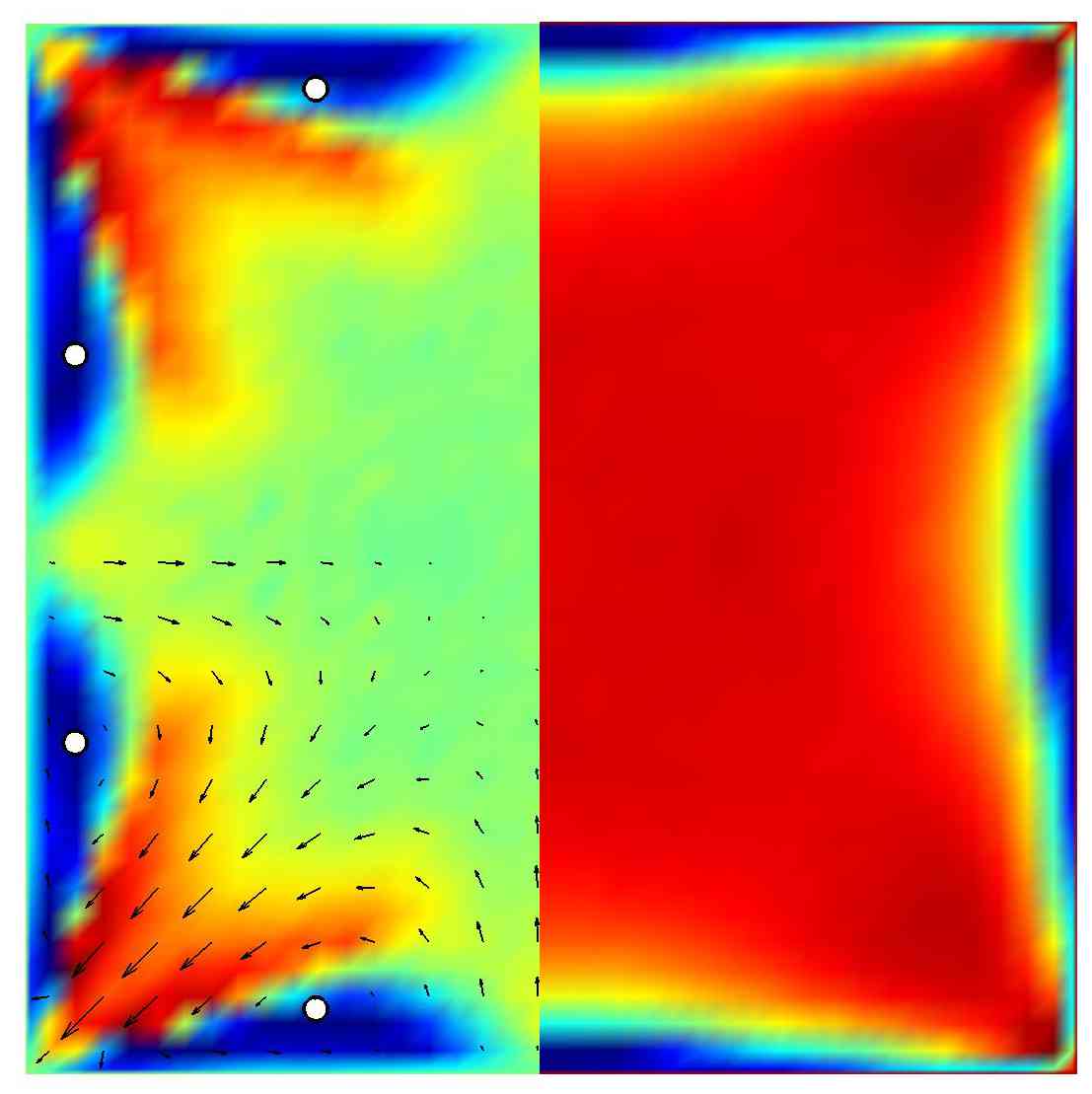}
              \includegraphics[width=0.6\linewidth]{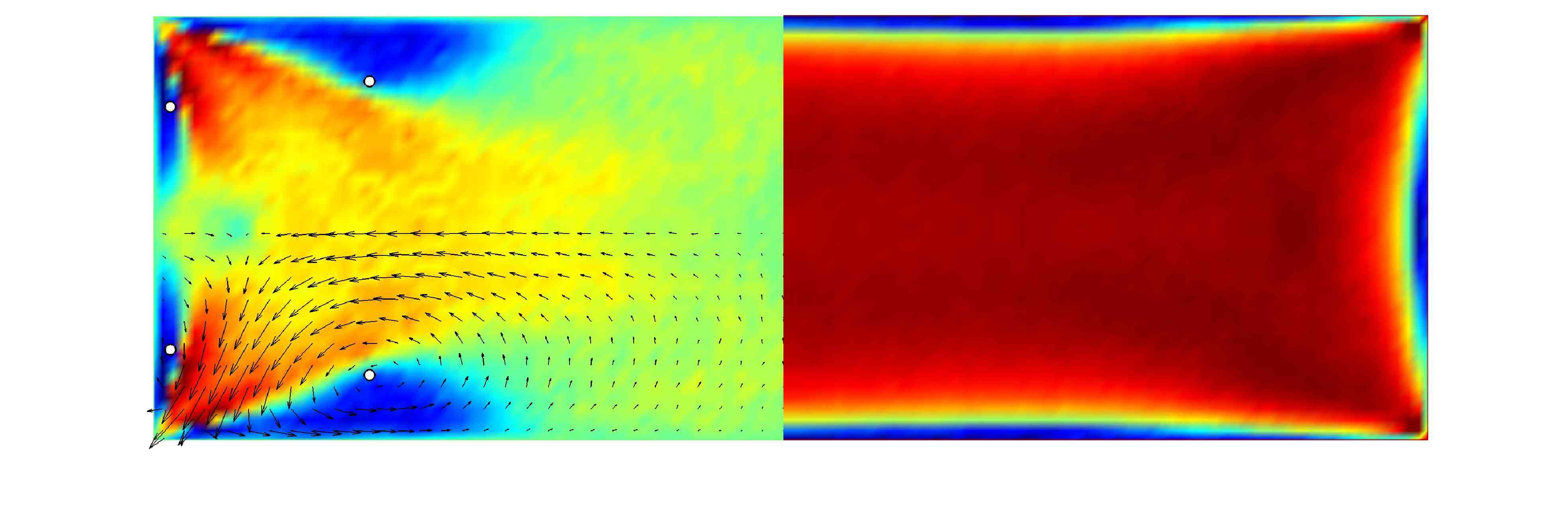}}
  \centerline{\includegraphics[width=0.167\textwidth,trim={0 -2.5in 0 0in}]{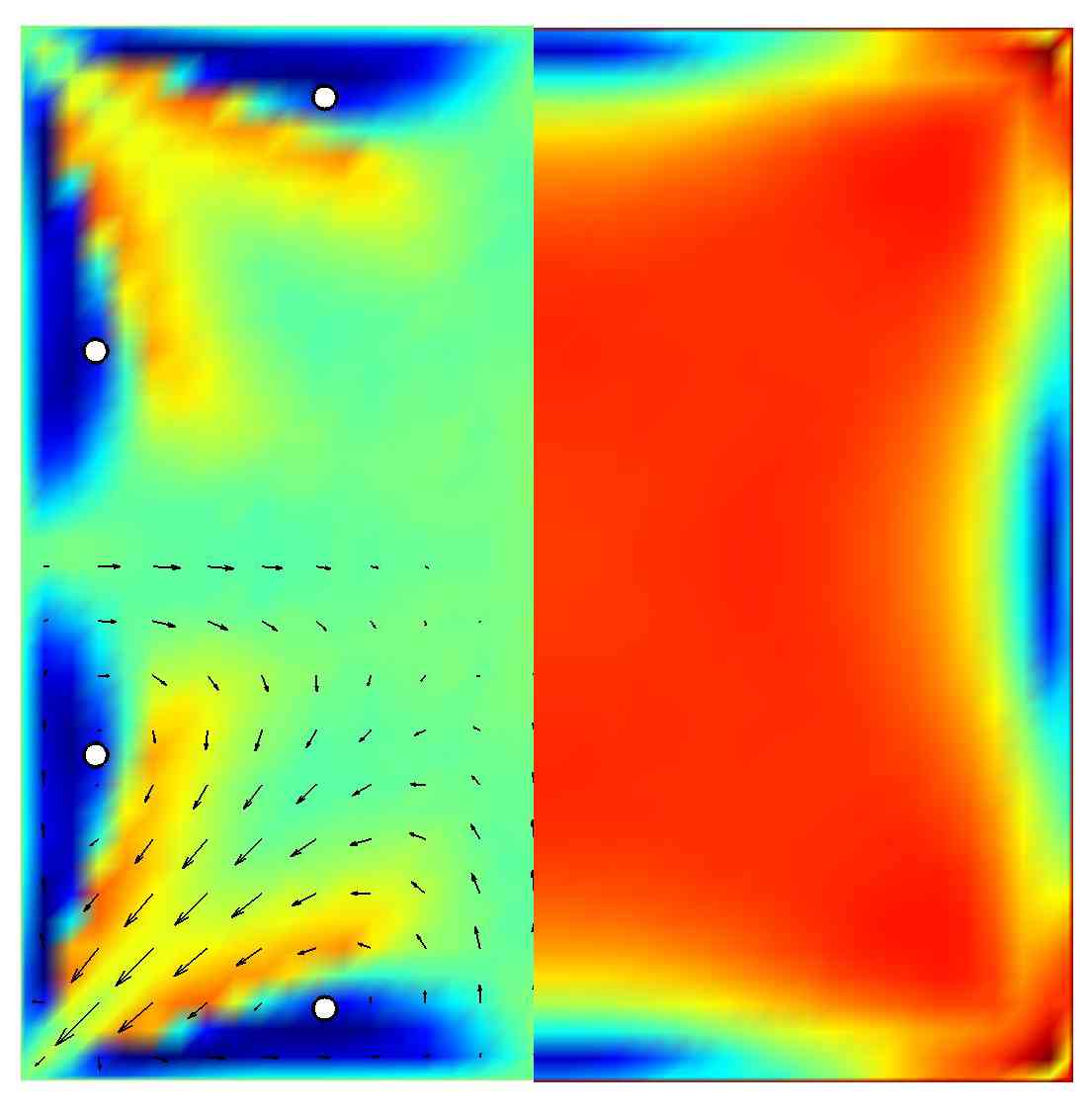}
              \includegraphics[width=0.6\textwidth]{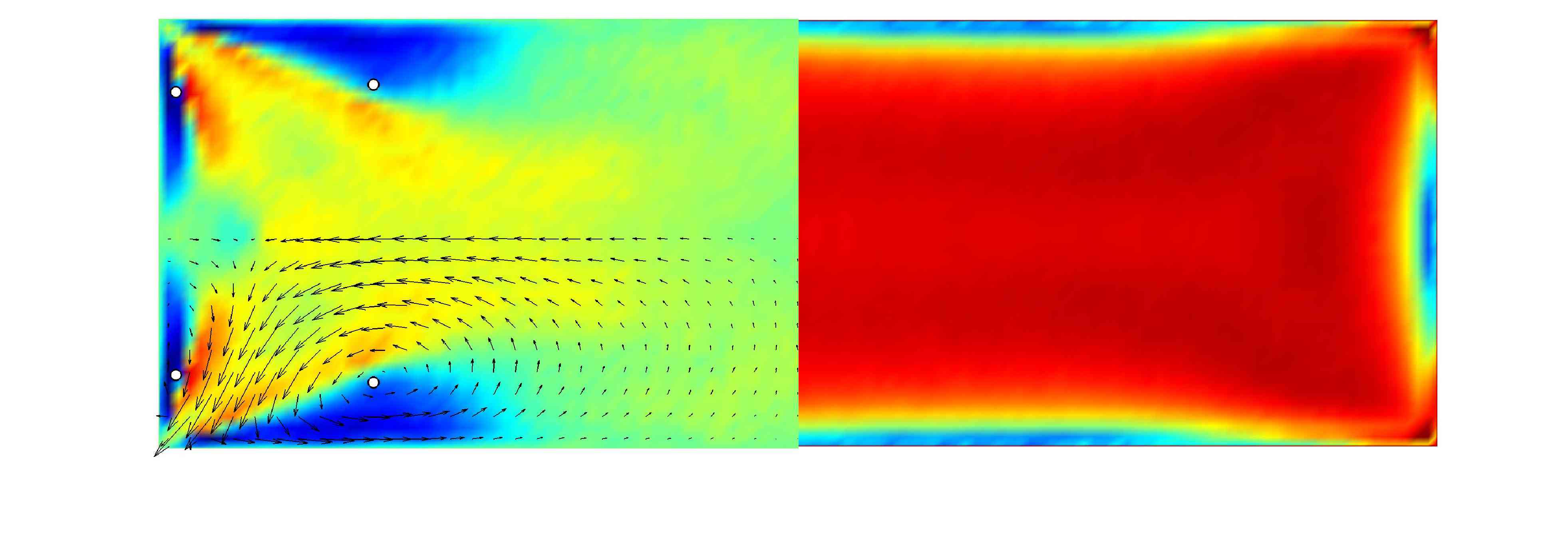}}
  \caption{\textcolor{black}{ (\emph{Left panel}) Pseudocolours of the difference between magnitude of mean in-plane velocities of the particulate phase and the fluid tracers in ducts (the vectors for the in-plane motion projected on top). (\emph{Right panel}) Pseudocolours of the difference between streamwise component of the mean particle velocity and fluid tracers. White dots indicate the core of the mean vortical structures associated with the particulate phase data. \emph{From top:}
    \textbf{Stp1},
    \textbf{Stp5},
    \textbf{Stp10},
    \textbf{Stp25},
    \textbf{Stp50},
    \textbf{Stp100}. 
  The velocity and length scales used in this figure are the bulk velocity $u_{b}$ and the duct half-height $h$, respectively.}}
\label{fig:fig11b}
\end{figure}

Fig.\ \ref{fig:fig11} displays the mean streamwise particle velocity, $\langle v_{x_p}\rangle$, and the magnitude of the in-plane velocities, $\sqrt(\langle v_{y_p} \rangle ^2+\langle v_{z_p}\rangle ^2)$, in the cross-section of the square and rectangular duct. The velocities are normalised with the bulk flow ($u_b$). \textcolor{black}{ In order to provide a clearer overview of the changes in inertial particle dynamics compared to fluid tracers, the difference between the velocity components of particles and tracers is shown in Fig.\ \ref{fig:fig11b}. From Figs.\ \ref{fig:fig11} and \ref{fig:fig11b} \emph{(right panels)} it can be seen that the mean particle velocities are very similar to the flow velocities although the streamwise velocity becomes less than that of the fluid in the buffer layer and the edge of outer-layer regions ($5 \lesssim y^+(z^+) \lesssim 100$). This difference increases with increasing $St_b$ up to the \textbf{Stp25} particles, after which the trend reverses. In the rest of the outer layer ($ y^+(z^+) \gtrsim 100$) the particle streamwise velocity becomes mildly faster than the flow with increasing particle inertia.} These are linked to preferential accumulation of inertial particles in low-speed regions of the turbulent field as already discussed in the context of canonical wall-bounded turbulent particle-laden flows.\citep[see][]{portela_etal_2002,marchioli_giusti_salvetti_soldati_2003}.
%%%%%%%%%%%%%%%%%%%%%%%%%%%%%%%%%%%%%%%%%%%%%%%%%%%%%%%%%%%%%%%%%%%%%%%%%%%%%%%%%%%%%%
%%%%%%%%%%%%%%%%%%%%%%%%%%%%%%%%%%%%%%%%%%%%%%%%%%%%%%%%%%%%%%%%%%%%%%%%%%%%%%%%%%%%%%

For both configurations the particle mean in-plane velocities largely deviate from that of the fluid flow. On the one hand, the magnitude of the in-plane motion of the particle phase adjacent to the walls significantly decreases with increasing particle inertia such that it is almost zero for heavier populations \textcolor{black}{ (\emph{c.f.}\ Figs.\ \ref{fig:fig11} and \ref{fig:fig11b} \emph{left panels}).} On the other hand, across the tangent line between the two corner vortices, this magnitude largely increases. The largest increase in this position can be found for \textbf{Stp25} particles in both configurations. As a result, the mean in-plane motion of the solid phase becomes the most intense along the corner bisectors. This is similar to the effect of the particle inertia on the secondary motion in mildly curved pipes (\emph{c.f.} Fig.\ $3$ of \citet{noorani_etal_2015}). Unlike the bent pipe, however, this effect does not alter the core of the mean secondary motion cells or the position of secondary motion extrema significantly. It is important to note that in the horizontal symmetry plane of the rectangular duct the secondary motion still effectively removes particles from the centre region of the duct. 
%%%%%%%%%%%%%%%%%%%%%%%%%%%%%%%%%%%%%%%%%%%%%%%%%%%%%%%%%%%%%%%%%%%%%%%%%%%%%%%%%%%%%%
%%%%%%%%%%%%%%%%%%%%%%%%%%%%%%%%%%%%%%%%%%%%%%%%%%%%%%%%%%%%%%%%%%%%%%%%%%%%%%%%%%%%%%

% %%%%%%%%%%%%%%%%%%%%%%%%%%%%%%%%%%%%%%%%%%%%%%%%%%%%%%%%%%%%%%%%%%%  
% \begin{figure} 
%    \begin{center}
%     \centerline{\includegraphics*[width=0.33\textwidth]{uz_comp_cut_R360AR1.jpg}\put(-125,80){$(a)$}
%                 \includegraphics*[width=0.33\textwidth]{uz_comp_vert_R360AR3.jpg}\put(-125,80){$(b)$}
%                 \includegraphics*[width=0.33\textwidth]{uz_comp_horiz_R360AR3.jpg}\put(-125,80){$(c)$}}
%     \caption{Mean streamwise velocity of particles at \emph{(a)} the midplane of the square duct (vertical cut), \emph{(b)} the midplane of the rectangular duct (vertical cut) and \emph{(c)} the midplane of the rectangular duct (horizontal cut).}
%    \label{fig:figure8}     
% \end{center}        
% \end{figure}
% %%%%%%%%%%%%%%%%%%%%%%%%%%%%%%%%%%%%%%%%%%%%%%%%%%%%%%%%%%%%%%%%%%%

\subsubsection{Particle velocity fluctuations}\label{section:partFluctuations}
\begin{figure}
  \centerline{\includegraphics*[width=0.29\textwidth]{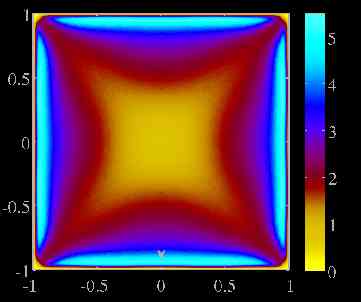}\put(-140,100){$(a)$}
              \includegraphics*[width=0.71\textwidth]{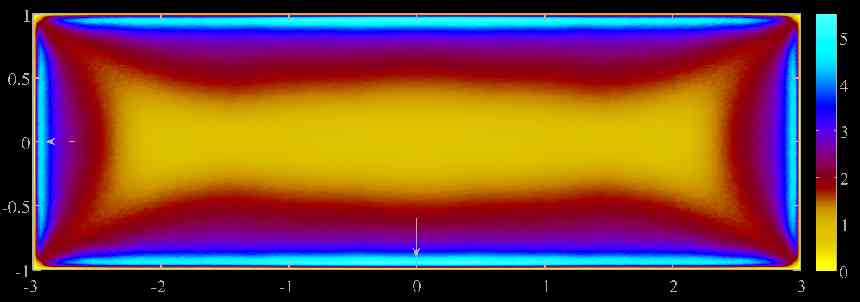}\put(-340,100){$(b)$}}
  \centerline{\includegraphics*[width=0.29\linewidth]{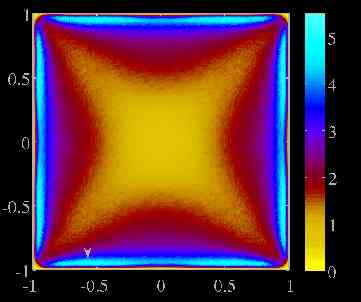}\put(-140,100){$(c)$}
              \includegraphics*[width=0.71\linewidth]{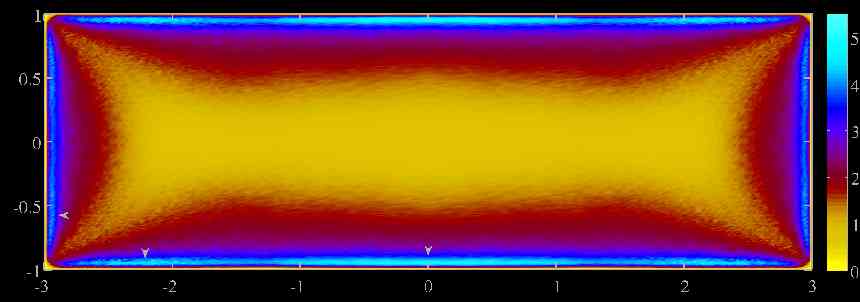}\put(-340,100){$(d)$}}
  \centerline{\includegraphics*[width=0.29\textwidth]{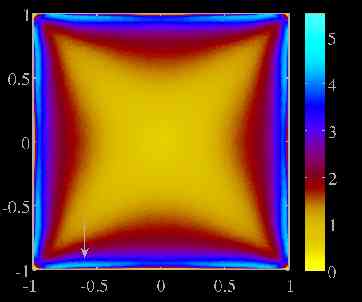}\put(-140,100){$(e)$}
              \includegraphics*[width=0.71\textwidth]{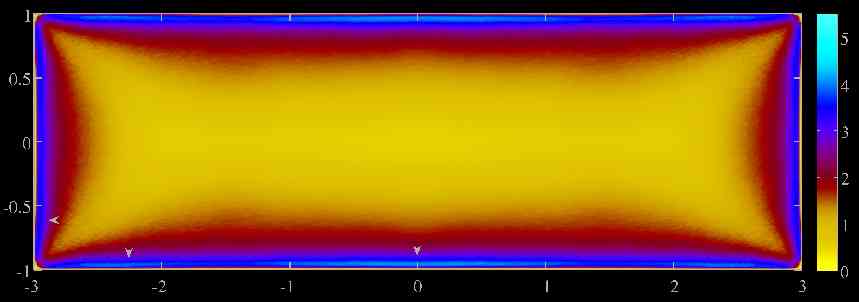}\put(-340,100){$(f)$}}
\caption{Pseudocolours of the turbulent kinetic energy (TKE) of the particulate phase normalised with the friction velocity $u_{\tau,c}^2$ of the fluid flow (at the midplane). \emph{From top:} \textbf{Stp5}, \textbf{Stp25}, \textbf{Stp100}. \textcolor{black}{ The positions of local TKE maxima are indicated by arrows.}}
\label{fig:fig12}
\end{figure}

The turbulent kinetic energy of the particle phase in the cross-section of the ducts is displayed in Fig.\ \ref{fig:fig12}. For the solid phase, the TKE is defined as $k_p=\langle v_{i_p}'v_{i_p}' \rangle/2$ where primes indicate particle fluctuating velocity. In the two ducts, the fluctuation energy $k$ decreases with increasing $St_b \geq$ \textbf{Stp5} particularly near the walls. This is similar to what was previously observed in turbulent canonical channels and pipes \citep{portela_etal_2002,picciotto_marchioli_soldati_2005}  and is caused by the damping effect of particle inertia that filters the fastest turbulent motions. %(check the results of straight pipe and compare to it, If it fits cite it!)
%%%%%%%%%%%%%%%%%%%%%%%%%%%%%%%%%%%%%%%%%%%%%%%%%%%%%%%%%%%%%%%%%%%%%%%%%%%%%%%%%%%%%%
%%%%%%%%%%%%%%%%%%%%%%%%%%%%%%%%%%%%%%%%%%%%%%%%%%%%%%%%%%%%%%%%%%%%%%%%%%%%%%%%%%%%%%

\begin{figure} 
   \begin{center}
    \centerline{\includegraphics*[width=0.43\textwidth]{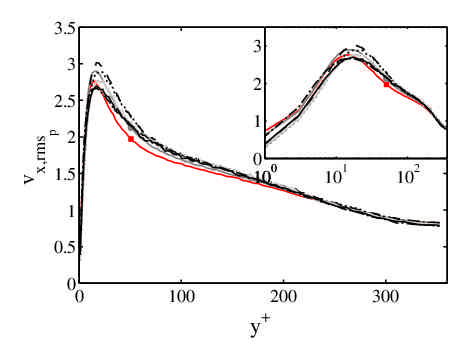}\put(-190,130){$(a)$}
                \includegraphics*[width=0.43\textwidth]{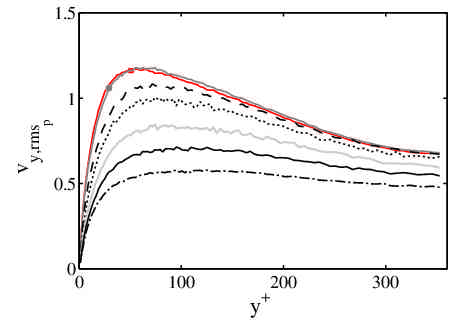}\put(-190,130){$(d)$}}
    \centerline{\includegraphics*[width=0.43\textwidth]{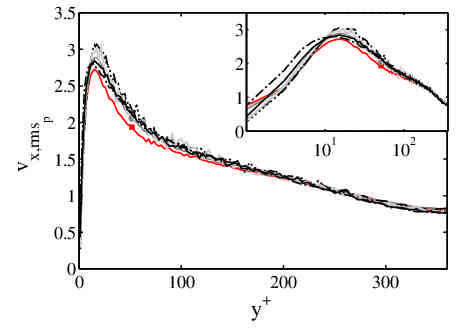}\put(-190,130){$(b)$}
                \includegraphics*[width=0.43\textwidth]{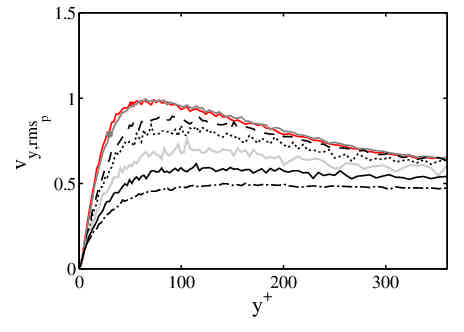}\put(-190,130){$(e)$}}
    \centerline{\includegraphics*[width=0.43\textwidth]{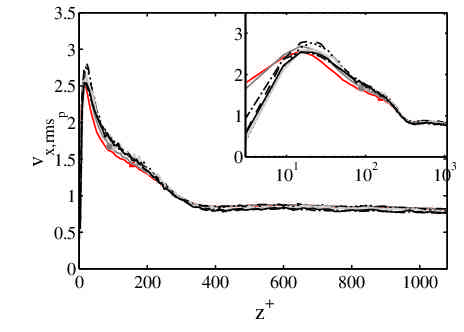}\put(-190,130){$(c)$}
                \includegraphics*[width=0.43\textwidth]{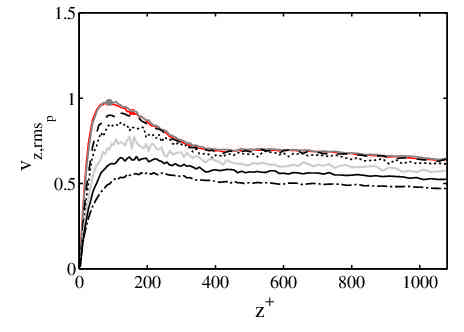}\put(-190,130){$(f)$}}
    \caption{Particle streamwise $r.m.s.$ velocity \emph{(left)} and wall-normal $r.m.s.$ velocity \emph{(right)} normalised with the flow friction velocity ($u_{\tau,c}^2$). \emph{(a,d)} in the horizontal/vertical plane of symmetry of the square duct; \emph{(b,e)} vertical plane of symmetry of the rectangular duct and \emph{(c,f)} horizontal plane of symmetry of rectangular duct.  The wall distance in inner units is presented in logarithmic scale in the plot insets.   
    {\color{red} \Gsolid ${\blacksquare}$ \Gsolid} \textbf{Stp0}, 
    {\color{dark-gray} \Gsolid $\bullet$ \Gsolid} \textbf{Stp1}, 
    \dashed \textbf{Stp5}, 
    \dotted \textbf{Stp10}, 
    {\color{light-gray} \solid} \textbf{Stp25}, 
    \solid \textbf{Stp50}, 
    \dotdashed  \textbf{Stp100}.}
   \label{fig:fig13}     
\end{center}        
\end{figure}

Further details on the individual components of the velocity fluctuations in the plane of symmetry of ducts are provided by the $r.m.s.$ of the streamwise and the wall normal velocity in Fig.\ \ref{fig:fig13}. \textcolor{black}{ The $r.m.s.$ value for a general particle characteristic ($\mathbb{X}_p$) in an individual Eulerian slab is computed from the following relation:
\begin{equation}
\mathbb{X}_{{rms}_p}^{2}=\frac{\mathfrak{N_{p,tot}} \displaystyle\sum_{j=1}^{n_t} \displaystyle\sum_{k=1}^{n_p(j)} \mathbb{X}^2_p(k,j) - \Big [ \displaystyle\sum_{j=1}^{n_t} \displaystyle\sum_{k=1}^{n_p(j)} \mathbb{X}_p(k,j) \Big ]^2}{\mathfrak{N_{p,tot}}(\mathfrak{N_{p,tot}-1})},
\end{equation}
where $\mathfrak{N_{p,tot}}$ is the total number of particles per slab for the total observation time, $n_t$ is the number of Lagrangian-field snapshots and $n_p$ is total number of particles per individual Eulerian slab for a specific time instance.} 

In both ducts, the streamwise particle velocity fluctuations, $v_{{x,rms}_p}$, are larger than those of the flow in the edge of outer layer ($y^+ (z^+) \approx 50$). This is similar to canonical particle-laden turbulent wall-bounded flows. As explained among others by \citet{portela_etal_2002} considering the mean fluid-velocity gradient, this larger $v_{{x,rms}_p}$ is due to the wall-normal motions between regions of high and low $\langle u_x \rangle$. A similar trend for $v_{{x,rms}_p}$ of different populations continues into the buffer layer of the rectangular duct (Fig.\ \ref{fig:fig13} \emph{b}), as the vicinity of the vertical symmetry plane is very similar to a canonical channel. However, the mean wall-normal flow towards the duct centreline alters the trend of $v_{{x,rms}_p}$ slightly as the results in the buffer layer of the square duct and the side wall of the rectangular duct are  different from the vertical symmetry plane of the rectangular geometry. In those cases (Fig.\ \ref{fig:fig13} \emph{a,c}) near the peak of the $v_{{x,rms}_p}$ the order with respect to different Stokes number, is altered and the peak for tracers (fluid particles) is situated in between the peaks associated with the solid phases. That is purely due to the outward weak secondary motion in this region. If the secondary motion is stronger, the effect of cross-flow motion will be further emphasised as is shown by \citet{noorani_etal_2015} (Fig.\ $8$ \emph{a,b} therein) where the flow peak is dominant in the inner bend position of the curved pipes.

For all cases, the wall-normal velocity fluctuations are lower than those of the fluid phase. Unlike in the canonical channel there is a mean wall-normal particle velocity in duct configurations towards the duct centre line in the plane of symmetry. This means the drag force due to the in-plane motion works against the turbophoretic drift towards the wall at these symmetry planes. 
%%%%%%%%%%%%%%%%%%%%%%%%%%%%%%%%%%%%%%%%%%%%%%%%%%%%%%%%%%%%%%%%%%%  
%%%%%%%%%%%%%%%%%%%%%%%%%%%%%%%%%%%%%%%%%%%%%%%%%%%%%%%%%%%%%%%%%%%

Another important and yet unexpected effect of the secondary motion on the dispersion of particulate phase can be deduced from the two-dimensional TKE map of particles in Fig.\ \ref{fig:fig12} and the profiles in Fig.\ \ref{fig:fig13}. It is readily seen that for lighter particles in the square duct and along the side wall of the rectangular geometry there is only one peak adjacent to the wall located on the plane of symmetry. This peak , however, disappears and the TKE is significantly reduced in the wall bisector with increasing particle inertia. In turn, two other peaks appear close to the corners in both cases. This process is different near the horizontal walls of the rectangular duct. The near-wall peak of TKE decreases half-way through the corner with increasing particle inertia. This generates two peaks close to the corner ($|z/h|\approx2.5$) while the one on the horizontal wall bisector endures.

\begin{figure}
   \begin{center}       
    \centerline{\includegraphics*[width=0.59\linewidth]{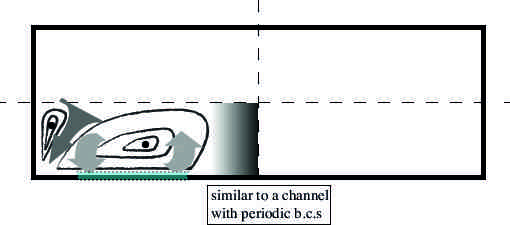}\put(-280,100){$(a)$}}
    \centerline{\includegraphics*[width=0.59\linewidth]{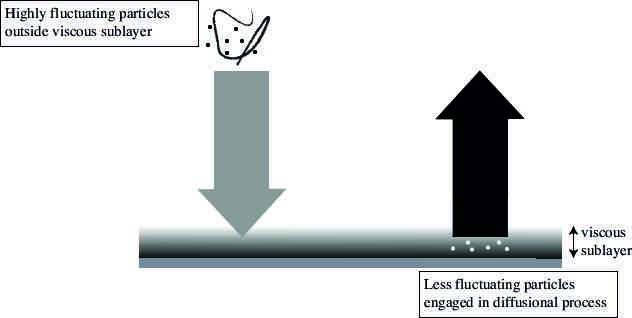}\put(-280,150){$(b)$}}  
    \caption{\emph{(a)} Schematic illustration of the secondary motion in a typical rectangular duct. The dark grey arrow represents the motion towards the corner while the light grey arrows show the wall-ward and outward transport of the inertial particles in the duct. The gradient grey area displays the region similar to spanwise-periodic channel where the influence of secondary cells is negligible. \emph{(b)} close-up view of the dotted region in \emph{(a)}. }
   \label{fig:fig15}       
\end{center}        
\end{figure}

The weak but resilient secondary motion in ducts can lead to the particle behaviour sketched in Fig.\ \ref{fig:fig15}. The streamlines adjacent to the corner bisectors may directly transport inertial particles towards the corners. However, the off-corner-bisector wall-ward streamlines push the particulate phase with large Stokes number (and thus long enough \emph{memory}) from a highly fluctuating region towards the quiescent region at the wall. Inertia prevents the particles to follow these curved streamlines and the particulate phase is inevitably centrifuged towards the wall viscous sublayer (grey arrow in \ref{fig:fig15} \emph{b}). The effect is larger for heavier particles as the peak appears closer to the corner for heavier populations (\emph{c.f.}\ Fig.\ \ref{fig:fig13}). On the  other hand, farther away from the corners, the less fluctuating particles engaged in a diffusive process deep in the viscous sublayer of the duct are extracted and directed towards the duct centre by the action of these outward (vertical) streamlines which in turn reduces the fluctuation value in the buffer layer (black arrow in Fig.\ \ref{fig:fig15} \emph{b}). This explains the maximum-minimum pattern in the velocity fluctuations away from the corners. In the rectangular duct (larger $AR$), however, the effect of the secondary cells near the horizontal wall bisector (vertical symmetry plane) of the duct is progressively attenuated and the state of the flow approaches to that of a spanwise-periodic channel. As a result, the TKE and velocity fluctuations in the buffer layer near the vertical symmetry plane increase again in the $AR=3$ duct, and consequently the peak of TKE is recovered in this region.
%%%%%%%%%%%%%%%%%%%%%%%%%%%%%%%%%%%%%%%%%%%%%%%%%%%%%%%%%%%%%%%%%%  
%%%%%%%%%%%%%%%%%%%%%%%%%%%%%%%%%%%%%%%%%%%%%%%%%%%%%%%%%%%%%%%%%
%%%%%%%%%%%%%%%%%%%%%%%%%%%%%%%%%%%%%%%%%%%%%%%%%%%%%%%%%%%%%%%%%%  
\begin{figure}[b!]
   \begin{center}       
    \centerline{\includegraphics*[width=0.5\linewidth]{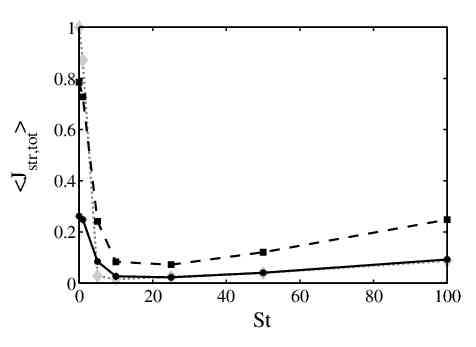}}
    \caption{Total mean axial particle flux in square duct (\dashed) and rectangular duct (\solid) and the straight pipe of \citet{noorani_etal_2015} (\dotted). Fluxes are normalised with the tracers' flux in the straight pipe.}
   \label{fig:fig15C}       
\end{center}        
\end{figure}
%\Gsolid ${\Box}$ \Gsolid $\kappa=0.00$; \Gsolid {\large{\Gcirc}} \Gsolid $\kappa=0.01$; \Gsolid \Diamond \Gsolid $\kappa=0.10$.
%%%%%%%%%%%%%%%%%%%%%%%%%%%%%%%%%%%%%%%%%%%%%%%%%%%%%%%%%%%%%%%%%

%\clearpage
\subsubsection{Dispersed phase flux}\label{sec:partFlux}
In order to further investigate transport phenomena in turbulent ducts, the particle mean streamwise flux $\langle J_{str,tot}\rangle$ is computed and compared to that in a circular duct (straight pipe), as illustrated in Fig.\ \ref{fig:fig15C}.  Although the Reynolds numbers of the two cases are slightly different, provided that the mass flux and characteristics of particles are the same, the values are comparable. The particle flux $J$ is defined as particle number density per unit volume multiplied by the corresponding velocities and normalised with the tracer flux in the straight pipe of \citet{noorani_etal_2015} It can be seen the $\langle J_{str,tot}\rangle$ of tracers in a square and in a rectangular duct are smaller than that in pipe purely due to the larger cross-sectional area of the former. For inertial particles when increasing the particle relaxation time, the axial particle flux first decreases drastically and then slowly recovers, independent of the geometry. This behaviour is similar to what is observed in bent pipes and is related to the efficiency of a population to remain a longer period in the viscous sublayer of the wall-bounded flow. More importantly, for \textbf{Stp10} particles and larger, the particle flux in the streamwise direction of the rectangular duct and pipe is almost identical while the flux in the square duct is larger. This is caused by the secondary motion that prevails over the whole cross-section of the square duct and transports the particles out from the wall towards regions with high streamwise velocity before sweeping them back to the near wall regions. During this process the particulate phase has time to speed up and the flux increases. The same mechanism exists also in the rectangular duct, though it is only significant near the corners. Considering the influence of the secondary cells is almost attenuated in the duct centre and in the vicinity of the horizontal wall bisector, the  turbophoresis is the only effect in this region and the flux of the particulate phase is reduced to approximately the level of a canonical pipe or channel.
%But I think in general there is a contact between the secondary region and channel like region and the channel like region should reach a steady state in receiving and giving the particles to the secondary region. This could be conjectured that the reason for so long time in reaching sssf is this.

%%%%%%%%%%%%%%%%%%%%%%%%%%%%%%%%%%%%%%%%%%%%%%%%%%%%%%%%%%%%%%%%%%%%%%%%%%%%%%%%%%%%%%
 \begin{figure}[b!]
   \centerline{\includegraphics*[width=0.29\textwidth]{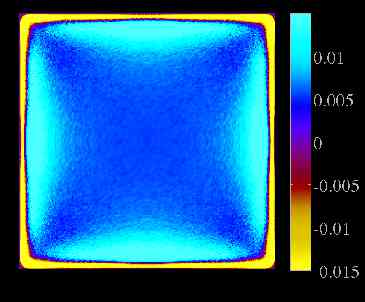}\put(-145,100){$(a)$}
               \includegraphics*[width=0.71\textwidth]{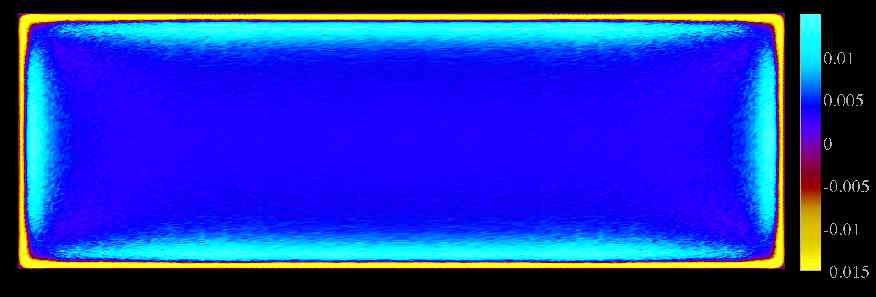}\put(-340,100){$(b)$}}
   \centerline{\includegraphics*[width=0.29\linewidth]{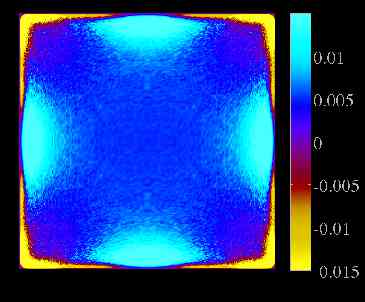}\put(-145,100){$(c)$}
               \includegraphics*[width=0.71\linewidth]{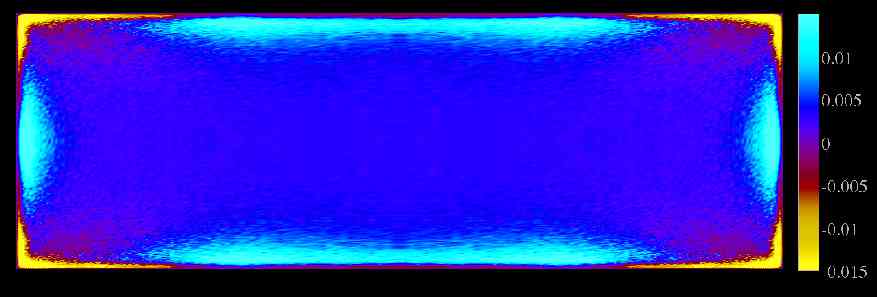}\put(-340,100){$(d)$}}
   \centerline{\includegraphics*[width=0.29\textwidth]{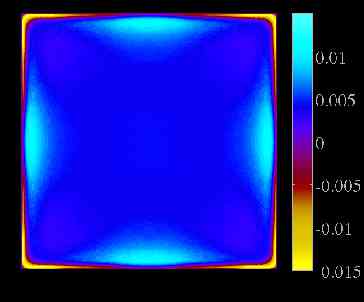}\put(-145,100){$(e)$}
               \includegraphics*[width=0.71\textwidth]{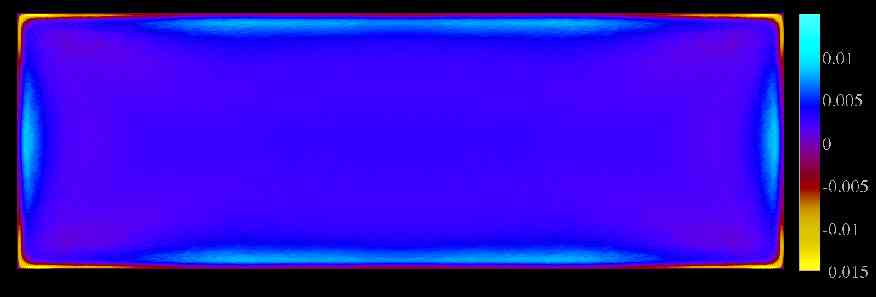}\put(-340,100){$(f)$}}
   \caption{Pseudocolours of mean streamwise particle acceleration $\langle a_{x_p} \rangle^{+}$ normalised with the $u_{\tau,c}^3/\nu$ value of the fluid flow in ducts. \emph{From top:} \textbf{Stp5}, \textbf{Stp25}, \textbf{Stp100}.}
 \label{fig:fig16}
 \end{figure}
 %%%%%%%%%%%%%%%%%%%%%%%%%%%%%%%%%%%%%%%%%%%%%%%%%%%%%%%%%%%%%%%%%%%%%%%%%%%%%%%%%%%%%%

 \subsubsection{Acceleration statistics}\label{sec:partAcceleration}
 From  the previous study by \citet{noorani_etal_2015} it is known that particle acceleration statistics are strongly affected by the secondary flow. An adequate description of the acceleration statistics is necessary for modelling the solid particulate dispersion in wall-bounded turbulent flows. The experimental and numerical studies of Gerashchenko et al.,\citep{gerashchenko_etal_2008} Yeo et al.,\citep{yeo_kim_lee_2010} \citet{lavezzo_etal_2010} and \citet{zamansky_2011} addressed the near-wall behaviour of the particle acceleration and its variance. From \citet{ayyalasomayajula_etal_2006} and Bec et al.,\citep{bec_etal_2006} it is known that in a particle-laden homogeneous isotropic turbulence (HIT)  the $r.m.s.$ of the acceleration decreases with increasing particle inertia. This is explained by \citet{bec_etal_2006} in terms of two concurrent mechanisms: the preferential concentration of inertial particles and their damping of the high-frequency turbulent fluctuations. While the filtering effect is more effective for larger particle inertia, the former mechanism (particle centrifugation out of intense vortices) is more important at lower $St$. According to \citet{eaton_fessler_1994} the preferential concentration can affect the particle acceleration variance by increasing the residence time of the particle in the high-strain (low-vorticity) regions.

 %%%%%%%%%%%%%%%%%%%%%%%%%%%%%%%%%%%%%%%%%%%%%%%%%%%%%%%%%%%%%%%%%%%  
 \begin{figure} 
    \centerline{\includegraphics*[width=0.5\textwidth]{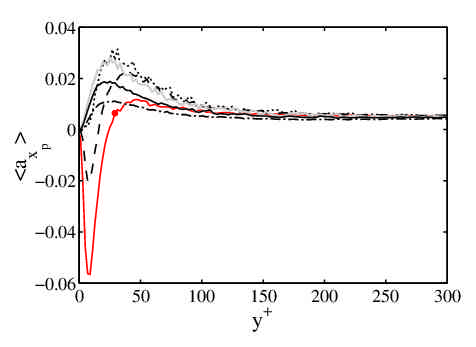}\put(-230,150){$(a)$}
                \includegraphics*[width=0.5\textwidth]{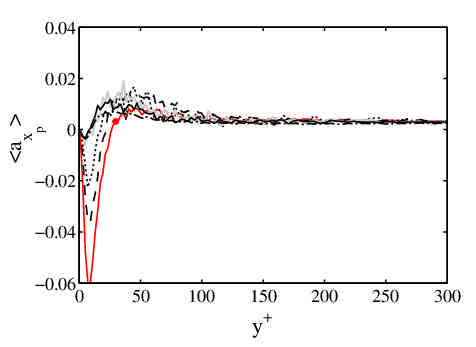}\put(-230,150){$(b)$}}
    \centerline{\includegraphics*[width=0.5\textwidth]{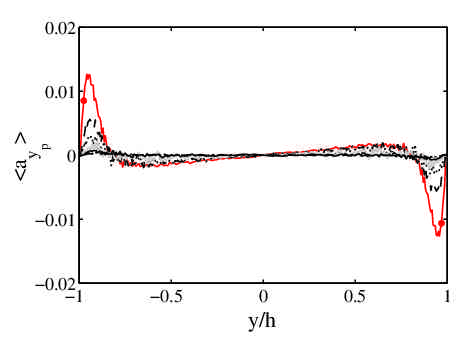}\put(-230,150){$(c)$}
                \includegraphics*[width=0.5\textwidth]{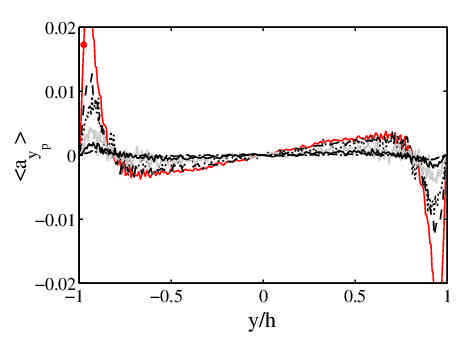}\put(-230,150){$(d)$}}
    \caption{Wall-normal distribution of \emph{(top)} mean streamwise particle acceleration; \emph{(bottom)} mean wall-normal particle acceleration in \emph{(a,c)} the square duct symmetry plane \emph{(b,d)} the vertical symmetry plane of rectangular duct. The profiles are normalised with the $u_{\tau,c}^3/\nu$ values of the fluid flow.
     {\color{red} \Gsolid $\bullet$ \Gsolid} \textbf{Stp1}, 
     \dashed \textbf{Stp5}, 
     \dotted \textbf{Stp10}, 
     {\color{light-gray} \solid} \textbf{Stp25}, 
     \solid \textbf{Stp50}, 
     \dotdashed  \textbf{Stp100}.}
 \label{fig:fig17}
 \end{figure}
 %%%%%%%%%%%%%%%%%%%%%%%%%%%%%%%%%%%%%%%%%%%%%%%%%%%%%%%%%%%%%%%%%%%
 % %%%%%%%%%%%%%%%%%%%%%%%%%%%%%%%%%%%%%%%%%%%%%%%%%%%%%%%%%%%%%%%%%%%%%%%%%%%%%%%%%%%%%%
 % \begin{figure}
 %  \centerline{\includegraphics*[width=0.28\textwidth]{St5_apinPlane_2DR360AR1.jpg}\put(-120,80){$(a)$}
 %              \includegraphics*[width=0.71\textwidth]{St5_apinPlane_2DR360AR3.jpg}\put(-280,80){$(b)$}}
 %  \centerline{\includegraphics*[width=0.28\linewidth]{St25_apinPlane_2DR360AR1.jpg}\put(-120,80){$(c)$}
 %              \includegraphics*[width=0.71\linewidth]{St25_apinPlane_2DR360AR3.jpg}\put(-280,80){$(d)$}}
 %  \centerline{\includegraphics*[width=0.28\textwidth]{St100_apinPlane_2DR360AR1.jpg}\put(-120,80){$(e)$}
 %              \includegraphics*[width=0.71\textwidth]{St100_apinPlane_2DR360AR3.jpg}\put(-280,80){$(f)$}}
 %  \caption{Pseudocolours of mean in-plane particle acceleration ($\sqrt(\langle a_{y_p} \rangle ^2+\langle a_{z_p}\rangle ^2)$) normalised with ${\overline{u_{\tau}}}^3/\nu$ of the fluid flow in both configurations. \emph{From top:} \textbf{Stp5}, \textbf{Stp25}, \textbf{Stp100}.}
 % \label{fig:fig17}
 % \end{figure}
 % %%%%%%%%%%%%%%%%%%%%%%%%%%%%%%%%%%%%%%%%%%%%%%%%%%%%%%%%%%%%%%%%%%%%%%%%%%%%%%%%%%%%%%
 
 The two-dimensional map of the mean streamwise particle acceleration $\langle a_{x_p} \rangle$ in ducts is illustrated in Fig.\ \ref{fig:fig16}. To offer a quantitative comparison of the streamwise and wall-normal particle acceleration between different populations the profiles along the wall bisector of the square duct and the vertical plane of symmetry of the rectangular geometry are also shown in Fig.\ \ref{fig:fig17}. Except in the near-wall regions, $\langle a_{x_p} \rangle$ is almost uniform across the duct section and takes similar values for the various populations. \textcolor{black}{ In the rectangular duct, the $\langle a_{x_p} \rangle$ profile reaches a minimum in all the populations within the buffer layer, at approximately $y^{+} \approx 10$. Farther from the wall, the various populations also reach a maximum at approximately the same location, $y^{+} \approx 40$, with values of around 0.01 in all the cases. Regarding the square duct, only the two lightest populations (\textbf{Stp1} and \textbf{Stp5}) exhibit a minimum within the buffer layer. Interestingly, this minimum is also located at $y^{+} \approx 10$, and whereas the minimum of the \textbf{Stp1} population takes approximately the same value of $\approx -0.06$ in the $AR=1$ and 3 cases, the minimum from the \textbf{Stp5} population is less pronounced in the square duct ($\approx -0.02$ instead of $\approx -0.035$). The heavier populations do not exhibit a minimum in the buffer layer, and show different maximum values ranging from around 0.02 to 0.03; note that these maxima are located between $y^{+} \simeq 25$ and 40, revealing a different acceleration behaviour in the square duct compared to the rectangular case. The trend observed in the latter is identical to the behaviour observed in canonical channel flows} (see \emph{e.g.}\ \citet{zamansky_2011}) where this deceleration is explained in terms of the dissipative nature of the Stokes drag. Although along the wall bisector of the square duct the lighter populations have a similar trend, the negative peak near the wall is significantly reduced for heavier particles while the maximum is doubled in magnitude. This is due to the anti-alignment of the flow shear and outward secondary motion of the carrier phase and is almost identical to what is observed in the inner bend of particle-laden mildly curved pipes (\emph{c.f.}\ \citet{noorani_etal_2015}). This mean wall-normal motion also affects $\langle a_{y_p} \rangle$ such that its near-wall peak in the wall bisector of the square duct  is almost half of that in the horizontal wall bisector of the rectangular duct (\emph{c.f.}\ Fig.\ \ref{fig:fig17} \emph{c,d}). 
 
A qualitative effect of the mean in-plane motion on the two-dimensional map of $\langle a_{x_p} \rangle$ can be observed in Fig.\ \ref{fig:fig16} where the stripe of the large positive values of $\langle a_{x_p} \rangle$ distributed over the walls of the square duct and the side wall of rectangular geometry are further localised around the wall bisectors and protruded towards the wall with increasing particle inertia. \textcolor{black}{ Close to the wall bisectors the wall-normal secondary motion transports the particles outward from the wall towards regions with high streamwise velocity. This velocity difference between the slow moving particles and relatively faster fluid causes the large positive values of $\langle a_{x_p} \rangle$. Comparison of Figs.\ \ref{fig:fig16} and  \ref{fig:fig11b} reveals that the heavier particles are affected more by the in-plane flow, particularly where the wall-normal secondary motion is stronger.} The single stripe of maximum acceleration for lighter particles near the horizontal walls of the rectangular duct is divided into two parts with increasing particle Stokes number. This behaviour is in excellent agreement with the mechanism previously mentioned in \S \ref{section:partFluctuations}. In Fig.\ \ref{fig:fig16}, the near-wall positions where the $\langle a_{x_p} \rangle$ is localised coincides exactly with the locations where the turbulent kinetic energy of particles (Fig.\ \ref{fig:fig12}) decreases. 
 
\begin{figure} 
    \centerline{\includegraphics*[width=0.4\textwidth]{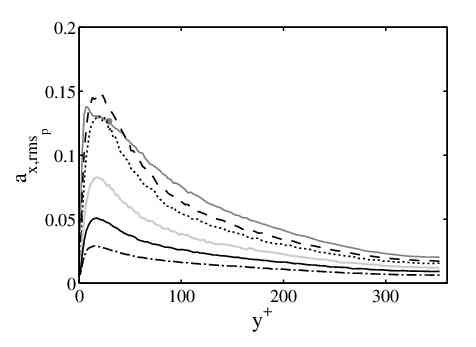}\put(-185,115){$(a)$}
                \includegraphics*[width=0.4\textwidth]{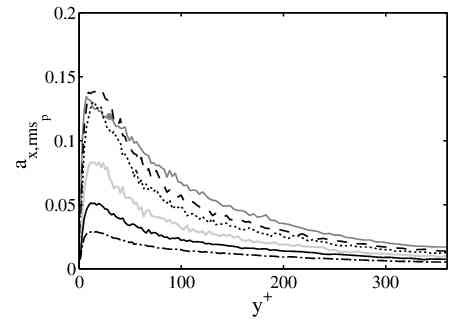}\put(-185,115){$(b)$}}
    \centerline{\includegraphics*[width=0.4\textwidth]{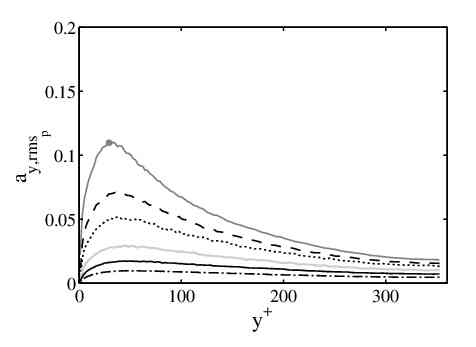}\put(-185,115){$(c)$}
                \includegraphics*[width=0.4\textwidth]{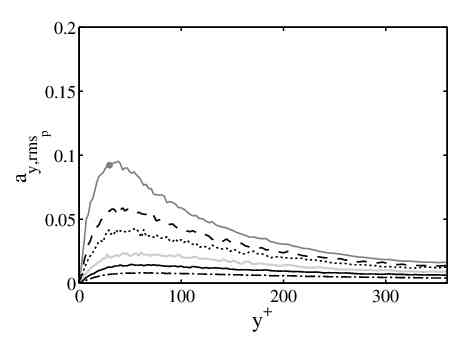}\put(-185,115){$(d)$}}
    \caption{Wall-normal distribution of particle acceleration $r.m.s.$ \emph{(top)} streamwise component; \emph{(bottom)} wall-normal component, in \emph{(a,c)} the square duct wall bisector and \emph{(b,d)} the vertical symmetry plane of rectangular duct. The profiles are normalised with the $u_{\tau,c}^3/\nu$ of the fluid flow. Same symbols as Fig.\ \ref{fig:fig13}.}
 \label{fig:fig18}
 \end{figure}

As reported for canonical particle-laden wall-bounded turbulent flows, the streamwise component of the acceleration variance reaches a maximum near the wall. This peak increases with the Stokes number, up to $St^+ \sim 5$, before decreasing with further increasing particle inertia.\citep{gerashchenko_etal_2008,zamansky_2011} The wall-normal acceleration $r.m.s.$, however, is reported to decrease with increasing $St$, and for all populations its value remains lower in magnitude compared to the streamwise component. To investigate this behaviour, the streamwise ($a_{{x,rms}_p}$) and wall-normal ($a_{{y,rms}_p}$) components of the acceleration variance are shown in the plane of symmetry of the square duct and vertical symmetry plane of $AR=3$ duct in Fig.\ \ref{fig:fig18}. The results in both configurations show the same \emph{trend} observed in straight channels;\citep{zamansky_2011} in addition, the magnitude of the peak values is directly comparable to the inner bend position of the curved pipes by \citet{noorani_etal_2015} (Fig.\ $13$ therein). This, in fact, shows that with increasing outward secondary motion the peak values will decrease significantly.
 %%%%%%%%%%%%%%%%%%%%%%%%%%%%%%%%%%%%%%%%%%%%%%%%%%%%%%%%%%%%%%%%%%%  
 %%%%%%%%%%%%%%%%%%%%%%%%%%%%%%%%%%%%%%%%%%%%%%%%%%%%%%%%%%%%%%%%%%%

\section{Conclusions} \label{sec:conclusions}

The aim of the present study is to determine and characterise the dispersion of dilute inertial heavy micro-particles in turbulent duct flow with width-to-height aspect ratios $AR$ of 1 and 3. The carrier phase is simulated in duct geometries and a spanwise-periodic channel with well-resolved DNSs in sufficiently long domains while seven populations of Lagrangian spherical particles, with inner-scaled Stokes numbers $St^{+}$ ranging from 0 (fluid tracers) to 100 (very heavy inertial particles), are tracked using one-way coupling. The main goal was to evaluate the effect of turbulence-driven secondary motion on the particle dynamics by computing the Eulerian statistics of particle data. 

From the mean in-plane velocity of the carrier phase it becomes obvious that the turbulent field at the centreplane of the rectangular duct approaches a spanwise-periodic canonical channel flow.  As a consequence, the Eulerian particle statistics in the rectangular duct significantly differ from the ones observed in the square duct. Note that the secondary motion vortices, in general, alter the cross-sectional map of the particle mean and fluctuation data, since they have a significant impact on the carrier phase through momentum transport from the duct centreplane towards the corner bisectors. Moreover, the secondary flow exhibits significantly different topologies in the $AR=1$ and 3 cases, where in the rectangular duct the vortex on the horizontal wall significantly expands in the horizontal direction. A manifestation of the convergence towards spanwise-periodic channel conditions can be observed in the buffer layer particle concentration at the duct centreplane, which is up to 3 times larger in the square duct than in the $AR=3$ case. On the other hand, the mean normalised particle concentration at the viscous sublayer along the square duct walls show that the largest concentration is in the duct centreplane, except for the heaviest populations. Note that another peak for particle accumulation is also found in the corner of the duct. Interestingly, the largest concentration at the horizontal plates of the rectangular duct did not occur on the plane of symmetry but at \textcolor{black}{$|z/h|\approx 1.25$} from the centreplane. In addition, the secondary cells change the in-plane map of the particle velocities significantly in comparison to fluid flow statistics. These directly affected the particle velocity fluctuations such that multiple peaks near the duct walls appear for the particle streamwise and wall-normal velocity fluctuations. We also observe an attenuation of the particle turbulent kinetic energy maxima at the buffer layer for progressively heavier populations: this value decreased by around $18\%$ from the \textbf{Stp5} to the \textbf{Stp100} populations. This effect, which can be explained by the action of the secondary motion on the solid phase velocity fluctuations, highlights the fact that the various particle populations respond differently to the secondary flow. The particle mean acceleration and its variance are also reported, and our results indicate that both decrease with increasing particle inertia.

The database of particle phase statistics in turbulent flow through square and rectangular ducts reported in the present study constitute an interesting platform to study the effect of turbulence-induced secondary flows on particle motions, especially when compared with similar data in bent pipes,\citep{noorani_etal_2014} where stronger skew-induced secondary motions are present.  The different natures of the two types of secondary flows, as well as the effect of the geometry, are of great relevance to achieve a deeper understanding in the fundamenal mechanisms present in particle-laden wall-bounded turbulent flows. Moreover, this comparison may facilitate modelling advancement for dispersed multiphase turbulent flow in wall-bounded flows with cross-flow motion.

\vspace{10pt}

\section*{Acknowledgments}
The authors gratefully acknowledge computer-time allocation from the Swedish National Infrastructure for Computing (SNIC). This research is also supported by the ERC grant `2013-CoG-616186, TRITOS' to L.B.

% %%%%%%%%%%%%%%%%%%%%%%%%%%%%%%%%%%%%%%%%%%%%%%%%%%%%%%%%%%%%%%%%%%%%%%%%%%%%%%%%%%%%%%
% \begin{figure}
%   \centerline{\includegraphics*[width=9.6cm]{./plots/machinary.jpg}}
%   \caption{The total (space--time averaged) mean and turbulent kinetic energy budget (MKE and TKE, respectively). Integral units ($R_a$, $U_b$) are used to non-dimensionalise the energy fluxes. $\mathcal{P}$ represents the total turbulent production. \highlight{The letters $\mathcal{DD}$ etc. are not explained in the text, and other terms are used $\varepsilon_{d,non-St}$. What should we use?}}
% \label{fig:fig6}
% \end{figure}
% %%%%%%%%%%%%%%%%%%%%%%%%%%%%%%%%%%%%%%%%%%%%%%%%%%%%%%%%%%%%%%%%%%%%%%%%%%%%%%%%%%%%%%
%%%%%%%%%%%%%%%%%%%%%%%%%%%%%%%%%%%%%%%%%%%%%%%%%%%%%%%%%%%%%%%%%%%%%%%%%%%%%%%%%%%%%%
% \begin{table}[bh!]
% \begin{tabular}{llllllllll}
% $L$&$T_{stat}$&$Re_\tau$&$f_{\langle \tau_{w} \rangle}$&$\epsilon_{mean}$&$\epsilon_{fluct}$&$\epsilon_{tot}$&$P$&$T_{dp/ds}$&$\langle dp/ds \rangle_{s,t}$\\[3pt]
% \hline
% 100 & 600     & 120.5 & 0.01005 & −0.008608 & −0.0014087 & −0.010017 & 0.0014139 & 600  & 0.01004\\[3pt]     
% \end{tabular}
% \end{table}
% %%%%%%%%%%%%%%%%%%%%%%%%%%%%%%%%%%%%%%%%%%%%%%%%%%%%%%%%%%%%%%%%%%%%%%%%%%%%%%%%%%%%%%

\bibliographystyle{apsper}

%\bibliography{ductPartPof}

\end{document}